\documentclass[12pt,english]{scrartcl}
\usepackage[T1]{fontenc}
\usepackage[latin1]{inputenc}
\usepackage{amsmath}
\usepackage{graphicx}
\usepackage{amssymb}

\makeatletter

\providecommand{\tabularnewline}{\\}

\usepackage{graphicx}
\usepackage{amsmath, amssymb, amsfonts}
\usepackage{bm}
\usepackage{graphicx}
\newif \ifpdf
\ifx \pdfoutput \undefined
      \pdffalse
   \else
      \pdftrue
\fi
\ifpdf

\usepackage{helvet}
\usepackage{ae,aecompl}
\usepackage[pdftex,urlcolor=blue,colorlinks=true]{hyperref}
\else
\usepackage[ps2pdf,urlcolor=blue,colorlinks=true]{hyperref}
\fi

\makeatother
\begin{document}

\title{Methods for 3-D vector microcavity problems involving a planar dielectric
mirror}

\author{David H. Foster and Jens U. Nöckel}

\date{
Department of Physics\\
University of Oregon\\
Eugene, OR 97403\\
\url{http://darkwing.uoregon.edu/~noeckel}
}

\publishers{{\em Published in} Opt. Commun. {\bf 234}, 351-383 (2004)}
\maketitle

\begin{abstract}
We develop and demonstrate two numerical methods for solving the class
of open cavity problems which involve a curved, cylindrically symmetric
conducting mirror facing a planar dielectric stack. Such dome-shaped
cavities are useful due to their tight focusing of light onto the
flat surface. The first method uses the Bessel wave basis. From this
method evolves a two-basis method, which ultimately uses a multipole
basis. Each method is developed for both the scalar field and the
electromagnetic vector field and explicit ``end user'' formulas are
given. All of these methods characterize the arbitrary dielectric
stack mirror entirely by its $2\times2$ transfer matrices for s-
and p-polarization. We explain both theoretical and practical limitations
to our method. Non-trivial demonstrations are gi ven, including one
of a stack-induced effect (the mixing of near-degenerate Laguerre-Gaussian
modes) that may persist arbitrarily far into the paraxial limit. Cavities
as large as 50$\lambda$ are treated, far exceeding any vectorial
solutions previously reported.
\end{abstract}
\tableofcontents{}

\section{\label{sec: intro}Introduction}

There is currently considerable interest in the nature of electromagnetic
(vector) modes both for free space propagation \cite{Hall,Hall2,Sheppard,Dorn}
and in cavity resonators \cite{Jens}. In particular, recent advances
in fabrication technology have given rise to optical cavities which
cannot be modeled by effectively two-dimensional, scalar or pseudo-vectorial
wave equations \cite{MEPchapter}. The resulting modes may exhibit
non-paraxial structure and nontrivial polarization, but this added
complexity also gives rise to desirable effects; an example from free-space
optics is the observation of enhanced focusing for radially polarized
beams \cite{Dorn}. This effect is shown here to arise in cavities
as well, among a rich variety of other modes that depend on the three-dimensional
geometry. The main goal of this paper is to present a set of numerical
techniques adapted to a realistic cavity design as described below.

Much work involving optical cavity resonators utilizes mirrors that
are composed of thin layers of dielectric material. These dielectric
stack mirrors offer both high reflectivity and a low ratio of loss
to transmission, which are desirable in many applications. The simplest
model of such cavities treats the mirrors as perfect conductors ($\bm{E}_{\text{tangential}}=\bm{0},H_{\text{normal}}=0$).
In applications involving paraxial modes and highly reflective mirrors,
it is often acceptable to use this treatment. The mature theory of
Gaussian modes (c.f. \!Siegman \cite{Siegman}) is applicable for
this class of cavity resonator. When an application requires going
beyond the paraxial approximation to describe the optical modes of
interest, the problem becomes significantly more involved and modeling
dielectric stack mirrors as conducting mirrors may become a poor approximation.
Also, one is often interested in the field inside the dielectric stack
and for this reason must include the stack structure into the problem.

In this paper we present a group of improved methods for resonators
with a cylindrically symmetric curved mirror facing a planar mirror.
The paraxial condition is not necessary for these methods; tightly
focused modes can be studied. Furthermore, the true vector electromagnetic
field is used, rather than a scalar field approximation. The planar
mirror is treated as infinite and is characterized by its polarization-dependent
reflection functions $r_{s}(\theta_{\text{inc}})$ and $r_{p}(\theta_{\text{inc}})$
for plane waves with incident angle $\theta_{\text{inc}}$. Our model
encompasses both cavities for which the planar mirror is an arbitrary
dielectric stack, and cavities for which the planar mirror is a simple
mirror (conducting with $r_{s/p}(\theta_{\text{inc}})=-1$ or {}``free''
with $r_{s/p}(\theta_{\text{inc}})=+1$). The opposing curved mirror
is always treated as a conductor. It should be noted that, for most
modes which are highly focused at the planar mirror, (modes which
are likely to be of interest in applications), this limitation can
be expected to cause little error because the local wave fronts at
the curved mirror are mostly perpendicular to its surface. (Of course,
for applications in which the curved mirror is indeed conducting,
our model is very well suited.) On the other hand, the correct treatment
of the \emph{planar} mirror can be a great improvement over the simplest
model. Applications with both dielectric and conducting curved mirrors
have been and are currently being used experimentally \cite{Jens,Raymer,Meissner}.

The methods described here belong to a class of methods which we refer
to as {}``basis expansion methods''. In basis expansion methods,
a complete, orthogonal basis (such as the basis of electromagnetic
plane waves) is chosen. Each basis function itself obeys Maxwell's
equations. The equations that determine the correct value of the basis
coefficients are boundary equations, resulting from matching appropriate
fields at dielectric interfaces, setting appropriate field components
to zero at conductor-dielectric interfaces, and setting certain fields
to be zero at the origin or infinity. In the usual application of
this method, each homogenous dielectric region is allocated its own
set of basis coefficients. Our methods use a single set of basis coefficients;
the matching between dielectric layers is handled by the $2\times2$
transfer matrices of the stack.

As dielectric stacks have nonzero transmission, optical cavities with
this type of mirror are necessarily open, or lossy. The methods described
here deal with the openness due to the stack and the solutions are
quasimodes, with discrete, isolated complex wavenumbers which denote
both the optimal driving frequency and the resonance width%
\footnote{For many modes, there is also loss due to lateral escape from the
sides of the cavity. While our model intrinsically incorporates the
openness due to lateral escape in the calculation of the fields (by
simply not closing the curved mirror surface, or extending its edge
into the dielectric stack), this loss is not included in the calculated
resonance width or quality factor, Q. Because a single set of basis
vectors is used to describe the field in the half-plane above the
planar mirror, this entire half-plane is the {}``cavity'' as far
as the calculation of resonance width is concerned.%
}. While the dielectric mirror is partially responsible for the openness
of our model system, the openness is not primarily responsible for
mode pattern changes resulting from replacing a dielectric stack mirror
with a simple mirror. The phase shifts of plane waves reflected off
a dielectric stack can vary with incident angle, and it is this variation
which can cause significant changes in the modes, even though reflectivities
may be greater than $0.99$. Generally speaking, the deviation of
$|r_{s/p}(\theta_{\text{inc}})|$ from 1 is not as important as the
deviation of $\arg(r_{s/p}(\theta_{\text{inc}}))$ from, say, $\arg(r_{s/p}(0))$.

We develop two general methods, the two-basis method and the Bessel
wave method. The scalar field versions of both methods are also developed
and are discussed first, acting as pedagogical stepping stones to
the vector field versions. The Bessel wave method uses the Bessel
wave basis which is the cylindrically symmetric version of the plane
wave basis. This method is described in Section \ref{sec: PWB}. The
two-basis method ultimately uses the vector or scalar multipole basis.
The multipole basis has an advantage in that it is the eigenbasis
of a conducting hemisphere, the {}``canonical'' dome-shaped cavity.
The unusual aspect of the two-basis method is the intermediate use
of the Bessel wave basis. The two-basis method is developed in Section
\ref{sec: MB}. We have implemented both methods and have used them
as numerical checks against each other. Various demonstrations and
comparisons are given in Section \ref{sec: demonstrations}. Our implementations
of all methods are programmed in C++, use the GSL, LAPACK, SLATEC,
and PGPlot numerical libraries, and run on a Macintosh G4 with OS
X. Limitations of our model and methods are discussed in Appendix
\ref{sec: model limitations}. Appendix \ref{sec: modetype} discusses
plotting modes that are associated with linear polarization and Appendix
\ref{sec: stacks} specifies dielectric stacks that are used in Section
\ref{sec: demonstrations}.

\section{\label{sec: model overview}Overview of the Model and Notation}

\begin{figure}[!htb]
\begin{center}\includegraphics[%
  width=0.75\columnwidth]{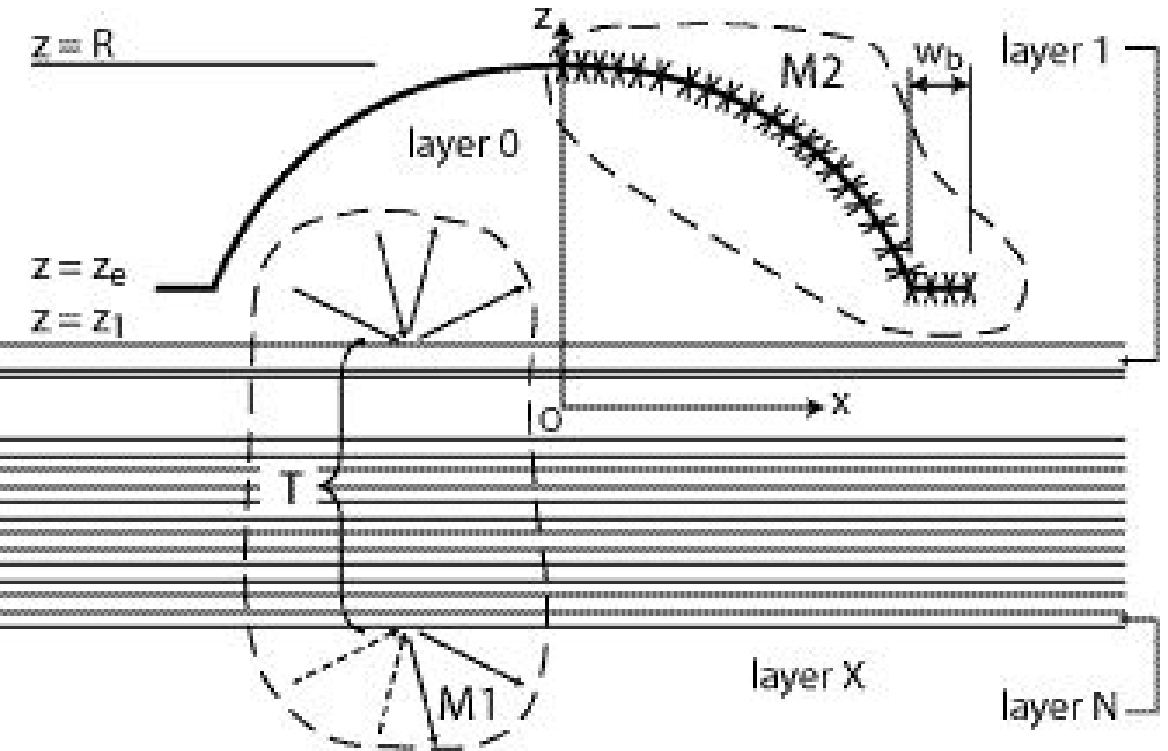}\hspace*{-1cm}\end{center}

\caption{The cavity model.}

\label{fig: model}
\end{figure}

A diagram of the model is shown in Figure \ref{fig: model}. The conducting
surface is indicated by the heavy line. The annular portion of this
surface extending horizontally from the dome edge will be referred
to as the {}``hat brim''. The dome is cylindrically symmetric with
maximum height $z=R$ and edge height $z=z_{e}$. The shape of the
dome is arbitrary, but in our demonstrations the dome will be a part
of an origin-centered sphere of radius $R_{s}=R$ unless otherwise
specified. The region surrounding the curved mirror will be referred
to as layer $0$. The dielectric interface between layer $0$ and
layer $1$ has height $z=z_{1}$. The last layer of the dielectric
stack is layer $N$ and the exit layer is called layer $X$. The depiction
of the stack layers in the figure suggests a design in which the stack
consists of some layers of experimental interest (perhaps containing
quantum wells, dots, or other structures \cite{Jens,Raymer}) at the
top of the stack where the field intensity is high, and a highly reflective
periodic structure below.

At the heart of the procedure to solve for the quasimodes is an overdetermined,
complex linear system of equations, $A\bm{y}=\bm{b}$. The column
vector $\bm{y}$ is made up of the coefficients of eigenmodes in some
basis B. The field in layer $0$ is given by expansion in B using
these coefficients. For a given wavenumber, $k$, a solution vector
$\bm{y}=\bm{y}_{\text{best}}$ can be found so that $|A(k)\,\bm{y}-\bm{b}|^{2}$
is minimized with respect to $\bm{y}$. Dips in the graph of the residual
quantity, $\Delta_{r}\equiv|A(k)\,\bm{y}_{\text{best}}-\bm{b}|$,
versus $k$ signify the locations of the isolated eigenvalues of $k$
(theoretically $\Delta_{r}$ should become 0 at the eigenvalues).
The solution vector $\bm{y}_{\text{best}}(k)$ at one of these eigenvalues
describes a quasimode. The system of equations is made up of three
parts (as shown below): M1 equations, M2 equations and an arbitrary
amplitude or {}``seed'' equation. \begin{align}
A\,\bm{y}=\left[\begin{matrix}\left[\begin{matrix}\\ & M1\\
\end{matrix}\right]\\
\left[\begin{matrix}\\ & M2\\
\end{matrix}\right]\\
\left[\begin{matrix}\text{s. eqn.}\end{matrix}\right]\end{matrix}\right]\bm{\cdot}\left[\begin{matrix}\\\bm{y}\\
\\\end{matrix}\right]=\left[\begin{matrix}0\\
\vdots\\
\\\vdots\\
0\\
1\end{matrix}\right].\label{eqn:Axb}\end{align}
 Henceforth M1 refers to the planar mirror and M2 to the curved mirror.

The M1 boundary condition for a plane wave basis is expressed simply
in terms of the $2\times2$ stack transfer matrices $T_{s}(\theta_{\text{inc}})$
and $T_{p}(\theta_{\text{inc}})$, as suggested by the M1 region (enclosed
by the dashed line) in Fig. \!\ref{fig: model}. In the scalar and
vector multipole bases, a sort of conversion to plane waves is required
as an intermediate step. The dashed $\bm{k}$ vectors in the figure
(incoming from the bottom of the stack) represent plane waves that
are given zero amplitude, in order to define a quasimode problem rather
than a scattering problem. The plane waves denoted by the solid $\bm{k}$
vectors have nonzero amplitude.

The M2 boundary condition is implemented as follows. A number of locations
on the curved mirror are chosen (the {}``X'' marks in Fig. \!\ref{fig: model}).
The width of the hat brim is $w_{b}$ as shown. An {}``infinitesimal''
hat brim ($w_{b}\ll\lambda$) is introduced to give the dome a diffractive
edge. An {}``infinite'' hat brim can be introduced theoretically
and can make the model more easily understandable in certain respects.
More about the model in relation to the hat brim is discussed in Appendix
\ref{sec: hat brim}. The M2 equations are the equations in basis
B setting the appropriate fields at these locations to zero. For a
problem not possessing cylindrical symmetry, these locations would
be points. The simplification due to this symmetry, however, allows
these locations to be entire rings about the $z$ axis, specified
by a single parameter such as the $\rho$ coordinate. Finally, the
seed equation sets some combination of basis coefficients equal to
one and is the only equation with a nonzero value on the right hand
side ($\bm{b}$).

The cylindrical symmetry of the boundary conditions allows one to
always find solutions which have a $\phi$ dependence of $\exp(\imath m\phi)$,
where $m$ is an integer. This in turn leads to a dimensionally reduced
version of the plane wave basis called the Bessel wave basis, in which
each basis function is a superposition of all the plane waves with
the same wavevector polar angle, $\theta_{k}$. The weight function
of the superposition is proportional to $\exp(\imath m\phi_{k})$.
We will refer to the non-reduced basis as the {}``simple plane wave
basis''. The unadorned phrase {}``plane wave basis'' (PWB; same
abbreviation for plural) will refer henceforth to either or both of
the Bessel wave and simple plane wave bases. When using the scalar
or vector multipole basis (MB; same abbreviation for plural), cylindrical
symmetry allows the problem to be solved separately for each quantum
number $m$ of interest%
\footnote{The modes with low $|m|$ are likely to be of practical interest since
they have the simplest transverse polarization structure. The $|m|=1$
family of vector eigenmodes is exceptional because the proportionality
of $E_{\rho}$ and $E_{\phi}$ to $\exp(\imath m\phi)$ means that
modes for $|m|\neq1$ have no average transverse electric field, even
instantaneously. (It is straightforward to show that $\langle\text{Re}E_{x}\rangle_{\phi}=\langle\text{Re}E_{y}\rangle_{\phi}=0$
if $|m|\neq1$.) Thus a uniformly polarized, focused beam centered
on the cavity axis can not couple to cavity modes with $|m|\neq1$
!%
}. The dimensional reduction in this case amounts to the removal of
a summation over $m$ in the basis expansion.

The refractive index in region (layer) $q$ is denoted $\tilde{n}_{q}$.
Layers are also denoted with an upper subscript in parenthesis: $\bm{E}^{(q)}$
means the electric vector field in layer $q$. Sometimes {}``fs''
is used as a value of $q$, meaning {}``in free space'' (e.g. \!$\tilde{n}_{\text{fs}}$),
whether or not any of the layers in the model we are considering actually
are free space. We note here that $\tilde{n}_{\text{fs}}=1$ only
in {}``cavity type I'' discussed below.

The symbol $k$, where not in a super/subscript and not bold nor having
any super/subscripts, always refers to what may be called the {}``wavenumber
in free space'', \emph{although it will have an imaginary part if
M1 is a dielectric mirror}. An imaginary part in wavenumber, refractive
index, and/or frequency is often introduced (as it is in this problem)
to turn open cavity problems into eigenvalue problems. The definition
of $k$ is as follows. Define $-k_{q}^{2}$ as the constant of separation
used to separate space and time equations from the wave equation for
layer $q$: \begin{align}
\nabla^{2}\text{X}^{(q)}=\frac{\tilde{n}_{q}^{2}}{c^{2}}\frac{\partial^{2}\text{X}^{(q)}}{\partial t^{2}}.\label{eqn: wave equation 1}\end{align}
 Here $\text{X}^{(q)}$ may be a vector or scalar field. In a few
steps, the selection of a global monochromatic time dependence $\exp(-\imath\omega t)$
reveals that the ratio $k_{q}/\tilde{n}_{q}$ is independent of $q$.
Then $k$ is defined as $k\equiv k_{\text{fs}}$, so that $k_{q}=n_{q}k$
where $n_{q}\equiv\tilde{n}_{q}/\tilde{n}_{\text{fs}}$. \emph{In
the model, the index ratios $n_{q}$ are assumed to be real.} The
single plane wave solution to (\ref{eqn: wave equation 1}) has the
form $\text{X}^{(q)}=C^{(q)}e^{\imath\bm{k}^{(q)}\bm{\cdot x}}e^{-\imath\omega t}$
where $C^{(q)}$ is a constant vector or scalar and the complex wave
vector $\bm{k}^{(q)}$ is given by $\bm{k}^{(q)}\equiv k_{q}\bm{\Omega}_{k}^{(q)}=kn_{q}\bm{\Omega}_{k}^{(q)}$
with $\bm{\Omega}_{k}^{(q)}$ being the unit direction vector of the
plane wave, specified by $\theta_{k}^{(q)}$ and $\phi_{k}$. The
generally complex frequency is given by $\omega=ck_{q}/\tilde{n}_{q}$.
At a refractive interface, the angle $\theta_{k}^{(q)}$ changes as
given by Snell's law.

To understand the meaning of a complex $k$, it is helpful to realize
that the spatial dependence of the quasimodes are \emph{identical}%
\footnote{There is the minor difference for the magnetic field first noted in
Eqn. (\ref{eqn: vector simple PWB expansion}). Since the discrepancy
is a constant factor multiplying $\bm{H}$, however, this difference
is not necessarily part of the spatial dependence.%
} in the following two physical cavities: \begin{align}
 & \text{I}: & \tilde{n}_{\text{fs}} & =1,\tilde{n}_{q}=n_{q},\notag\\
 & \text{II}: & \tilde{n}_{\text{fs}} & =\Upsilon,\tilde{n}_{q}=\Upsilon n_{q},\Upsilon\in\mathbb{C}.\label{eqn: cavity I and II}\end{align}
 (The shape and size of each dielectric and conducting region are
the same for cavities I and II.) Cavity I is composed of of conductors
and zero-gain regions of real refractive index. Cavity II is constructed
by taking cavity I and multiplying the refractive index of each region,
including free space, by an arbitrary complex number $\Upsilon$.
The congruence of the spatial quasimodes follows from separating the
variables in (\ref{eqn: wave equation 1}). The values of $k$, $k_{q}$,
and $n_{q}$ for a given quasimode are the same in cavities I and
II. The frequency in cavity II is $\omega_{\text{II}}=\omega_{\text{I}}/\Upsilon$.
If we henceforth consider only the specific cavity II for which $\Upsilon$
= ($1-\imath g$) where $g$ is tuned to be the ratio $(-\text{Im}k/\text{Re}k)$
(for a given quasimode), we see that $\omega_{\text{II}}=c\text{Re}k=\text{Re}\omega_{\text{I}}$.
While in cavity I the quasimode decays in time, in cavity II the quasimode
is a steady state because the gain exactly offsets the loss. Either
of the two cavity types may be imagined to be the case in our treatment.
The only difference is the existence of the decay factor $\exp(c\text{Im}(k)t)$
for cavity I. (The inequality $\text{Im}k\leq0$ always turns up for
an eigenvalue problem with conducting and/or dielectric interface
boundary conditions.) We note that the relation of $k$ to the free
space wavelength (always real) of a plane wave is $k=(1-\imath g)2\pi/\lambda$.
The quality factor, $Q$, of the quasimode is $\text{Re}k/(2|\text{Im}k|)=1/(2g)$.

We note that Snell's law, $\tilde{n}_{q}\sin\theta_{k}^{(q)}=\tilde{n}_{q+1}\sin\theta_{k}^{(q+1)}$,
is independent of whether we have a cavity of type I or II because
the quantity $(1-\imath g)$, if present, divides out. One of the
limitations of our method is the omission of evanescent waves in layer
$0$ and in layers where $n_{q}\geq n_{0}$ (see Appendix \ref{sec: exclusion high-angle}).
\emph{Snell's law may cause $\theta_{k}^{(q)}$ to become complex
for layers with index ratios $n_{q}$ less than $n_{0}$.} In this
case $\sin\theta_{k}^{(q)}>1$ and $\cos\theta_{k}^{(q)}=\imath{\textrm{sgn}}(\cos\theta_{k}^{(0)})[\sin^{2}\theta_{k}^{(q)}-1]^{1/2}$.

In most cases the symbols $\psi$, $\bm{E}$, and $\bm{H}$ stand
for complex-valued fields. The time dependence is $\exp(-\imath\omega t)$
and it is usually suppressed. Physical fields are obtained by multiplying
by the time dependence and then taking the real part.

Throughout this paper, the common functions denoted by $Y_{lm}$,
$P_{l}^{m}$, $P_{l}$, $J_{n}$, $j_{l}$, and $n_{l}$ are defined
as they are in the book by Jackson \cite{Jackson}.

In the implementation, $c$ and the related constants $\epsilon_{0}$,
$\mu_{0}$, and $Z_{0}$ are all unity, and they will usually be dropped
in our treatment. We also assume non-magnetic materials so that $\mu_{q}=\mu_{0}=1$.

\section{\label{sec: PWB} Plane Wave Bases and the Bessel Wave Method}

Although this major section describes the Bessel wave method, much
of what is discussed here is applicable to the two-basis method with
little alteration. The discussion in Section \ref{sec: MB} is greatly
shortened due to this overlap of concepts and procedures.

\subsection{\label{sec: simple PWB}The Field Expansion in the Simple Plane Wave
Bases}

\subsubsection{Scalar basis}

A single scalar plane wave in layer $q$ has the form $\psi=C\exp{(\imath\bm{k}^{(q)\
}\bm{\cdot x}-\imath\omega t)}$. For a general monochromatic field, $k$ and $\omega$ are fixed
and the field can be expressed (due to the completeness of the scalar
PWB) uniquely (due to the orthogonality of the scalar PWB) as a sum
over plane waves in different directions. In our treatment however,
we omit plane waves in layer $q$ which would only exist as evanescent
waves when refracted into layer $0$. The expansion for the field
in layer $q$ is \begin{align}
\psi^{(q)}(\bm{x})=\int_{0}^{2\pi}\text{d}\phi_{k}\int_{0}^{\pi}\text{d}\theta_{k}^{(0)}\sin(\theta_{k}^{(0)})\tilde{\psi}_{k}^{(q)}e^{\imath\bm{k}^{(q)}\bm{\cdot x}}.\label{eqn: scalar expansion}\end{align}
 Here the basis expansion coefficients are the $\tilde{\psi}_{k}^{(q)}$
(continuous coefficients in the integral, and discrete coefficients
in implementation). \emph{The above expansion effectively propagates
each plane wave existing in the cavity down (whether forward or backward)
into the stack layers, and adds up all of their contributions.} In
order to express the $\tilde{\psi}_{k}^{(q)}$ in terms of $\tilde{\psi}_{k}^{(0)}$,
it is first necessary to separate the coefficients with $k_{z}>0$
from those with $k_{z}<0$ and write the above expansion as \begin{align}
\psi^{(q)}= & \;\int_{0}^{2\pi}\text{d}\phi_{k}\int_{0}^{\pi/2}\text{d}\alpha_{k}^{(0)}\sin(\alpha_{k}^{(0)})\notag\\
 & \times\left(\tilde{\psi}_{u}^{(q)}e^{\imath\bm{k}_{u}^{(q)}\bm{\cdot x}}+\tilde{\psi}_{d}^{(q)}e^{\imath\bm{k}_{d}^{(q)}\bm{\cdot x}}\right).\end{align}
 The $u$ and $d$ refer to the plane waves going upward or downward,
e.i. \!$\tilde{\psi}_{u}^{(q)}$ is the expansion coefficient $\tilde{\psi}_{k}^{(q)}$
for $k_{z}^{(q)}>0$ (or $k_{z}^{(0)}>0$, since ${\textrm{sgn}}(k_{z}^{(q)})={\textrm{sgn}}(k_{z}^{(0)})$)
and $\tilde{\psi}_{d}^{(q)}$ takes the place of $\tilde{\psi}_{k}^{(q)}$
for $k_{z}^{(0)}<0$. The wavevector $\bm{k}^{(q)}$ in cylindrical
coordinates is $(k_{\rho}^{(q)},k_{\phi}^{(q)},k_{z}^{(q)})$, for
which the following relationships hold: \begin{align}
k_{\rho}^{(q)} & =k_{q}\sin\theta_{k}^{(q)}=kn_{0}\sin\theta_{k}^{(0)}=kn_{0}\sin\alpha_{k}^{(0)}=k_{\rho}^{(0)},\notag\\
k_{z}^{(q)} & =k_{q}\cos\theta_{k}^{(q)}=kn_{q}{\textrm{sgn}}(\cos\theta_{k}^{(0)})\cos\alpha_{k}^{(q)}\notag\\
 & =kn_{q}{\textrm{sgn}}(\cos\theta_{k}^{(0)})\sqrt{1-\left(\frac{n_{0}}{n_{q}}\sin\alpha_{k}^{(0)}\right)^{2}},\notag\\
k_{\phi}^{(q)} & =\phi_{k}\text{ (indep. of $q$).}\end{align}
 This leads to \begin{align}
\psi^{(q)}= & \;\int_{0}^{2\pi}\text{d}\phi_{k}\int_{0}^{\pi/2}\text{d}\alpha_{k}^{(0)}\sin(\alpha_{k}^{(0)})e^{\imath\varphi_{\rho}}\notag\\
 & \times\left(\tilde{\psi}_{u}^{(q)}e^{\imath\varphi_{z}}+\tilde{\psi}_{d}^{(q)}e^{-\imath\varphi_{z}}\right),\end{align}
 where \begin{align}
\varphi_{\rho} & \equiv\rho kn_{0}\sin(\alpha_{k}^{(0)})\cos(\phi-\phi_{k}),\notag\\
\varphi_{z} & \equiv zkn_{q}\cos\alpha_{k}^{(q)}.\end{align}
 From standard theory regarding plane waves and layered media\cite{Yeh},
one can calculate the $2\,\times\,2$ complex transfer matrix, $T_{s}^{(q)}$,
that obeys the following equation \begin{equation}
\left(\begin{matrix}\tilde{\psi}_{d}^{(q)}e^{-\imath\varphi_{z}}\\
\tilde{\psi}_{u}^{(q)}e^{\imath\varphi_{z}}\end{matrix}\right)=T_{s}^{(q)}\bm{\cdot}\left(\begin{matrix}\tilde{\psi}_{d}^{(0)}\\
\tilde{\psi}_{u}^{(0)}\end{matrix}\right).\label{eqn: Ts equation, scalar}\end{equation}
 Defining the column sums $\,_{+}\beta_{s}^{(q)}\equiv T_{s,12}^{(q)}+T_{s,22}^{(q)}$
and $\,_{+}\gamma_{s}^{(q)}\equiv T_{s,11}^{(q)}+T_{s,21}^{(q)}$
allows us to write the expansion of the scalar wave in the layers
as \begin{align}
\psi^{(q)}= & \;\int_{0}^{2\pi}\text{d}\phi_{k}\int_{0}^{\pi/2}\text{d}\alpha_{k}^{(0)}\sin(\alpha_{k}^{(0)})\notag\\
 & \times e^{\imath\varphi_{\rho}}\left(\,_{+}\beta_{s}^{(q)}\tilde{\psi}_{u}^{(0)}+\,_{+}\gamma_{s}^{(q)}\tilde{\psi}_{d}^{(0)}\right).\label{eqn: scalar simple PWB expansion, using T matrix}\end{align}
 The reason for the notation with the subscripts {}``+'' and {}``s''
will become apparent in the vector discussion. The {}``s'' refers
to s-polarization.

The variables in the simple PWB for scalar fields are the complex
$\tilde{\psi}_{u}^{(0)}$ and $\tilde{\psi}_{d}^{(0)}$ (the superscript
will often be dropped). Next we consider the simple PWB for vector
fields.

\subsubsection{Vector basis}

We assume that for our purposes a general monochromatic electromagnetic
field can be expressed uniquely as a sum of vector (electromagnetic)
plane waves. For every given frequency and wavevector direction $\bm{\Omega}_{k}$
there are two orthogonally polarized plane waves (as opposed to a
single plane wave in the scalar case). Instead of a single coefficient
$\tilde{\psi}_{k}$ for each spatial direction we need two, $\tilde{S}_{k}$
and $\tilde{P}_{k}$, which we can define as follows. $\tilde{S}_{k}$
is the amplitude of the vector plane wave propagating in direction
$\bm{\Omega}_{k}$ which has its electric field polarized in the $x$-$y$
plane ($E_{z}=0$). Thus, this plane wave is an {}``s-wave'' with
regard to the planar mirror. $\tilde{P}_{k}$ is the amplitude of
the {}``p-wave'', the vector plane wave in direction $\bm{\Omega}_{k}$
which has its electric field polarized in the plane of incidence ($E_{\phi}=0$).
The coefficients $\tilde{S}_{k}$ and $\tilde{P}_{k}$ will be separated
into $\tilde{S}_{u}$, $\tilde{S}_{d}$, $\tilde{P}_{u}$, and $\tilde{P}_{d}$.

To specify the polarization of the fields we will use unit vectors
denoted by $\bm{\epsilon}$. The unit vector $\bm{\epsilon}_{s,k}^{(q)}$
denotes the direction of the electric field associated with the plane
wave with wavevector $\bm{k}^{(q)}$ and s-polarization. We take the
direction of the unit vectors to be: \begin{align}
\bm{\epsilon}_{s,k}^{(q)} & =-\hat{\bm{\phi}}_{k}\nonumber \\
 & =\hat{\bm{x}}\sin\phi_{k}-\hat{\bm{y}}\cos\phi_{k},\nonumber \\
\bm{\epsilon}_{p,k}^{(q)} & =\hat{\bm{\theta}}_{k}^{(q)}{\textrm{sgn}}(\cos\theta_{k}^{(q)})\notag\\
 & =\hat{\bm{\rho}}_{k}{\textrm{sgn}}(\cos\theta_{k}^{(0)})\cos\theta_{k}^{(q)}-\hat{\bm{z}}{\textrm{sgn}}(\cos\theta_{k}^{(0)})\sin\theta_{k}^{(q)}.\label{eqn: epsilon definitions}\end{align}
 In this phase convention (used by Yeh\cite{Yeh}), the projections
of the $\bm{\epsilon}_{p,k}^{(q)}$ vectors for the incident and reflected
waves onto the $x$-$y$ plane are equal. The other common phase convention
has these projections being in opposite directions.

The entire electric and magnetic field can be broken up into two parts:
$\bm{E}^{(q)}=\bm{E}_{s}^{(q)}+\bm{E}_{p}^{(q)}$ and $\bm{H}^{(q)}=\bm{H}_{s}^{(q)}+\bm{H}_{p}^{(q)}$
where $\bm{H}_{s}^{(q)}$ is the field with magnetic s-polarization
($H_{z}=0$) and $\bm{H}_{p}^{(q)}$ is the field with magnetic p-polarization
($H_{\phi}=0$). We can now write down the most compact expansion
of the vector field. \begin{align}
\bm{E}_{s}^{(q)} & =\int\textbf{d}\bm{\Omega}_{k}^{(0)}\tilde{S}_{k}^{(q)}\bm{\epsilon}_{s,k}^{(q)}e^{\imath\bm{k}^{(q)}\bm{\cdot x}},\nonumber \\
\bm{E}_{p}^{(q)} & =\int\textbf{d}\bm{\Omega}_{k}^{(0)}\tilde{P}_{k}^{(q)}\bm{\epsilon}_{p,k}^{(q)}e^{\imath\bm{k}^{(q)}\bm{\cdot x}},\displaybreak[0]\nonumber \\
\bm{H}_{s}^{(q)} & =-\tilde{n}_{q}\int\textbf{d}\bm{\Omega}_{k}^{(0)}\tilde{P}_{k}^{(q)}\bm{\epsilon}_{s,k}^{(q)}{\textrm{sgn}}(\cos\theta_{k}^{(0)})e^{\imath\bm{k}^{(q)}\bm{\cdot x}},\nonumber \\
\bm{H}_{p}^{(q)} & =\tilde{n}_{q}\int\textbf{d}\bm{\Omega}_{k}^{(0)}\tilde{S}_{k}^{(q)}\bm{\epsilon}_{p,k}^{(q)}{\textrm{sgn}}(\cos\theta_{k}^{(0)})e^{\imath\bm{k}^{(q)}\bm{\cdot x}}.\label{eqn: vector simple PWB expansion}\end{align}
 The factors of $\tilde{n}_{q}$ in the $H$ equations come from the
physical relation of $H$ to $E$ for a plane wave. Note that $\tilde{n}_{q}$
is different for cavity types I and II (as given in Eqn. (\ref{eqn: cavity I and II})).
Separating up and down coefficients yields \begin{align}
\bm{E}_{s}^{(q)}= & \;\int_{0}^{2\pi}\text{d}\phi_{k}\,\bm{\epsilon}_{s,k}^{(q)}\int_{0}^{\pi/2}\text{d}\alpha_{k}^{(0)}\sin(\alpha_{k}^{(0)})e^{\imath\varphi_{\rho}}\notag\\
 & \times\left(\tilde{S}_{u}^{(q)}e^{\imath\varphi_{z}}+\tilde{S}_{d}^{(q)}e^{-\imath\varphi_{z}}\right),\notag\displaybreak[0]\\
\bm{E}_{p}^{(q)}= & \;\int_{0}^{2\pi}\text{d}\phi_{k}\int_{0}^{\pi/2}\text{d}\alpha_{k}^{(0)}\sin(\alpha_{k}^{(0)})\, e^{\imath\varphi_{\rho}}\notag\\
 & \times\Bigl[\hat{\bm{\rho}}_{k}\cos(\alpha_{k}^{(q)})\left(\tilde{P}_{u}^{(q)}e^{\imath\varphi_{z}}+\tilde{P}_{d}^{(q)}e^{-\imath\varphi_{z}}\right)\notag\\
 & +\hat{\bm{z}}\sin(\alpha_{k}^{(q)})\left(-\tilde{P}_{u}^{(q)}e^{\imath\varphi_{z}}+\tilde{P}_{d}^{(q)}e^{-\imath\varphi_{z}}\right)\Bigr].\end{align}
 These expressions explicitly use coordinate vectors only where necessary
due to a dependence of the $\bm{\epsilon}_{k}$ vectors on the sign
of $\cos\theta_{k}^{(0)}$. The expressions for $\bm{H}^{(q)}$ are
omitted for brevity.

To relate $\tilde{S}_{u/d}^{(q)}$ and $\tilde{P}_{u/d}^{(q)}$ to
$\tilde{S}_{u/d}^{(0)}$ and $\tilde{P}_{u/d}^{(0)}$ we can use the
transfer matrices: $T_{s}$ for s-polarized light and $T_{p}$ for
p-polarized light. The transfer matrix used for the scalar field in
Eqn. (\ref{eqn: Ts equation, scalar}) is the same matrix we will
use here for s-polarization. These matrices perform the following
transformations \begin{align}
\left(\begin{matrix}\tilde{S}_{d}^{(q)}e^{-\imath\varphi_{z}}\\
\tilde{S}_{u}^{(q)}e^{\imath\varphi_{z}}\end{matrix}\right) & =T_{s}^{(q)}\,\bm{\cdot}\,\left(\begin{matrix}\tilde{S}_{d}^{(0)}\\
\tilde{S}_{u}^{(0)}\end{matrix}\right),\nonumber \\
\left(\begin{matrix}\tilde{P}_{d}^{(q)}e^{-\imath\varphi_{z}}\\
\tilde{P}_{u}^{(q)}e^{\imath\varphi_{z}}\end{matrix}\right) & =T_{p}^{(q)}\,\bm{\cdot}\,\left(\begin{matrix}\tilde{P}_{d}^{(0)}\\
\tilde{P}_{u}^{(0)}\end{matrix}\right).\end{align}
 We define \begin{align}
\,_{\pm}\beta_{s}^{(q)} & \equiv T_{s,12}^{(q)}\pm T_{s,22}^{(q)},\notag\\
\,_{\pm}\gamma_{s}^{(q)} & \equiv T_{s,21}^{(q)}\pm T_{s,11}^{(q)},\notag\displaybreak[0]\\
\,_{\pm}\beta_{p}^{(q)} & \equiv T_{p,12}^{(q)}\pm T_{p,22}^{(q)},\notag\\
\,_{\pm}\gamma_{p}^{(q)} & \equiv T_{p,21}^{(q)}\pm T_{p,11}^{(q)}.\end{align}
 Note $\,_{+}\beta_{s}^{(q)}$ and $\,_{+}\gamma_{s}^{(q)}$ are defined
as before. The $\beta$ and $\gamma$ quantities are functions of
$z$ and $z_{1}$ and not of $\rho$ or $\phi$. They are functions
of $k$ and $\alpha_{k}^{(0)}$ but not of $\phi_{k}$.

Now the field expansions become \begin{align}
\bm{E}_{s}^{(q)}= & \;\int_{0}^{2\pi}\text{d}\phi_{k}\,\bm{\epsilon}_{s,k}^{(q)}\int_{0}^{\pi/2}\text{d}\alpha_{k}^{(0)}\sin(\alpha_{k}^{(0)})e^{\imath\varphi_{\rho}}\notag\\
 & \times\left(\,_{+}\beta_{s}^{(q)}\tilde{S}_{u}^{(0)}+\,_{+}\gamma_{s}^{(q)}\tilde{S}_{d}^{(0)}\right),\notag\displaybreak[0]\\
\bm{E}_{p}^{(q)}= & \;\int_{0}^{2\pi}\text{d}\phi_{k}\int_{0}^{\pi/2}\text{d}\alpha_{k}^{(0)}\sin(\alpha_{k}^{(0)})\, e^{\imath\varphi_{\rho}}\notag\\
 & \times\Bigl[\hat{\bm{\rho}}_{k}\cos(\alpha_{k}^{(q)})\left(\,_{+}\beta_{p}^{(q)}\tilde{P}_{u}^{(0)}+\,_{+}\gamma_{p}^{(q)}\tilde{P}_{d}^{(0)}\right)\notag\\
 & +\hat{\bm{z}}\sin(\alpha_{k}^{(q)})\left(\,_{-}\beta_{p}^{(q)}\tilde{P}_{u}^{(0)}-\,_{-}\gamma_{p}^{(q)}\tilde{P}_{d}^{(0)}\right)\Bigr],\notag\displaybreak[0]\\
\bm{H}_{s}^{(q)}= & \;\tilde{n}_{q}\int_{0}^{2\pi}\text{d}\phi_{k}\,\bm{\epsilon}_{s,k}^{(q)}\int_{0}^{\pi/2}\text{d}\alpha_{k}^{(0)}\sin(\alpha_{k}^{(0)})e^{\imath\varphi_{\rho}}\notag\\
 & \times\left(\,_{-}\beta_{p}^{(q)}\tilde{P}_{u}^{(0)}-\,_{-}\gamma_{p}^{(q)}\tilde{P}_{d}^{(0)}\right),\notag\displaybreak[0]\\
\bm{H}_{p}^{(q)}= & \;\tilde{n}_{q}\int_{0}^{2\pi}\text{d}\phi_{k}\int_{0}^{\pi/2}\text{d}\alpha_{k}^{(0)}\sin(\alpha_{k}^{(0)})\, e^{\imath\varphi_{\rho}}\notag\\
 & \times\Bigl[\hat{\bm{\rho}}_{k}\cos(\alpha_{k}^{(q)})\left(-\,_{-}\beta_{s}^{(q)}\tilde{S}_{u}^{(0)}+\,_{-}\gamma_{s}^{(q)}\tilde{S}_{d}^{(0)}\right)\notag\\
 & -\hat{\bm{z}}\sin(\alpha_{k}^{(q)})\left(\,_{+}\beta_{s}^{(q)}\tilde{S}_{u}^{(0)}+\,_{+}\gamma_{s}^{(q)}\tilde{S}_{d}^{(0)}\right)\Bigr].\label{eqn: vector simple PWB expansion, using T matrix}\end{align}

The variables in the simple PWB for vector fields are the complex
$\tilde{S}_{u}^{(0)}$, $\tilde{S}_{d}^{(0)}$, $\tilde{P}_{u}^{(0)}$,
and $\tilde{P}_{d}^{(0)}$ (the superscript will often be dropped).

\subsection{\label{sec: conical PWB}The Field Expansion in the Bessel Wave Bases}

\subsubsection{Scalar basis}

We have already assumed a time dependence of $\exp(-\imath\omega t)$.
As mentioned in the Overview, a cylindrically symmetric set of boundary
conditions allows us to assume an azimuthal dependence of $\exp(\imath m\phi)$
with $m$ being an integer. Consider the expansion (\ref{eqn: scalar expansion}).
We wish to find the conditions on $\tilde{\psi}_{k}^{(q)}$ which
cause the entire dependence of $\psi(\bm{x})$ on $\phi$ to be $\exp(\imath m\phi)$.

The general Fourier series expansion of $\tilde{\psi}_{k}^{(q)}$
is \begin{equation}
\tilde{\psi}_{k}^{(q)}(\theta_{k}^{(0)},\phi_{k})=\sum_{n}f_{n}^{(q)}(\theta_{k}^{(0)})\, e^{\imath n\phi_{k}}.\end{equation}
 We can then write (\ref{eqn: scalar expansion}) as \begin{align}
\psi^{(q)}= & \;\sum_{n}\int_{0}^{\pi}\text{d}\theta_{k}^{(0)}\sin(\theta_{k}^{(0)})e^{\imath\tilde{\varphi}_{z}}f_{n}^{(q)}(\theta_{k}^{(0)})\nonumber \\
 & \times\int_{0}^{2\pi}\,\text{d}\phi_{k}\; e^{\imath\rho kn_{0}\sin(\theta_{k}^{(0)})\cos(\phi-\phi_{k})}e^{\imath n\phi_{k}},\end{align}
 where \begin{align}
\tilde{\varphi}_{z}\equiv zkn_{q}\cos\theta_{k}^{(q)}.\end{align}
 The last integral is of the solved form \begin{equation}
\int_{0}^{2\pi}e^{\imath y\cos(\phi^{'}-\phi)}e^{\imath n\phi^{'}}\text{d}\phi^{'}=2\pi(\imath)^{n}J_{n}(y)e^{\imath n\phi},\label{eqn: Bessel integral}\end{equation}
 where $J_{n}$ denotes the regular Bessel function of order $n$
($n$ can be negative). This yields \begin{align}
\psi^{(q)}= & \;2\pi\sum_{n}(\imath)^{n}e^{\imath n\phi}\int_{0}^{\pi}\text{d}\theta_{k}^{(0)}\sin(\theta_{k}^{(0)})\nonumber \\
 & \times e^{\imath\tilde{\varphi}_{z}}J_{n}(\rho kn_{0}\sin\theta_{k}^{(0)})f_{n}^{(q)}(\theta_{k}^{(0)}).\end{align}
 In order to have only $\exp(\imath m\phi)$ dependence on $\phi$,
the integral on the right hand side must be zero for $n\neq m$. Because
$f_{n}^{(q)}(\theta_{k}^{(0)})$ cannot be a function of $z$ or $\rho$,
the only way to have this for all $z$ and $\rho$ is to pick $f_{n}=0$
for $n\neq m$. Thus $\tilde{\psi}_{k}^{(q)}(\theta_{k}^{(0)},\phi_{k})=f_{m}^{(q)}(\theta_{k}^{(0)})e^{\imath m\phi_{k}}$.
At this point we define the symbol $\psi_{k}^{(q)}$ to mean the coefficient
$f_{m}^{(q)}$. The cylindrically symmetric expansion is \begin{align}
\psi^{(q)}= & \;\xi\int_{0}^{\pi}\!\text{d}\theta_{k}^{(0)}\sin(\theta_{k}^{(0)})e^{\imath\tilde{\varphi}_{z}}J_{m}(\rho kn_{0}\sin\theta_{k}^{(0)})\psi_{k}^{(q)}(\theta_{k}^{(0)}),\label{eqn: scalar symmetric expansion, f}\end{align}
 where \begin{align}
\xi\equiv2\pi(\imath)^{m}e^{\imath m\phi}.\label{eqn: xi definition}\end{align}
 This is an expansion in scalar Bessel waves, defined to be \begin{align}
\xi\exp(\imath zkn_{q}\cos\theta_{k}^{(q)})J_{m}(\rho kn_{0}\sin\theta_{k}^{(0)}),\label{eqn: scalar Bessel wave}\end{align}
 with $\{\psi_{k}^{(q)}(\theta_{k}^{(0)})\}$ being the set of coefficients.
Each Bessel wave is a set of simple plane waves with fixed polar angle
but having the full range (0 to $2\pi$) of azimuthal angles, $\phi_{k}$.
The weight factors of the plane waves are proportional to $\exp(\imath m\phi_{k})$.
The final cylindrically symmetric scalar expansion with up and down
separated is \begin{align}
\psi^{(q)}(\bm{x})= & \;\xi\int_{0}^{\pi/2}\text{d}\alpha_{k}^{(0)}\sin(\alpha_{k}^{(0)})J_{m}(\rho kn_{0}\sin\alpha_{k}^{(0)})\notag\\
 & \times\left(\,_{+}\beta_{s}^{(q)}\psi_{u}^{(0)}+\,_{+}\gamma_{s}^{(q)}\psi_{d}^{(0)}\right).\label{eqn: scalar symmetric PWB expansion, using T matrix}\end{align}
 The $\psi_{u}^{(0)}$ and $\psi_{d}^{(0)}$ are the (complex) variables
in the Bessel wave method for scalar fields; they make up the solution
vector $\bm{y}$ in (\ref{eqn:Axb}).

\subsubsection{Vector basis}

For cylindrically symmetric boundary conditions, the $\phi$ dependence
of $E_{\rho}$, $E_{z}$, $E_{\phi}$, $H_{\rho}$, $H_{z}$, and
$H_{\phi}$ can be taken (for a single mode) to be $\exp(\imath m\phi)$.

Consider doing the $\phi_{k}$ integrations in (\ref{eqn: vector simple PWB expansion})
or (\ref{eqn: vector simple PWB expansion, using T matrix}). The
unit vectors $\bm{\epsilon}_{s/p,k}^{(q)}$ and $\hat{\bm{\rho}}_{k}$
depend on $\phi_{k}$. The $z$ components do not depend on $\phi_{k}$
so we will look at these first. There is no contribution to the $z$
component of the electric field from $\bm{E}_{s}^{(q)}$ nor is there
any contribution to the $z$ component of the magnetic field from
$\bm{H}_{s}^{(q)}$. We define \begin{align}
\,_{z}P^{(q)}(\bm{x}) & \equiv\bm{E}^{(q)}\bm{\cdot}\hat{\bm{z}}=\bm{E}_{p}^{(q)}\bm{\cdot}\hat{\bm{z}},\notag\\
\,_{z}^{H}\! P^{(q)}(\bm{x}) & \equiv\bm{H}^{(q)}\bm{\cdot}\hat{\bm{z}}=\bm{H}_{p}^{(q)}\bm{\cdot}\hat{\bm{z}}.\label{eqn:Pzdefinitions}\end{align}
 Requiring that these quantities have an $\exp(\imath m\phi)$ dependence
produces results similar to the scalar case. Defining \begin{align}
S_{k}^{(q)}e^{\imath m\phi_{k}}\equiv\tilde{S}_{k}^{(q)},\notag\\
P_{k}^{(q)}e^{\imath m\phi_{k}}\equiv\tilde{P}_{k}^{(q)},\label{eqn: S/P sub k definitions}\end{align}
 and using (\ref{eqn: vector simple PWB expansion, using T matrix}),
we have the final, useful expansions for $_{z}P^{(q)}$ and $_{z}^{H}\! P^{(q)}$:
\begin{align}
\,_{z}P^{(q)}= & \;\xi\int_{0}^{\pi/2}\text{d}\alpha_{k}^{(0)}\sin(\alpha_{k}^{(0)})\sin(\alpha_{k}^{(q)})\nonumber \\
 & \times J_{m}(\rho kn_{0}\sin\alpha_{k}^{(0)})\left(\,_{-}\beta_{p}^{(q)}P_{u}^{(0)}-\,_{-}\gamma_{p}^{(q)}P_{d}^{(0)}\right),\displaybreak[0]\nonumber \\
\,_{z}^{H}\! P^{(q)}= & \;-\xi\tilde{n}_{q}\int_{0}^{\pi/2}\text{d}\alpha_{k}^{(0)}\sin(\alpha_{k}^{(0)})\sin(\alpha_{k}^{(q)})\nonumber \\
 & \times J_{m}(\rho kn_{0}\sin\alpha_{k}^{(0)})\left(\,_{+}\beta_{s}^{(q)}S_{u}^{(0)}+\,_{+}\gamma_{s}^{(q)}S_{d}^{(0)}\right).\label{eqn:PzexpansionsusingTmatrix}\end{align}

To deal with the transverse part of the electromagnetic field it is
helpful to use quantities related to circular polarization. We define
\begin{align}
\,_{\pm}S^{(q)} & \equiv\pm\imath\bm{E}_{s}^{(q)}\bm{\cdot}\hat{\bm{x}}-\bm{E}_{s}^{(q)}\bm{\cdot}\hat{\bm{y}}\notag\\
 & =e^{\pm\imath\phi}(\pm\imath\bm{E}_{s}^{(q)}\bm{\cdot}\hat{\bm{\rho}}-\bm{E}_{s}^{(q)}\bm{\cdot}\hat{\bm{\phi}}),\notag\displaybreak[0]\\
\,_{\pm}^{H}\! S^{(q)} & \equiv\pm\imath\bm{H}_{s}^{(q)}\bm{\cdot}\hat{\bm{x}}-\bm{H}_{s}^{(q)}\bm{\cdot}\hat{\bm{y}}\notag\\
 & =e^{\pm\imath\phi}(\pm\imath\bm{H}_{s}^{(q)}\bm{\cdot}\hat{\bm{\rho}}-\bm{H}_{s}^{(q)}\bm{\cdot}\hat{\bm{\phi}}),\notag\displaybreak[0]\\
\,_{\pm}P^{(q)} & \equiv\bm{E}_{p}^{(q)}\bm{\cdot}\hat{\bm{x}}\pm\imath\bm{E}_{p}^{(q)}\bm{\cdot}\hat{\bm{y}}\notag\\
 & =e^{\pm\imath\phi}(\bm{E}_{p}^{(q)}\bm{\cdot}\hat{\bm{\rho}}\pm\imath\bm{E}_{p}^{(q)}\bm{\cdot}\hat{\bm{\phi}}),\notag\displaybreak[0]\\
\,_{\pm}^{H}\! P^{(q)} & \equiv\bm{H}_{p}^{(q)}\bm{\cdot}\hat{\bm{x}}\pm\imath\bm{H}_{p}^{(q)}\bm{\cdot}\hat{\bm{y}}\notag\\
 & =e^{\pm\imath\phi}(\bm{H}_{p}^{(q)}\bm{\cdot}\hat{\bm{\rho}}\pm\imath\bm{H}_{p}^{(q)}\bm{\cdot}\hat{\bm{\phi}}).\label{eqn: plus/minus S/P definitions}\end{align}
 Inverting (\ref{eqn: plus/minus S/P definitions}) yields \begin{align}
\bm{E}_{s}^{(q)}\bm{\cdot}\hat{\bm{\rho}} & =\frac{-\imath}{2}(\,_{+}S^{(q)}e^{-\imath\phi}-\,_{-}S^{(q)}e^{\imath\phi}),\notag\\
\bm{E}_{s}^{(q)}\bm{\cdot}\hat{\bm{\phi}} & =\frac{-1}{2}(\,_{+}S^{(q)}e^{-\imath\phi}+\,_{-}S^{(q)}e^{\imath\phi}),\notag\displaybreak[0]\\
\bm{E}_{p}^{(q)}\bm{\cdot}\hat{\bm{\rho}} & =\frac{1}{2}(\,_{+}P^{(q)}e^{-\imath\phi}+\,_{-}P^{(q)}e^{\imath\phi}),\notag\\
\bm{E}_{p}^{(q)}\bm{\cdot}\hat{\bm{\phi}} & =\frac{-\imath}{2}(\,_{+}P^{(q)}e^{-\imath\phi}-\,_{-}P^{(q)}e^{\imath\phi}),\label{eqn:ErhoandEphi}\end{align}
 with the magnetic field quantities having similar relations. Now
we use (\ref{eqn: vector simple PWB expansion}) and (\ref{eqn: epsilon definitions})
with (\ref{eqn: plus/minus S/P definitions}). The resulting electric
field quantities are \begin{align}
\,_{\pm}S^{(q)}= & \;\int_{0}^{\pi}\,\text{d}\theta_{k}^{(0)}\sin(\theta_{k}^{(0)})e^{\imath\tilde{\varphi}_{z}}\notag\\
 & \times\int_{0}^{2\pi}\,\text{d}\phi_{k}e^{\pm\imath\phi_{k}}e^{\imath\rho kn_{0}\sin(\theta_{k}^{(0)})\cos(\phi-\phi_{k})}\tilde{S}_{k}^{(q)},\notag\displaybreak[0]\\
\,_{\pm}P^{(q)}= & \;\int_{0}^{\pi}\,\text{d}\theta_{k}^{(0)}\sin(\theta_{k}^{(0)})\cos(\theta_{k}^{(q)}){\textrm{sgn}}(\cos\theta_{k}^{(0)})e^{\imath\tilde{\varphi}_{z}}\notag\\
 & \times\int_{0}^{2\pi}\,\text{d}\phi_{k}e^{\pm\imath\phi_{k}}e^{\imath\rho kn_{0}\sin(\theta_{k}^{(0)})\cos(\phi-\phi_{k})}\tilde{P}_{k}^{(q)}.\label{eqn: plus/minus S/P expansions, intermediate}\end{align}
 It is the $\exp(\pm\imath\phi_{k})$ factors in the integrands here
that motivated the definitions of $\,_{\pm}S/P$ (\ref{eqn: plus/minus S/P definitions}).
We see that the substitution of (\ref{eqn: S/P sub k definitions})
into (\ref{eqn: plus/minus S/P expansions, intermediate}) results
in $\phi_{k}$ integrals of the form (\ref{eqn: Bessel integral}).
Performing this substitution, doing the integrals, and separating
the up and down parts gives the final expansions: \begin{align}
\,_{\pm}S^{(q)}= & \;\xi_{\pm}\int_{0}^{\pi/2}\,\text{d}\alpha_{k}^{(0)}\sin(\alpha_{k}^{(0)})J_{m\pm1}(\rho kn_{0}\sin\alpha_{k}^{(0)})\notag\\
 & \times\left(\,_{+}\beta_{s}^{(q)}S_{u}^{(0)}+\,_{+}\gamma_{s}^{(q)}S_{d}^{(0)}\right),\notag\displaybreak[0]\\
\,_{\pm}P^{(q)}= & \;\xi_{\pm}\int_{0}^{\pi/2}\,\text{d}\alpha_{k}^{(0)}\sin(\alpha_{k}^{(0)})J_{m\pm1}(\rho kn_{0}\sin\alpha_{k}^{(0)})\notag\\
 & \times\cos(\alpha_{k}^{(q)})\left(\,_{+}\beta_{p}^{(q)}P_{u}^{(0)}+\,_{+}\gamma_{p}^{(q)}P_{d}^{(0)}\right),\notag\displaybreak[0]\\
\,_{\pm}^{H}\! S^{(q)}= & \;\xi_{\pm}\tilde{n}_{q}\int_{0}^{\pi/2}\,\text{d}\alpha_{k}^{(0)}\sin(\alpha_{k}^{(0)})J_{m\pm1}(\rho kn_{0}\sin\alpha_{k}^{(0)})\notag\\
 & \times\left(\,_{-}\beta_{p}^{(q)}P_{u}^{(0)}-\,_{-}\gamma_{p}^{(q)}P_{d}^{(0)}\right),\notag\displaybreak[0]\\
\,_{\pm}^{H}\! P^{(q)}= & \;\xi_{\pm}\tilde{n}_{q}\int_{0}^{\pi/2}\,\text{d}\alpha_{k}^{(0)}\sin(\alpha_{k}^{(0)})J_{m\pm1}(\rho kn_{0}\sin\alpha_{k}^{(0)})\notag\\
 & \times\cos(\alpha_{k}^{(q)})\left(-\,_{-}\beta_{s}^{(q)}S_{u}^{(0)}+\,_{-}\gamma_{s}^{(s)}S_{d}^{(0)}\right),\label{eqn; plus/minus S/P expansions, using T matrix}\end{align}
 where \begin{align}
\xi_{\pm}\equiv2\pi(\imath)^{m\pm1}e^{\imath(m\pm1)\phi}.\label{eqn: xi plus/minus definition}\end{align}
 At this point one can quickly verify, using (\ref{eqn:ErhoandEphi})
and the above equations, that $E_{\rho}$, $E_{\phi}$, $H_{\rho}$,
and $H_{\phi}$ do indeed have a $\phi$-dependence of $\exp(\imath m\phi)$.

The $S_{u}^{(0)}$, $S_{d}^{(0)}$, $P_{u}^{(0)}$, and $P_{d}^{(0)}$
are the (complex) variables in the Bessel wave method and make up
the solution vector $\bm{y}$ in (\ref{eqn:Axb}). They are essentially
coefficients of electromagnetic Bessel waves, although we need not
explicitly combine (\ref{eqn; plus/minus S/P expansions, using T matrix}),
(\ref{eqn:ErhoandEphi}), (\ref{eqn:PzexpansionsusingTmatrix}), and
(\ref{eqn:Pzdefinitions}) to obtain an explicit expression for the
E and H vector Bessel waves as we did for the scalar case (\ref{eqn: scalar Bessel wave}).

\subsection{\label{sec: PWB Ax=3D3Db}The Linear System of Equations for the
PWB}

Until now the PWB coefficients have been treated as continuous, when
in practice they must be chosen discrete. Let us keep in mind this
discrete nature in the following subsections. We denote by $N_{\text{dirs}}$
the number of directions $\alpha_{k}^{(0)}$ we choose. Thus there
are $2N\!_{\text{dirs}}$ coefficient variables for a scalar problem
and $4N\!_{\text{dirs}}$ coefficient variables for a vector problem.
The distribution of the $\alpha_{k}^{(0)}$ on $[0,\pi/2]$ need not
be uniform, and the effect of distribution choice will be briefly
mentioned later.

\subsubsection{\label{sec: PWB M1}The planar mirror (M1) boundary equations}

The M1 equations (planar mirror boundary condition equations) in the
PWB are very simple. In fact, because of this simplicity, the M1 equations
in the MB are basically a transformation to and from the PWB with
the M1 equations for the PWB sandwiched between. The reflection of
a plane wave off of a layered potential is a well known problem. For
the purpose of determining the field in layer $0$, the entire dielectric
stack is characterized by the complex $r_{s}$ and $r_{p}$ coefficients
acting at the first surface of the stack. For the scalar case with
the \mbox{layer 0}--\mbox{layer 1} interface at $z_{1}=0$, the
boundary condition is just $\tilde{\psi}_{u}^{(0)}=r_{s}(\alpha_{k}^{(0)})\tilde{\psi}_{d}^{(0)}$
where $r_{s}(\alpha)$ is the stack reflection function. Since Bessel
waves are linear superpositions of many plane waves with the same
$\theta_{k}^{(0)}$ parameter, this equations is true for Bessel waves:
\begin{align}
\psi_{u}^{(0)}=r_{s}(\alpha_{k}^{(0)})\psi_{d}^{(0)}.\end{align}
 For a conducting mirror, set $r_{s}=-1$ and for a free mirror set
$r_{s}=1$.

If the interface is at a general height $z_{1}$, then the same $r_{s}$
function is used at this surface yielding \begin{align}
\psi_{u}^{(0)}e^{\imath z_{1}kn_{0}\cos\alpha_{k}^{(0)}}=r_{s}(\alpha_{k}^{(0)})\psi_{d}^{(0)}e^{-\imath z_{1}kn_{0}\cos\alpha_{k}^{(0)}},\notag\end{align}
 or \begin{align}
\psi_{u}^{(0)}-\bar{r}_{s}(\alpha_{k}^{(0)})\psi_{d}^{(0)}=0,\label{eqn: M1 eqn, PWB, scalar}\end{align}
 where \begin{align}
\bar{r}_{s/p}(\alpha_{k}^{(0)})\equiv r_{s/p}(\alpha_{k}^{(0)})e^{-\imath2z_{1}kn_{0}\cos\alpha_{k}^{(0)}}.\label{eqn: r s/p bar definition}\end{align}
 (The equation is given for both s- and p-polarization since we will
shortly be using the $r_{p}$ quantities.) The quantities $r_{s/p}(\alpha_{k}^{(0)})$
are independent of $z$ and $z_{1}$. When M2 is a dielectric stack
mirror, the $\bar{r}_{s/p}(\alpha)$ are determined by $T_{s/p}^{(q=\text{layer X})}$
according to \begin{align}
\bar{r}_{s/p}(\alpha)=-T_{s/p,21}^{(X)}/T_{s/p,22}^{(X)}.\label{eqn: r s/p bar relation to T}\end{align}

For the vector case the equation for the s-polarized plane waves is
\begin{align}
\bm{E}_{u}^{(0)}\bm{\cdot\epsilon}_{s,u}^{(0)}=\bar{r}_{s}(\alpha_{k}^{(0)})\bm{E}_{d}^{(0)}\bm{\cdot\epsilon}_{s,d}^{(0)},\label{eqn: M1 eqn, PWB, vector S raw}\end{align}
 where $\bm{E}_{u/d}^{(0)}$ is the total electric field of the two
plane waves going in the direction specified by $\alpha_{k}$, $\phi_{k}$,
and $u$ or $d$. Shifting to our current notation and to Bessel waves,
the equation becomes \begin{align}
S_{u}^{(0)}-\bar{r}_{s}(\alpha_{k}^{(0)})S_{d}^{(0)}=0.\label{eqn: M1 eqn, PWB, vector S}\end{align}
 For p-polarization there is an arbitrary conventional sign. In our
phase convention (chosen in (\ref{eqn: epsilon definitions})) the
equation is \begin{align}
\bm{E}_{u}^{(0)}\bm{\cdot\epsilon}_{p,u}^{(0)}=\bar{r}_{p}(\alpha_{k}^{(0)})\bm{E}_{d}^{(0)}\bm{\cdot\epsilon}_{p,d}^{(0)},\label{eqn: M1 eqn, PWB, vector P raw}\end{align}
 or \begin{align}
P_{u}^{(0)}-\bar{r}_{p}(\alpha_{k}^{(0)})P_{d}^{(0)}=0.\label{eqn: M1 eqn, PWB, vector P}\end{align}
 The vector $\bm{\epsilon}_{p,u/d}^{(0)}$ is $\bm{\epsilon}_{p,\tilde{k}}^{(0)}$
with $\tilde{\bm{k}}$ being $\bm{k}$ forced into the up/down version
of itself.

For a simple mirror, set $r_{p}=r_{s}=$ \{$-1$ for conducting, $+1$
for free\} and use (\ref{eqn: r s/p bar definition}) instead of (\ref{eqn: r s/p bar relation to T})
to determine $\bar{r}_{s/p}$. Sometimes we will use $\cos\alpha_{k}$
as the explicit argument to $\bar{r}_{s/p}$ or $r_{s/p}$ instead
of $\alpha_{k}$.

Equation (\ref{eqn: M1 eqn, PWB, scalar}) or Eqns. (\ref{eqn: M1 eqn, PWB, vector S})
and (\ref{eqn: M1 eqn, PWB, vector P}), given for each discrete $\alpha_{k}^{(0)}$,
form the M1 equations.

\subsubsection{\label{sec: PWB M2} The curved mirror (M2) boundary equations}

As mentioned in Section \ref{sec: model overview}, the M2 equations
come from setting the appropriate field components equal to zero at
some number of locations on the curved mirror and the hat brim. If
we chose individual points on the two dimensional surface, the $\phi$-dependence
factor, $\exp(\imath m\phi)$, could be divided away. Thus picking
locations with the same $\rho$ and $z$ but different $\phi$ yields
identical boundary equations. Therefore we simply set $\phi=0$ (and
$t=0$) and pick equations by incrementing a single parameter (such
as $\rho$) on the one dimensional curve given by the intersection
of the conducting mirror and the $x$-$z$ plane. The number of locations,
$N_{\text{M2 loc}}$, determines the number of M2 boundary equations.
All of the locations are taken to lie in layer $0$.

We obtain the M2 equations by, in effect, doing the $\alpha_{k}^{(0)}$
integrals in (\ref{eqn: scalar symmetric PWB expansion, using T matrix})
or in (\ref{eqn:PzexpansionsusingTmatrix}) and (\ref{eqn; plus/minus S/P expansions, using T matrix}).
Before making the integrals discrete, the distribution of representative
directions must be chosen. If the interval $\Delta\alpha_{k}^{(0)}$
between successive directions is not constant, the integral over $\alpha_{k}^{(0)}$
must be transformed to an integral over a new variable, $x$, where
$\Delta x$ \emph{is} constant. Such a transformation will generate
a new integration factor. At this point the integral is turned into
a sum according to: $\int_{a}^{b}\rightarrow\sum_{j}\;,\;\text{d}x\rightarrow(b-a)/N$.
Choosing the direction distribution to be uniform in $\alpha_{k}^{(0)}$
requires no change in integration factor and yields a {}``summation
factor'' of $\pi/(2N\!_{\text{dirs}})$.

Using (\ref{eqn: scalar symmetric PWB expansion, using T matrix})
and (\ref{eqn: xi definition}), the M2 equations for the scalar problem
are \begin{align}
2\pi\left(\frac{\pi}{2N\!_{\text{dirs}}}\right)\!\sum_{j=1}^{N\!_{\text{dirs}}}\Bigl[J_{m}(\rho_{*}kn_{0}\sin\alpha_{k_{j}}^{(0)})\notag\\
\times\,\sin(\alpha_{k_{j}}^{(0)})\left(\,_{+}\beta_{s}^{(0)}\bigl|_{z=z_{*}}\psi_{u_{j}}+\,_{+}\gamma_{s}^{(0)}\bigl|_{z=z_{*}}\psi_{d_{j}}\right)\Bigr]=0.\label{eqn: scalar PWB M2 eqn}\end{align}
 An equation is added to the linear system for each chosen location
specified by $(\rho_{*},z_{*})$. All phase factors have been divided
out of (\ref{eqn: scalar PWB M2 eqn}) but the scale factor $\pi^{2}/N\!_{\text{dirs}}$
has been kept for representative weighting. Of course there is also
an effective weight produced by the distribution of the evaluation
locations on the conducting mirror. In our implementation, we choose
equal steps of $\theta$ to cover the dome and equal steps of $\rho$
to cover the hat brim (see Fig. \!\ref{fig: model}).

For the vector problem there are three equations associated with each
location: $E_{\phi}=0$, $E_{\parallel}=0$, and $H_{\perp}=0$. (Here
the subscript {}``$\parallel$'' refers to the direction that is
both tangential to the M2 surface and perpendicular to $\hat{\bm{\phi}}$.)
From (\ref{eqn:ErhoandEphi}) the $E_{\phi}=0$ equation is \begin{align}
\frac{-1}{2}\left(\,_{+}S^{(0)}+\,_{-}S^{(0)}+\imath(\,_{+}P^{(0)}-\,_{-}P^{(0)})\right)=0.\label{eqn:Ephi0}\end{align}
 The expansions for $\,_{\pm}S/P$ in terms of the unknowns $S/P_{u/d}^{(0)}$
in equation (\ref{eqn; plus/minus S/P expansions, using T matrix})
must now be used, along with the identical integral-to-sum conversion
used in the scalar problem (\ref{eqn: scalar PWB M2 eqn}). It is
probably not beneficial to work out the long form of this boundary
equation, as its computer implementation can be done with substitutions.

The $E_{\parallel}=0$ equation depends on the shape of the mirror.
If $\eta$ is the angle that the outward-oriented surface normal makes
with the $z$ axis, then $E_{\parallel}$ is given by \begin{align}
E_{\parallel}=E_{\rho}\cos\eta-E_{z}\sin\eta,\end{align}
 where $E_{z}=\,_{z}P^{(0)}$ and, using (\ref{eqn:ErhoandEphi}),
\begin{align}
E_{\rho}=\frac{1}{2}\left(\imath(\,_{-}S^{(0)}-\,_{+}S^{(0)})+\,_{+}P^{(0)}+\,_{-}P^{(0)}\right).\end{align}
 Again equation (\ref{eqn; plus/minus S/P expansions, using T matrix})
and the integral-to-sum conversion must be used to obtain the explicit
row equation. The $H_{\perp}=0$ equation is obtained by doing the
same type of substitutions with \begin{align}
H_{\perp}=H_{\rho}\sin\eta+H_{z}\cos\eta.\end{align}
 Here $H_{z}=\,_{z}^{H}\! P^{(0)}$ and \begin{align}
H_{\rho}=\frac{1}{2}\left(\imath(\,_{-}^{H}\! S^{(0)}-\,_{+}^{H}\! S^{(0)})+\,_{+}^{H}\! P^{(0)}+\,_{-}^{H}\! P^{(0)}\right).\end{align}
 For locations on the hat brim, $\eta=0$.

\subsubsection{The seed equation}

All the M1 and M2 equations have no constant term. Thus the best numerical
solution will be the trivial solution $\bm{y}_{\text{best}}=\bm{0}$.
To prevent $\bm{0}$ from being a solution, an equation with a constant
term must be added. One simple type of equation sets a single variable
equal to 1, for instance $S_{u(j=5)}=1$. Another simple type sets
the sum of all of the coefficients equal to 1. A more complicated
type sets the field (or a field component) at a certain point in space
equal to a constant. No one type of seed equation is always best.

\subsubsection{Solution of \textrm{\emph{Ay=b}}}

As depicted in (\ref{eqn:Axb}), the matrix $A$ is made up of the
left hand sides of the M1, M2, and seed equations. For the scalar
case there are $2N_{\text{dirs}}$ columns and $(N_{\text{dirs}}+N_{\text{M2 loc}}+1)$
rows. For the vector case there are $4N_{\text{dirs}}$ columns and
$(2N_{\text{dirs}}+3N_{\text{M2 loc}}+1)$ rows. The value of $N_{\text{M2 loc}}$
is picked so that $A$ has several times as many rows as columns.
A value of $k$ is picked and the overdetermined system of equations
is {}``solved'' as well as possible by a linear least squares method.
The best such methods rely on a technique known as singular value
decomposition \cite{recipes}. Our implementation relies on the function
\texttt{zgelsd} of the LAPACK fortran library. To find the eigenvalues
of $k$, the imaginary part of $k$ is set to zero and the real part
of $k$ is scanned. As mentioned in the Overview, this results in
dips in the value of $\Delta_{r}$. Using Brent's method \cite{recipes}
for minimization, the minimum of the dip is found. The real part of
$k$ is now fixed and Brent's method is used again to find the best
imaginary part of $k$. Then Brent's method may again be used on the
real part of $k$. By this alternating method, the complex eigenvalue
of $k$ is found, along with the eigenmode, $\bm{y}$. In practice
Brent's method need only be used two to four times per scan dip to
get an accurate complex $k$. We usually normalize each row of $A$
to 1 so that the normalized error per equation in the system can be
expected to be around $\Delta_{n}\equiv\Delta_{r}/[|\bm{y}|\times(\text{number of rows in }A)^{1/2}]$.
$\Delta_{n}$ is one indicator of the accuracy of the solution.

\subsubsection{Calculating the field from \textrm{\emph{y}}}

Once $\bm{y}$ is calculated for a quasimode, the values of the field
in any layer can found by using the expansion (\ref{eqn: scalar symmetric PWB expansion, using T matrix})
for the scalar field and equations (\ref{eqn; plus/minus S/P expansions, using T matrix}),
(\ref{eqn:ErhoandEphi}), (\ref{eqn:PzexpansionsusingTmatrix}), and
(\ref{eqn:Pzdefinitions}) for the vector field. Of course the integrals
over $\alpha_{k}^{(0)}$ must be made discrete as discussed previously.
Appendix \ref{sec: modetype} explains more regarding the plotting
of the fields.

\section{\label{sec: MB}Multipole Bases and the Two-Basis Method}

As mentioned in the Introduction, the two-basis method ultimately
uses the scalar or vector MB. The MB is the eigenbasis for the closed,
conducting hemisphere or sphere. Both the vector and scalar multipole
bases have known forms which we will use but not derive. The basis
functions already possess an azimuthal dependence of $\exp(\imath m\phi)$
and the dimensional reduction due to cylindrical symmetry is accomplished
by picking a value for $m$ instead of summing over basis functions
with many $m$.

\begin{figure}[!htb]
\begin{center}\includegraphics[%
  width=0.55\columnwidth]{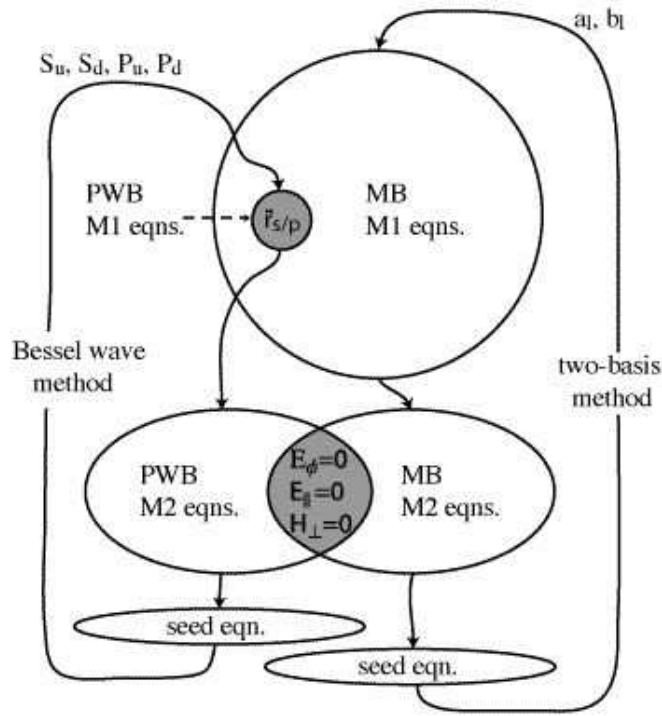}\hspace*{-2cm}\end{center}

\caption{Diagram for the two primary methods. The closed loops suggest the
self-consistency or {}``constructive interference'' of the quasimode
solutions. Grey regions indicate intersection between PWB and MB methods.
Size roughly indicates the {}``post-basis-derivation work'' required
to get the equations. The variable coefficients for the vector problem
are shown. }

\label{fig: circuit}
\end{figure}

The method of stepping along a one-parameter location curve to obtain
the M2 equations is the same as for the PWB. Explicit formulas in
the MB of course will be completely different and will be given in
this major section. The methods of solution to the linear system of
equations are the same as for the PWB. The development of the M1 equations
in the MB, however, requires considerable work. After the system of
equations has been solved, using the resulting solution vector $\bm{y}$
to calculate/plot the fields in layers other than layer $0$ also
requires significant work. We use the term {}``two-basis method''
because of the role of plane waves in these two calculations. Figure
\ref{fig: circuit} represents the linear system of equations of the
two primary methods and how they are related.

\subsection{The Scalar Multipole Basis}

The scalar MB functions we use are the $\psi_{lm}=j_{l}(kn_{0}r)Y_{lm}(\theta,\phi)$
where $j_{l}$ denotes the spherical Bessel function of the first
kind and $Y_{lm}$ is the spherical harmonic function. The scalar
MB functions, like the scalar PWB functions, satisfy the wave equation.
We assume the field in layer $0$, in a region large enough to encompass
the cavity, can be expanded uniquely in terms of the scalar MB functions.
Using the cylindrical symmetry of the cavity to solve the problem
separately for each value of $m$, we expand the field in the cavity
as \begin{align}
\psi^{(0)}(\bm{x})=\sum_{l=|m|}^{l_{\text{max}}}c_{l}j_{l}(kn_{0}r)Y_{lm}(\theta,\phi).\label{eqn: scalar MB expansion}\end{align}
 The expansion coefficients, $c_{l}$, are complex. One should never
need to choose $l_{\text{max}}$ much larger than $\text{Re}(k)n_{0}r_{\text{max}}$
where $r_{\text{max}}$ is the maximum radial extent of the dome (not
the hat brim). (Semiclassically, the maximum angular momentum a sphere
or hemisphere of radius $R_{s}$ can support (for a given $k$) is
$\sim\text{Re}(k)n_{0}R_{s}$, which corresponds to a whispering-gallery
mode.)

The scalar MB functions are the exact eigenfunctions of the problem
of a hypothetical spherical conductor specified by $r=R_{s}=R$, with
eigenvalues given by the zeros of $j_{l}(kn_{0}R)$. The basis functions
for which $(l+m)$ is odd are the eigenfunctions of the closed hemispherical
conductor. This is because \begin{align}
Y_{lm}\text{ has parity }(-1)^{l+m}\text{ in }\cos\theta\label{eqn: Ylm parity}\end{align}
 and thus is zero at $\theta=\pi/2$. It can also be shown, using
the power series expansion of $P_{l}(x)$ given in Ref. \!\cite{Arfken},
that $Y_{lm}(\pi/2,\phi)$ is nonzero if $(l+m)$ is even.

\subsection{The M1 Equations in the Scalar MB}

The expansion of a plane wave in terms of the (monochromatic) scalar
MB functions is \cite{Jackson}: \begin{align}
e^{\imath\bm{k\cdot x}}=4\pi\sum_{l=0}^{\infty}\sum_{m=-l}^{l}(\imath)^{l}Y_{lm}^{*}(\theta,\phi)j_{l}(kr)Y_{lm}(\theta_{k},\phi_{k}).\label{eqn: expansion of scalar PWB in MB}\end{align}
 The inverse of this relation, the expansion of a scalar MB function
in terms of monochromatic plane waves, is \begin{align}
j_{l}(kr)Y_{lm}(\theta,\phi)=\int{\textbf{d}}\bm{\Omega}_{k}\left(\frac{(-\imath)^{l}}{4\pi}Y_{lm}(\theta_{k},\phi_{k})\right)e^{\imath\bm{k\cdot x}}.\label{eqn: expansion of scalar MB in PWB}\end{align}
 This equation is easy to verify by inserting (\ref{eqn: expansion of scalar PWB in MB})
into the right hand side. The use of $Y_{lm}=(-1)^{m}Y_{l,-m}^{*}$
and the orthogonality relation for the $Y_{lm}$ leads directly to
the left hand side. The expansion (\ref{eqn: expansion of scalar MB in PWB}),
applied to layer 0 ($k\rightarrow k_{0}=kn_{0}$), is the foundation
for this section and its counterpart for vector fields.

Using (\ref{eqn: scalar MB expansion}) and (\ref{eqn: expansion of scalar MB in PWB})
yields the scalar field expansion for a given $m$: \begin{align}
\psi^{(0)}= & \;\int\text{d}\bm{\Omega}_{k}^{(0)}\, e^{\imath\bm{k}^{(0)}\bm{\cdot x}}\biggl(\sum_{l=|m|}c_{l}\frac{(-\imath)^{l}}{4\pi}Y_{lm}(\theta_{k}^{(0)},\phi_{k})\biggr).\label{eqn: hybrid expansion}\end{align}
 The quantity in the curved brackets is $\psi_{k}$, the simple plane
wave coefficient (compare to Eqn. (\ref{eqn: scalar expansion}) with
$q$ set to 0). We can now use (\ref{eqn: M1 eqn, PWB, scalar}).
For $k_{z}>0$, $\psi_{u}=\psi_{k}$ and $\psi_{d}=\psi_{k^{'}}$,
where in cylindrical coordinates $\bm{k^{'}}=\bm{k^{'}}(\bm{k})\equiv(k_{\rho},k_{\phi},-k_{z})$.
For $k_{z}<0$, $\psi_{d}=\psi_{k}$ and $\psi_{u}=\psi_{k^{'}}$.
Using (\ref{eqn: Ylm parity}), one gets $\psi_{k^{'}}=\sum_{l}c_{l}Y_{lm}(\theta_{k}^{(0)},\phi_{k})(-1)^{l+m}(-\imath)^{l}/(4\pi)$.
We can now derive an equation that expresses (\ref{eqn: M1 eqn, PWB, scalar})
and holds for all $\bm{\Omega}_{k}^{(0)}$: \begin{multline}
\sum_{l=|m|}c_{l}\frac{(-\imath)^{l}}{4\pi}Y_{lm}(\theta_{k}^{(0)},\phi_{k})\\
\times\left\{ \begin{matrix}1-\bar{r}_{s}, & l+m\text{ is even}\\
{\textrm{sgn}}(\cos\theta_{k}^{(0)})\left(1+\bar{r}_{s}\right), & l+m\text{ is odd}\end{matrix}\right\} =0.\label{eqn: M1 MB, scalar, general}\end{multline}
 At this point there are two ways to proceed.

\subsubsection{Variant 1}

One way to construct the M1 portion of the matrix $A$ is to pick
some number $N_{\text{M1 \! dirs}}$ of discrete directions, $\theta_{k}^{(0)}$,
and use (\ref{eqn: M1 MB, scalar, general}) for each direction (the
$\phi_{k}$ dependence divides out). Inspection of the equation reveals
that it is even in $\cos\theta_{k}^{(0)}$; thus only polar angles
in the domain $[0,\pi/2]$ are needed. Using $\alpha_{k}^{(0)}$ as
before to denote this reduced domain, the M1 equations in this variant
become: \begin{multline}
\sum_{l=|m|}^{l_{\text{max}}}c_{l}\frac{(-\imath)^{l}}{4\pi}Y_{lm}(\alpha_{k}^{(0)},0)\\
\times\left(1-(-1)^{l+m}\, r_{s}(\cos\alpha_{k}^{(0)})e^{-\imath2z_{1}kn_{0}\cos\alpha_{k}^{(0)}}\right)=0.\label{eqn: M1 MB, scalar, variant 1}\end{multline}
 We note for later comparison that for each $\alpha_{k}^{(0)}$ we
need only evaluate a single associated Legendre function (inside the
$Y_{lm}$), because the recursive calculation technique for computing
$P_{l_{\text{max}}}^{m}(x)$ also computes $P_{l}^{m}(x)$ for $|m|\leq l\leq l_{\text{max}}$.
Since each \emph{step} of this recursive calculation involves a constant
number of floating point operations, we may say that the complexity
associated with the $P_{l}^{m}$ calculations for each $\alpha_{k}^{(0)}$
is $O(l_{\text{max}}^{2})$. Generally $N_{\text{M1 \! dirs}}\sim l_{\text{max}}$
so the $P_{l}^{m}$ complexity is $O(l_{\text{max}}^{3})$ and the
overall number of evaluations of $\bar{r}_{s}$ is $O(l_{\text{max}})$.

\subsubsection{Variant 2}

Rather than picking discrete directions to turn (\ref{eqn: M1 MB, scalar, general})
into many equations, we could project the entire left hand side onto
the spherical harmonic basis, $\{ Y_{l^{'}m^{'}}\}$, (that is, integrate
(\ref{eqn: M1 MB, scalar, general}) against $Y_{l^{'}m^{'}}^{*}(\bm{\Omega}_{k}^{(0)})$
in $\bm{\Omega}_{k}^{(0)}$). Due to the uniqueness of the basis expansion,
the integral for each $(l^{'},m^{'})$ pair must be zero. The integrals
for $m^{'}\neq m$ are zero and can be neglected. Thus the number
of equations generated is $N_{l^{'}}=l_{\text{max}}^{'}-|m|+1$. We
set $l_{\text{max}}^{'}=l_{\text{max}}$.

Since (\ref{eqn: M1 MB, scalar, general}) is even in $\cos\theta_{k}^{(0)}$,
integrating against $Y_{l^{'}m}^{*}$ gives $0$ if $l^{'}+m$ is
odd, halving the number of equations. When $l^{'}+m$ is even the
integration must be done numerically. The most analytically simplified
version of the coefficients of $c_{l}$ in the M1 equation corresponding
to $l^{'}$ (for $l^{'}+m$ even) is: \begin{align}
\text{M1}_{l^{'},l}=\zeta\int_{0}^{1}P_{l}^{|m|}(x)P_{l^{'}}^{|m|}(x)\left(1-(-1)^{l+m}\bar{r}_{s}(x)\right)\text{d}x,\end{align}
 where \begin{align}
\zeta\equiv\frac{(-\imath)^{l}}{4\pi}\!\left((2l+1)(2l^{'}+1)\frac{(l-|m|)!}{(l+|m|)!}\frac{(l^{'}-|m|)!}{(l^{'}+|m|)!}\right)^{1/2}.\end{align}
 This comes from using the properties of $Y_{lm}$ under a sign change
of $m$, using the definition of $Y_{lm}$, doing the $\phi_{k}$
integral, and halving the domain of the even integral over $x=\cos\alpha_{k}^{(0)}$.
Noting that the number of unknowns in the solution vector $\bm{y}$
is $N_{l}=l_{\text{max}}-|m|+1$, the number of (complex) coefficients
that must be calculated is about $N_{l}^{2}/2$. Noting that $\text{M1}_{l,l^{'}}=(-\imath)^{l^{'}-l}\text{M1}_{l^{'},l}$
when $l+m$ and $l^{'}+m$ are both even, the calculation is reduced
to about $3N_{l}^{2}/4$ real-valued integrations.

While this variant of the method is in some ways the most elegant,
the numerical integrals are extremely computationally intensive. Our
implementation used an adaptive Gaussian quadrature function, \texttt{gsl\_integration\_qag}
of the GSL. (Adaptive algorithms choose a different set of evaluation
points for each integration and achieve a prescribed accuracy; in
a sense the integral is done in a continuous rather than a discrete
manner.) For an adaptive algorithm, the complexity%
\footnote{The exponent $\nu$ is defined so that the number of integration points
that must be sampled (for the most complicated integrals where $l\sim l^{'}\sim l_{\text{max}}$)
to maintain a constant accuracy goes like $l_{\text{max}}^{\nu}$.
(This definition may not be rigorous if $\nu$ is a function of $l_{\text{max}}$.)
Since $P_{l}^{0}(x)$ has $O(l)$ zeros and the $P_{l}^{m}$ for $m>0$
are even more complicated, it is reasonably certain that $\nu\geq1$.%
} associated with the $P_{l}^{m}$ calculation is $O(l_{\text{max}}^{3+\nu})$,
where $\nu\geq1$. The number of evaluations of $\bar{r}_{s}$ is
$O(l_{\text{max}}^{2+\nu})$. In practice, variant 2 is much slower
than variant 1 even for cavities as small as $R/\lambda\approx5$
with simple mirrors. Checking between the two variants has generally
shown very good agreement.

\subsection{The Linear System of Equations in the Scalar MB}

The M1 equations have been given in the previous section. Using (\ref{eqn: scalar MB expansion})
yields an M2 equation \begin{align}
\sum_{l=|m|}^{l_{\text{max}}}c_{l}j_{l}(kn_{0}r_{*})Y_{lm}(\theta_{*},0)=0,\end{align}
 for each location $(r_{*},\theta_{*})$. The discussions from Section
\ref{sec: PWB Ax=3D3Db} regarding the M2 equations, the seed equation,
and the method of solution to $A\bm{y}=\bm{b}$ apply here. The number
of columns in $A$ is $N_{l}$ and the number of rows is $(N_{\text{M1 dirs}}+N_{\text{M2 loc}}+1)$
for variant 1 or about $(N_{l}/2+N_{\text{M2 loc}}+1)$ for variant
2.

\subsection{Calculating the Field in the Layers with the Scalar MB}

To calculate the complex field anywhere in layer $0$ once a quasimode
solution $\bm{y}$ has been found, Eqn. (\ref{eqn: scalar MB expansion})
can be used directly. To calculate the field in the layers below the
cavity ($q>0$) where the expansion does not apply, the Bessel waves
must be used, along with the $T_{s}^{(q)}$ matrix that propagates
them. Performing the $\phi_{k}$ integration in (\ref{eqn: hybrid expansion}),
using (\ref{eqn: Ylm parity}), and then comparing with (\ref{eqn: Ts equation, scalar})
and (\ref{eqn: scalar symmetric PWB expansion, using T matrix}) with
$q=0$ in both of these yields the Bessel wave coefficients: \begin{align}
\psi_{u}^{(0)} & =\sum_{l}c_{l}\frac{(-\imath)^{l}}{4\pi}Y_{lm}(\alpha_{k}^{(0)},0),\notag\\
\psi_{d}^{(0)} & =\sum_{l}c_{l}\frac{(-\imath)^{l}}{4\pi}Y_{lm}(\alpha_{k}^{(0)},0)(-1)^{l+m}.\label{eqn: scalar MB to PWB conversion}\end{align}
 Now (\ref{eqn: scalar symmetric PWB expansion, using T matrix})
can be used with $q>0$ yielding \begin{align}
\psi^{(q)}(\bm{x})= & \;\frac{(\imath)^{m}}{2}e^{\imath m\phi}\int_{0}^{\pi/2}\text{d}\alpha_{k}^{(0)}\sin(\alpha_{k}^{(0)})\notag\\
 & \times J_{m}(\rho kn_{0}\sin\alpha_{k}^{(0)})\Bigl[\sum_{l}c_{l}(-\imath)^{l}Y_{lm}(\alpha_{k}^{(0)},0)\notag\\
 & \times\left(\,_{+}\beta_{s}^{(q)}+(-1)^{l+m}\,_{+}\gamma_{s}^{(q)}\right)\Bigr].\label{eqn: scalar MB expansion, layers, using T matrix}\end{align}

This is a costly numerical integration to do with an adaptive algorithm.
For every sampled $\alpha_{k}^{(0)}$, the quantities $T_{s}$, $P_{l_{\text{max}}}^{m}$
and $J_{m}$ must be calculated once. For displaying large regions
of the field in the stack, it is sufficiently accurate to simply convert
the solved MB solution vector $\bm{y}$ into a PWB solution vector
(with $N_{\text{dirs}}\sim l_{\text{max}}$) by means of a separate
program using (\ref{eqn: scalar MB to PWB conversion}), and then
use the discrete form of (\ref{eqn: scalar symmetric PWB expansion, using T matrix})
to plot.

\subsection{The Vector Multipole Basis}

We expand the electromagnetic field in layer $0$ (at least in a finite
region surrounding the cavity) in the vector multipole basis using
spherical Bessel functions of the first kind ($j_{l}$). The multipole
basis uses the vector spherical harmonics (VSH; same abbreviation
for singular), which are given by\cite{Videen}\begin{align}
\bm{M}_{lm}(\bm{x}) & =-j_{l}(kn_{0}r)\,\bm{x\times\nabla}Y_{lm}(\theta,\phi),\notag\\
\bm{N}_{lm}(\bm{x}) & =\frac{1}{kn_{0}}\bm{\nabla\times M}_{lm}.\end{align}
 The VSH are not defined for $l=0$. The nature of the electromagnetic
multipole expansion is developed in section 9.7 of the book by Jackson\cite{Jackson}
using somewhat different terminology. The multipole expansion of the
electromagnetic fields is \begin{align}
\bm{E}^{(0)}(\bm{x}) & =\sum_{l=l_{\text{min}}}^{l_{\text{max}}}\left(-a_{l}\bm{N}_{lm}+\imath b_{l}\bm{M}_{lm}\right),\notag\\
\bm{H}^{(0)}(\bm{x}) & =\tilde{n}_{0}\sum_{l=l_{\text{min}}}^{l_{\text{max}}}\left(\imath a_{l}\bm{M}_{lm}+b_{l}\bm{N}_{lm}\right),\label{eqn: VSH expansion of E, H}\end{align}
 where cylindrical symmetry has been invoked to remove the sum over
$m$, and $l_{\text{min}}=\max(1,|m|)$. The $a_{l}$ and $b_{l}$
are complex coefficients and there are $N_{l}=l_{\text{max}}-l_{\text{min}}+1$
of each of them. The $a_{l}$ coefficients correspond to electric
multipoles and the $b_{l}$ coefficients correspond to magnetic multipoles.
The explicit forms of the VSH are \begin{align}
\bm{M}_{lm}(\bm{x})= & \;\hat{\bm{\theta}}\left(\frac{\imath m}{\sin\theta}j_{l}(kn_{0}r)Y_{lm}(\theta,\phi)\right)\nonumber \\
 & +\hat{\bm{\phi}}\left(-j_{l}(kn_{0}r)\frac{\partial}{\partial\theta}Y_{lm}(\theta,\phi)\right),\notag\displaybreak[0]\\
\bm{N}_{lm}(\bm{x})= & \;\hat{\bm{r}}\left(\frac{l(l+1)}{kn_{0}r}j_{l}(kn_{0}r)Y_{lm}(\theta,\phi)\right)\nonumber \\
 & +\hat{\bm{\theta}}\left(\frac{1}{kn_{0}r}\frac{\partial}{\partial r}\left(rj_{l}(kn_{0}r)\right)\frac{\partial}{\partial\theta}Y_{lm}(\theta,\phi)\right)\nonumber \\
 & +\hat{\bm{\phi}}\left(\frac{\imath m}{kn_{0}r\sin\theta}\frac{\partial}{\partial r}\left(rj_{l}(kn_{0}r)\right)Y_{lm}(\theta,\phi)\right).\label{eqn: M and N explicit}\end{align}

\subsection{The M1 Equations in the Vector MB}

The goal of the calculation here is to transform the M1 relations,
(\ref{eqn: M1 eqn, PWB, vector S raw}) and (\ref{eqn: M1 eqn, PWB, vector P raw}),
into equations such that each dot product is written in the form $\sum_{l}(a_{l}f_{a}(\bm{k})+b_{l}f_{b}(\bm{k}))$.
The resulting (two) equations will be the vector analogues of (\ref{eqn: M1 MB, scalar, general})
and will hold for all $\bm{\Omega}_{k}^{(0)}$, leading to two variants
of solution method in the same manner as before. First we must Fourier
expand $\bm{E}^{(0)}$ as we expanded $\psi^{(0)}$ to get (\ref{eqn: hybrid expansion}).
The only Fourier relation we have is (\ref{eqn: expansion of scalar MB in PWB})
which expands the quantity $j_{l}Y_{lm}(\bm{x})$. We must therefore
break $\bm{M}_{lm}$ into terms containing this quantity.

Here we can make use of the orbital angular momentum operator $\bm{L}=-\imath(\bm{x\times\nabla})$.
Using the ladder operators, $L_{\pm}=L_{x}\pm\imath L_{y}$, the VSH
$\bm{M}_{lm}=-\imath j_{l}\bm{L}Y_{lm}$ can be shown to be \begin{align}
\bm{M}_{lm}= & \;\left(-\imath j_{l}\right)\times\Bigl[\hat{\bm{x}}\frac{1}{2}\left(d_{lm}^{+}Y_{lm+1}+d_{lm}^{-}Y_{lm-1}\right)\notag\\
 & +\hat{\bm{y}}\frac{\imath}{2}\left(-d_{lm}^{+}Y_{lm+1}+d_{lm}^{-}Y_{lm-1}\right)+\hat{\bm{z}}\left(mY_{lm}\right)\Bigr],\end{align}
 where \begin{align}
d_{lm}^{\pm}\equiv\sqrt{(l\mp m)(l\pm m+1)}.\end{align}
 Using (\ref{eqn: expansion of scalar MB in PWB}) multiple times
yields \begin{align}
\bm{M}_{lm}=\int{\textbf{d}}\bm{\Omega}_{k}^{(0)}\tilde{\bm{M}}_{lm}e^{\imath\bm{k}^{(0)}\bm{\cdot x}},\label{eqn: M Fourier}\end{align}
 with \begin{align}
\tilde{\bm{M}}_{lm}= & \;\hat{\bm{x}}\frac{(-\imath)^{l}}{8\pi}(-\imath)e_{lm}^{+}+\hat{\bm{y}}\frac{(-\imath)^{l}}{8\pi}(-1)e_{lm}^{-}\notag\\
 & +\hat{\bm{z}}\frac{(-\imath)^{l}}{4\pi}mY_{lm}(\bm{\Omega}_{k}^{(0)}),\label{eqn: explicit Mtilde}\end{align}
 where \begin{align}
e_{lm}^{\pm}\equiv d_{lm}^{+}Y_{lm+1}(\bm{\Omega}_{k}^{(0)})\pm d_{lm}^{-}Y_{lm-1}(\bm{\Omega}_{k}^{(0)}).\end{align}
 We can avoid the task of expanding $\bm{N}_{lm}$ into $j_{l}Y_{lm}$
terms by just using \begin{align}
\bm{N}_{lm}=\frac{\bm{\nabla\times M}_{lm}}{kn_{0}}=\int\!{\textbf{d}}\bm{\Omega}_{k}^{(0)}\frac{\imath}{kn_{0}}\bm{k}^{(0)}\bm{\times}\tilde{\bm{M}}_{lm}e^{\imath\bm{k}^{(0)}\bm{\cdot x}}.\end{align}
 Performing the cross product and substituting into (\ref{eqn: VSH expansion of E, H})
using (\ref{eqn: M Fourier}-\ref{eqn: explicit Mtilde}) yields

\begin{align}
E_{x}(\bm{x})= & \;\int{\textbf{d}}\bm{\Omega}_{k}^{(0)}\, e^{\imath\bm{k}^{(0)}\bm{\cdot x}}\biggl\{\sum_{l}\frac{(-\imath)^{l}}{4\pi}\biggl[a_{l}\Bigl(-m\sin\theta_{k}^{(0)}\sin\phi_{k}Y_{lm}-\frac{\imath}{2}\cos\theta_{k}^{(0)}\, e_{lm}^{-}\Bigr)+b_{l}\Bigl(\frac{1}{2}e_{lm}^{+}\Bigr)\biggr]\biggr\},\notag\displaybreak[0]\\
E_{y}(\bm{x})= & \;\int{\textbf{d}}\bm{\Omega}_{k}^{(0)}\, e^{\imath\bm{k}^{(0)}\bm{\cdot x}}\biggl\{\sum_{l}\frac{(-\imath)^{l}}{4\pi}\biggl[a_{l}\Bigl(m\sin\theta_{k}^{(0)}\cos\phi_{k}Y_{lm}-\frac{1}{2}\cos\theta_{k}^{(0)}\, e_{lm}^{+}\Bigr)+b_{l}\Bigl(\frac{-\imath}{2}e_{lm}^{-}\Bigr)\biggr]\biggr\},\notag\displaybreak[0]\\
E_{z}(\bm{x})= & \;\int{\textbf{d}}\bm{\Omega}_{k}^{(0)}\, e^{\imath\bm{k}^{(0)}\bm{\cdot x}}\biggl\{\sum_{l}\frac{(-\imath)^{l}}{4\pi}\biggl[a_{l}\Bigl(\frac{\imath}{2}\sin\theta_{k}^{(0)}\cos\phi_{k}\, e_{lm}^{-}+\frac{1}{2}\sin\theta_{k}^{(0)}\sin\phi_{k}\, e_{lm}^{+}\Bigr)+b_{l}\Bigl(mY_{lm}\Bigr)\biggr]\biggr\},\label{eqn: E cartesian}\end{align}
 where $Y_{lm}$ is understood to mean $Y_{lm}(\bm{\Omega}_{k}^{(0)})$.
\emph{The quantities inside the curly braces in (\ref{eqn: E cartesian})
will be denoted by} $\tilde{E}_{x}(\bm{k})$, $\tilde{E}_{y}(\bm{k})$,
and $\tilde{E}_{z}(\bm{k})$, where the {}``(0)'' superscript on
$\bm{k}$ has been dropped. We will also need $\tilde{\bm{E}}(\bm{k^{'}})$
where $\bm{k^{'}}(\bm{k})=(k_{\rho},k_{\phi},-k_{z})$ as before.
Again using (\ref{eqn: Ylm parity}) we find \begin{align}
\tilde{E}_{x}(\bm{k^{'}})= & \;\sum_{l}\frac{(-\imath)^{l}}{4\pi}(-1)^{l+m}\biggl[a_{l}\Bigl(-m\sin\theta_{k}^{(0)}\sin\phi_{k}Y_{lm}-\frac{\imath}{2}\cos\theta_{k}^{(0)}\, e_{lm}^{-}\Bigr)+b_{l}\Bigl(\frac{-1}{2}e_{lm}^{+}\Bigr)\biggr],\notag\displaybreak[0]\\
\tilde{E}_{y}(\bm{k^{'}})= & \;\sum_{l}\frac{(-\imath)^{l}}{4\pi}(-1)^{l+m}\biggl[a_{l}\Bigl(m\sin\theta_{k}^{(0)}\cos\phi_{k}Y_{lm}-\frac{1}{2}\cos\theta_{k}^{(0)}\, e_{lm}^{+}\Bigr)+b_{l}\Bigl(\frac{\imath}{2}e_{lm}^{-}\Bigr)\biggr],\notag\displaybreak[0]\\
\tilde{E}_{z}(\bm{k^{'}})= & \;\sum_{l}\frac{(-\imath)^{l}}{4\pi}(-1)^{l+m}\biggl[a_{l}\Bigl(\frac{-\imath}{2}\sin\theta_{k}^{(0)}\cos\phi_{k}\, e_{lm}^{-}+\frac{-1}{2}\sin\theta_{k}^{(0)}\sin\phi_{k}\, e_{lm}^{+}\Bigr)+b_{l}\Bigl(mY_{lm}\Bigr)\biggr].\end{align}

\subsubsection{The M1 equations for s-polarization}

We will now define $f_{lm}$ and $g_{lm}$ by \begin{align}
\tilde{\bm{E}}(\bm{k})\bm{\cdot\epsilon}_{s,k}\equiv\sum_{l}\frac{(-\imath)^{l}}{4\pi}\left(a_{l}g_{lm}+b_{l}f_{lm}\right).\label{eqn: E dot epsilon S, MB}\end{align}
 Performing the dot product using (\ref{eqn: epsilon definitions})
yields \begin{align}
f_{lm}= & \;\frac{\imath}{2}\left(d_{lm}^{+}Y_{lm+1}e^{-\imath\phi_{k}}-d_{lm}^{-}Y_{lm-1}e^{\imath\phi_{k}}\right),\notag\displaybreak[0]\\
g_{lm}= & \;\frac{1}{2}\cos\theta_{k}^{(0)}\left(d_{lm}^{+}Y_{lm+1}e^{-\imath\phi_{k}}+d_{lm}^{-}Y_{lm-1}e^{\imath\phi_{k}}\right)\notag\\
 & -m\sin\theta_{k}^{(0)}Y_{lm}\notag\\
= & \;\frac{1}{2}\sqrt{\frac{2l+1}{2l-1}}\biggl(\sqrt{(l-m)(l-m-1)}Y_{l-1,m+1}e^{-\imath\phi_{k}}\notag\\
 & +\sqrt{(l+m)(l+m-1)}Y_{l-1,m-1}e^{\imath\phi_{k}}\biggr).\end{align}
 The simplification in the last step has been done using recursion
relations for the $P_{l}^{m}$. Performing the dot product for $\tilde{\bm{E}}(\bm{k^{'}})$
yields \begin{align}
\tilde{\bm{E}}(\bm{k^{'}})\bm{\cdot\epsilon}_{s,k}=\sum_{l}\frac{(-\imath)^{l}}{4\pi}(-1)^{l+m}\left(a_{l}g_{lm}-b_{l}f_{lm}\right).\label{eqn: E dot epsilon S, k-prime, MB}\end{align}
 The M1 relation (from (\ref{eqn: M1 eqn, PWB, vector S raw})) is
$\tilde{\bm{E}}(\bm{k})\bm{\cdot\epsilon}_{s,k}=\bar{r}_{s}\tilde{\bm{E}}(\bm{k^{'}})\bm{\cdot\epsilon}_{s,k}$
for $k_{z}>0$ and $\tilde{\bm{E}}(\bm{k^{'}})\bm{\cdot\epsilon}_{s,k}=\bar{r}_{s}\tilde{\bm{E}}(\bm{k})\bm{\cdot\epsilon}_{s,k}$
for $k_{z}<0$. These lead to an M1 equation that holds for all $\bm{\Omega}_{k}^{(0)}$:
\begin{align}
\sum_{l=l_{\text{min}}}^{l_{\text{max}}}\frac{(-\imath)^{l}}{4\pi}\left\{ \begin{matrix}a_{l}s_{1}g_{lm}+b_{l}s_{2}f_{lm}, & l+m\text{ even}\\
a_{l}s_{2}g_{lm}+b_{l}s_{1}f_{lm}, & l+m\text{ odd}\end{matrix}\right\} =0,\label{eqn: MB, vector S M1 eqn}\end{align}
 where $s_{1}\equiv1-\bar{r}_{s}(|\cos\theta_{k}^{(0)}|)$ and $s_{2}\equiv{\textrm{sgn}}(\cos\theta_{k}^{(0)})(1+\bar{r}_{s}(|\cos\theta_{k}^{(0)}|))$.

\subsubsection{The M1 equations for p-polarization}

The M1 equation for p-polarization is obtained the same way as above.
The calculation is simplified by temporarily switching phase conventions.
We define the unit vector $\tilde{\bm{\epsilon}}_{p,k}\equiv{\textrm{sgn}}(\cos\theta_{k}^{(0)})\bm{\epsilon}_{p,k}$
(using (\ref{eqn: epsilon definitions})) so that \begin{align}
\tilde{\bm{\epsilon}}_{p,k}= & \;\hat{\bm{x}}\cos\theta_{k}^{(0)}\cos\phi_{k}+\hat{\bm{y}}\cos\theta_{k}^{(0)}\sin\phi_{k}-\hat{\bm{z}}\sin\theta_{k}^{(0)},\notag\\
\tilde{\bm{\epsilon}}_{p,k^{'}}= & \;-\hat{\bm{x}}\cos\theta_{k}^{(0)}\cos\phi_{k}-\hat{\bm{y}}\cos\theta_{k}^{(0)}\sin\phi_{k}-\hat{\bm{z}}\sin\theta_{k}^{(0)}.\end{align}
 The M1 relation is now $\tilde{\bm{E}}(\bm{k})\bm{\cdot}\tilde{\bm{\epsilon}}_{p,k}=(-\bar{r}_{p})\tilde{\bm{E}}(\bm{k^{'}})\bm{\cdot}\tilde{\bm{\epsilon}}_{p,k^{'}}$
for $k_{z}>0$ and $\tilde{\bm{E}}(\bm{k^{'}})\bm{\cdot}\tilde{\bm{\epsilon}}_{p,k^{'}}=(-\bar{r}_{p})\tilde{\bm{E}}(\bm{k})\bm{\cdot}\tilde{\bm{\epsilon}}_{p,k}$
for $k_{z}<0$, where $\bar{r}_{p}$ has not changed in any way. Performing
the dot products, we find that \begin{align}
\tilde{\bm{E}}(\bm{k})\bm{\cdot}\tilde{\bm{\epsilon}}_{p,k}= & \;\sum_{l}\frac{(-\imath)^{l}}{4\pi}(-a_{l}f_{lm}+b_{l}g_{lm}),\notag\\
\tilde{\bm{E}}(\bm{k^{'}})\bm{\cdot}\tilde{\bm{\epsilon}}_{p,k^{'}}= & \;\sum_{l}\frac{(-\imath)^{l}}{4\pi}(-1)^{l+m}(a_{l}f_{lm}+b_{l}g_{lm}).\label{eqn: E dot epsilon P, MB}\end{align}
 This leads to the M1 equation for p-polarization which holds for
all $\bm{\Omega}_{k}^{(0)}$: \begin{align}
\sum_{l=l_{\text{min}}}^{l_{\text{max}}}\frac{(-\imath)^{l}}{4\pi}\biggl\{\begin{matrix}-a_{l}p_{2}f_{lm}+b_{l}p_{1}g_{lm}, & l+m\text{ even}\\
-a_{l}p_{1}f_{lm}+b_{l}p_{2}g_{lm}, & l+m\text{ odd}\end{matrix}\biggr\}=0,\label{eqn: MB, vector P M1 eqn}\end{align}
 where $p_{1}\equiv1+\bar{r}_{p}(|\cos\theta_{k}^{(0)}|)$ and $p_{2}\equiv{\textrm{sgn}}(\cos\theta_{k}^{(0)})(1-\bar{r}_{p}(|\cos\theta_{k}^{(0)}|))$.

\subsubsection{Variant 1}

Variant 1 for the vector case proceeds exactly as it does for the
scalar case. The M1 equations (\ref{eqn: MB, vector S M1 eqn}) and
(\ref{eqn: MB, vector P M1 eqn}) are seen to be even in $\cos\theta_{k}^{(0)}$,
allowing us to pick directions only in the domain $[0,\pi/2]$. Thus
for each discrete direction $\alpha_{k}^{(0)}$ we get one equation
(in the M1 portion of $A$) for each polarization. No modification
of (\ref{eqn: MB, vector S M1 eqn}) or (\ref{eqn: MB, vector P M1 eqn})
is needed, except that we can take $\phi_{k}=0$ and $\cos\theta_{k}^{(0)}=\cos\alpha_{k}^{(0)}\geq0$.

The complexity of calculating the M1 portion of $A$ is of the same
order in $l_{\text{max}}$ as it is for the scalar MB. However, numerous
constant factors make the computation time an order of magnitude longer,
as there are more $P_{l}^{m}$ functions to compute and $\bar{r}_{p}$
must be computed in addition to $\bar{r}_{s}$.

\subsubsection{Variant 2}

In this minor section, we assume that $m\geq0$ (see Appendix \ref{sec: modetype}
for plotting negative $m$ modes from positive $m$ solutions). We
follow the same procedure used in variant 2 for the scalar field.
Integrating (\ref{eqn: MB, vector S M1 eqn}) and (\ref{eqn: MB, vector P M1 eqn})
against $Y_{l^{'}m^{'}}(\bm{\Omega}_{k}^{(0)})$ yields zero if either
$m^{'}\neq m$ or $l^{'}+m$ is odd. For $m^{'}=m$ and $l^{'}+m$
even, the M1 equation associated with each $l^{'}$ in $\{|m|,|m|+1,\ldots,l_{\text{max}}\}$
for s-polarization is \begin{multline}
\sum_{l=l_{\text{min}}}^{l_{\text{max}}}-\frac{1}{2}\zeta\,\biggl[a_{l}\int_{0}^{1}\Bigl(1-(-1)^{l+m}\,\bar{r}_{s}\Bigr)\Bigl(P_{l-1}^{m+1}(x)+(l+m)(l+m-1)P_{l-1}^{m-1}(x)\Bigr)P_{l^{'}}^{m}(x)\,\text{d}x\\
+b_{l}(\imath)\int_{0}^{1}\Bigl(1+(-1)^{l+m}\,\bar{r}_{s}\Bigr)\Bigl(P_{l}^{m+1}(x)-(l+m)(l-m+1)P_{l}^{m-1}(x)\Bigr)P_{l^{'}}^{m}(x)\,\text{d}x\biggr]=0,\label{eqn: MB vector S M1 eqn, variant2}\end{multline}
 and the M1 equation for p-polarization is \begin{multline}
\sum_{l=l_{\text{min}}}^{l_{\text{max}}}-\frac{1}{2}\zeta\,\biggl[a_{l}(-\imath)\int_{0}^{1}\Bigl(1-(-1)^{l+m}\,\bar{r}_{p}\Bigr)\Bigl(P_{l}^{m+1}(x)-(l+m)(l-m+1)P_{l}^{m-1}(x)\Bigr)P_{l^{'}}^{m}(x)\,\text{d}x\\
+b_{l}\int_{0}^{1}\Bigl(1+(-1)^{l+m}\,\bar{r}_{p}\Bigr)\Bigl(P_{l-1}^{m+1}(x)+(l+m)(l+m-1)P_{l-1}^{m-1}(x)\Bigr)P_{l^{'}}^{m}(x)\,\text{d}x\biggr]=0,\label{eqn: MB vector P M1 eqn, variant2}\end{multline}
 where $P_{\mu}^{n}=0$ for $|n|>\mu$. Note that if $m=0$ the equations
for $l^{'}=0$ \emph{are} included. We implemented this by calculating
an array of each of the following types of integrals: \begin{align}
\begin{matrix}\int_{0}^{1}P_{\mu}^{m+1}P_{\nu}^{m}\,\text{d}x, & \int_{0}^{1}P_{\mu}^{m-1}P_{\nu}^{m}\,\text{d}x,\\
\int_{0}^{1}P_{\mu}^{m+1}P_{\nu}^{m}\bar{r}_{s}\,\text{d}x, & \int_{0}^{1}P_{\mu}^{m-1}P_{\nu}^{m}\bar{r}_{s}\,\text{d}x,\\
\int_{0}^{1}P_{\mu}^{m+1}P_{\nu}^{m}\bar{r}_{p}\,\text{d}x, & \int_{0}^{1}P_{\mu}^{m-1}P_{\nu}^{m}\bar{r}_{p}\,\text{d}x.\end{matrix}\notag\end{align}
 The integrals in the first row can be calculated ahead of time and
stored in a data file. The other integrals must be calculated each
time the matrix $A$ is calculated. The complexity of the calculation
is of the same order in $l_{\text{max}}$ as for variant 2 in the
scalar case. These integrals are very numerically intensive and in
practice variant 2 takes much longer than variant 1. We have found
good agreement between the two variants, indicating that variant 1
can be used all or most of the time.

\subsection{The Linear System of Equations in the Vector MB}

The M2 equations are found in the usual way, choosing locations $(\rho_{*},\theta_{*})$
with $\phi=0$ and constructing the equations $E_{\phi}=0$, $E_{\parallel}=E_{\rho}\cos\eta-E_{z}\sin\eta=0$,
and $H_{\perp}=H_{\rho}\sin\eta+H_{z}\cos\eta=0$. The fields $\bm{E}$
and $\bm{H}$ are found by substituting (\ref{eqn: M and N explicit})
into (\ref{eqn: VSH expansion of E, H}). The derivatives in (\ref{eqn: M and N explicit})
are \begin{align}
\frac{\partial}{\partial r}(rj_{l}(kn_{0}r))=kn_{0}rj_{l-1}(kn_{0}r)-lj_{l}(kn_{0}r),\label{eqn: d (rj) /dr}\end{align}
 and, for $m\geq0$, \begin{multline}
\frac{\partial}{\partial\theta}Y_{lm}(\theta,\phi)=\frac{1}{2}\left[\frac{(2l+1)(l-m)!}{4\pi(l+m)!}\right]^{1/2}\\
\times\left(P_{l}^{m+1}-(l+m)(l-m+1)P_{l}^{m-1}\right)e^{\imath m\phi},\label{eqn: d (Y) /dtheta}\end{multline}
 where $P_{l}^{l+1}=0$. (For negative $m$ use $Y_{l,-m}=(-1)^{m}Y_{lm}^{*}$.)
The number of columns in $A$ is $2N_{l}$ and the number of rows
is $(2N_{\text{M1 dirs}}+3N_{\text{M2 loc}}+1)$ for variant 1 and
about $(N_{l}+3N_{\text{M2 loc}}+1)$ for variant 2.

\subsection{Calculating the Field in the Layers with the Vector MB}

To calculate the field in layer $0$, the expansion (\ref{eqn: VSH expansion of E, H})
is used directly, with (\ref{eqn: M and N explicit}), (\ref{eqn: d (rj) /dr}),
and (\ref{eqn: d (Y) /dtheta}) being used to obtain the explicit
form. As for the scalar case, there is no direct expansion for the
layers $q>0$ in the MB, and we must use the Bessel wave basis. Using
(\ref{eqn: E dot epsilon S, MB}), (\ref{eqn: E dot epsilon S, k-prime, MB}),
and (\ref{eqn: E dot epsilon P, MB}), the conversion is seen to be
\begin{align}
S_{u} & =\sum_{l=l_{\text{min}}}^{l_{\text{max}}}\frac{(-\imath)^{l}}{4\pi}\left(a_{l}g_{lm}+b_{l}f_{lm}\right)\biggl|_{\theta_{k}^{(0)}=\alpha_{k}^{(0)},\phi_{k}=0},\notag\displaybreak[0]\\
S_{d} & =\sum_{l}\frac{(-\imath)^{l}}{4\pi}(-1)^{l+m}\left(a_{l}g_{lm}-b_{l}f_{lm}\right)\biggl|_{\theta_{k}^{(0)}=\alpha_{k}^{(0)},\phi_{k}=0},\notag\displaybreak[0]\\
P_{u} & =\sum_{l}\frac{(-\imath)^{l}}{4\pi}\left(-a_{l}f_{lm}+b_{l}g_{lm}\right)\biggl|_{\theta_{k}^{(0)}=\alpha_{k}^{(0)},\phi_{k}=0},\notag\displaybreak[0]\\
P_{d} & =\sum_{l}\frac{(-\imath)^{l}}{4\pi}(-1)^{l+m}\left(a_{l}f_{lm}+b_{l}g_{lm}\right)\biggl|_{\theta_{k}^{(0)}=\alpha_{k}^{(0)},\phi_{k}=0}.\label{eqn: vector MB to PWB conversion}\end{align}
 This is the vector case analogue of Eqn. (\ref{eqn: scalar MB to PWB conversion}).

At this point the Bessel wave coefficients $S_{u}$, etc., are continuous
functions of $\alpha_{k}^{(0)}$, not discrete as they were in the
Bessel wave method. These continuous coefficients are substituted
into the integrals (\ref{eqn; plus/minus S/P expansions, using T matrix}),
and finally (\ref{eqn:ErhoandEphi}) is used to give the vector fields
in the layers. As in the scalar case, however, doing the integrals
with an adaptive algorithm is impractical if one is plotting the vector
fields in a sizable region. In this case, it is best to make the Bessel
wave coefficients discrete and use a simple integration method, or,
as was suggested in the scalar case, to pick a $N_{\text{dirs}}$
and create a PWB solution vector that can then be plotted via a PWB
plotting routine.

\section{\label{sec: demonstrations}Demonstrations and Comparisons}

In this section we will demonstrate, but not thoroughly analyze, several
results that we have obtained using our model and methods. The modes
shown here all have $\Delta_{n}<0.0002$ and do not contain large
amounts of high-angle plane waves, which in principle can cause problems
(see Appendix \ref{sec: exclusion high-angle}). The authors expect
to publish a separate paper focusing on the modes themselves. In this
section we will also compare our implementations of the two-basis
method and the Bessel wave method. Before reading this section it
may be helpful to look at Appendix \ref{sec: modetype}.

\subsection{\label{sec: V} The {}``V'' Mode: A Stack Effect}

\begin{figure}[!htb]
\begin{center}\includegraphics[%
  width=0.60\columnwidth]{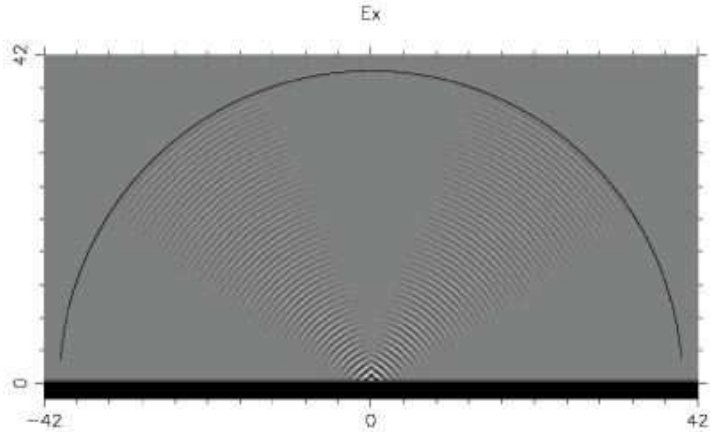}\end{center}

\caption{\label{fig: V mode} Contrast-enhanced plot of $\text{Re}E_{x}$
in the $y$-$z$ plane for a {}``V'' mode ($m=1,k=8.1583-\imath0.000019$).
M2 is spherical and origin-centered; $R=40$; $z_{1}=0.3$; $z_{e}=3$;
M1 = stack II; $k_{s}=8.1600$; $N_{s}=20$. In this and other side
view plots, the field outside the cavity has been set to zero. }
\end{figure}

\begin{figure}[!htb]
\begin{center}\includegraphics[%
  width=0.75\columnwidth]{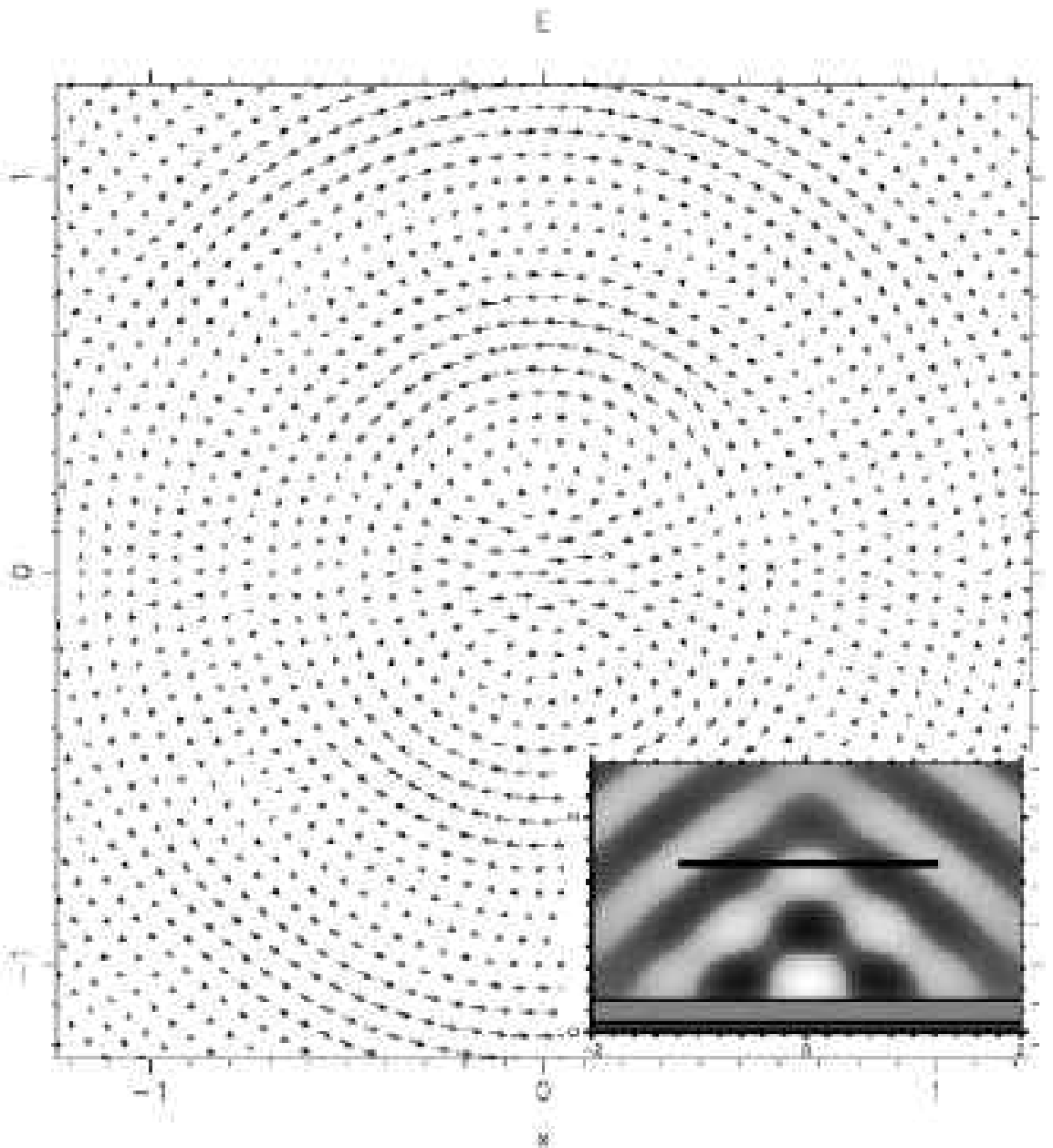}\hspace*{-1cm}\end{center}

\caption{\label{fig: V mode, XY} An $x$-$y$ cross section of the mode in
Fig. \!\ref{fig: V mode} at $z=1.58$. Units are in $\mu$m as usual.
For $m=+1$, the forward time evolution is simply the counter-clockwise
rotation of the entire vector field. If $m=+1$ and $m=-1$ are mixed
to create a cosine mode, the arrow directions remain fixed and their
lengths simply oscillate sinusoidally in time so that the mode is
associated with $x$ polarization. The inset shows the location of
the cross section in the view of Fig. \!\ref{fig: V mode}. (The
field is plotted only in layer 0.) }
\end{figure}

\begin{figure}[!htb]
\begin{center}\includegraphics[%
  width=0.65\columnwidth]{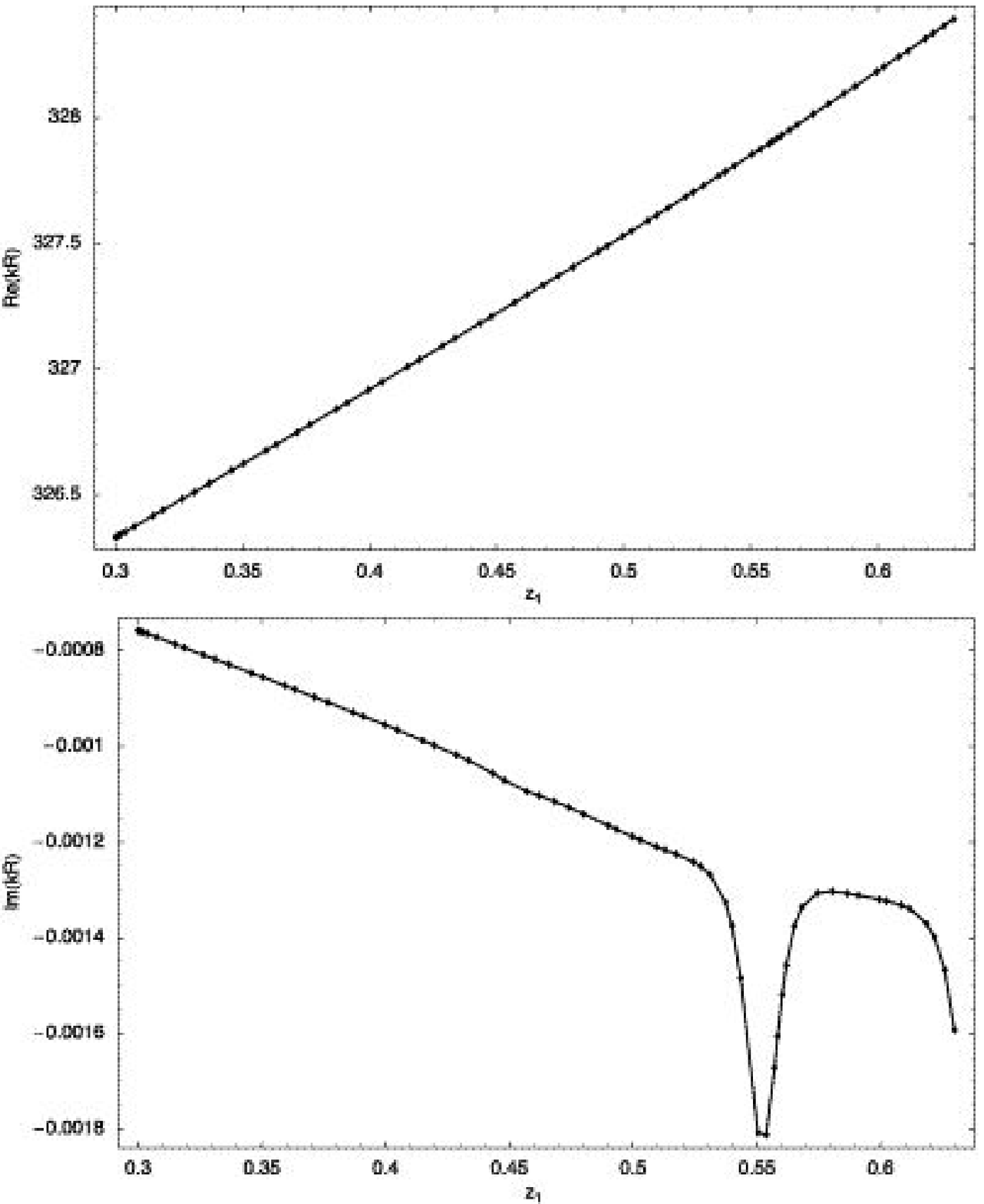}\hspace*{-1.4cm}\end{center}

\caption{\label{fig: folgraphs} Following $kR$ as $z_{1}$ is varied. The
points of the two graphs correspond one-to-one, with the parameter
$z_{1}$ increasing from left to right in both graphs. The peaks in
resonance width have not yet been analyzed.}
\end{figure}

\begin{figure}[!htb]
\includegraphics[%
  width=0.85\columnwidth]{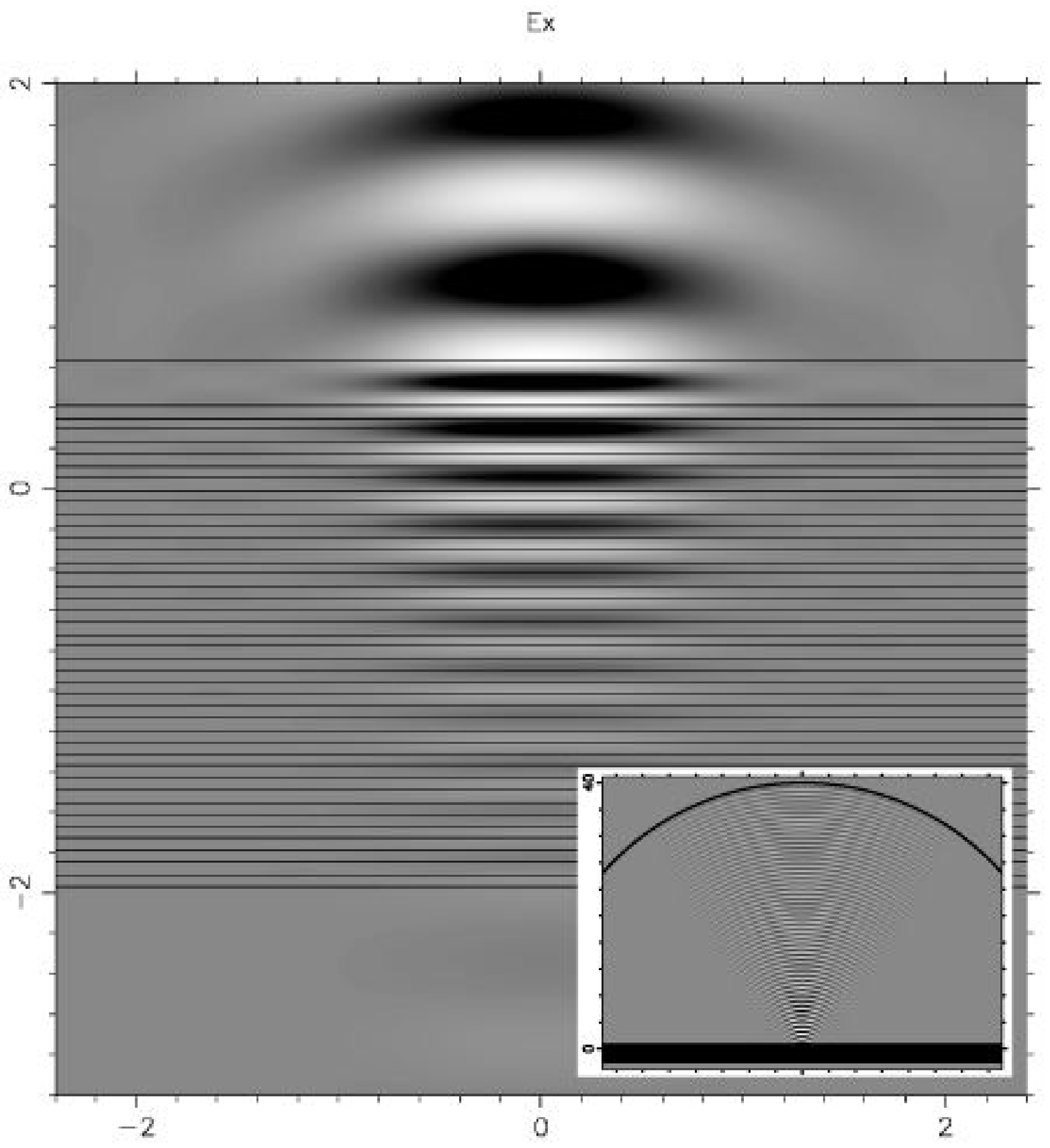}\hspace*{-1cm}

\caption{\label{fig: zoom} The field in layers 0--X for the mode at $z_{1}=0.63$.
The inset shows the entire mode. The {}``V'' nature of the mode
has been lost as it has become more paraxial and more like the fundamental
Gaussian. $\text{Re}\bm{E}_{T}$ everywhere lies nearly in a single
direction at any instant (here it is in the $x$ direction). The $x$-$z$
plane is shown here, although the plots of $E_{x}$ in any plane containing
the $z$ axis are very similar. Here $k=8.2098-\imath0.0000401$. }
\end{figure}

As mentioned several times in the Introduction, the correct treatment
of a dielectric stack can be essential for getting results that are
even qualitatively correct%
\footnote{It is not always essential however, see Note Added in Proof.%
}. One of the most remarkable differences that we have observed when
switching from simple to dielectric M1 mirrors occurs right where
someone looking for highly-focused, drivable modes would be interested:
the widening behavior of the fundamental Gaussian mode (the 00 mode).
A Gaussian mode of a near-hemispherical cavity will become more and
more focused (wide at the curved mirror and narrow at the flat mirror)
as $z_{1}$ is decreased (from a starting value for which the Gaussian
mode is paraxial). As the mode becomes more focused, of course, the
paraxial approximation becomes less valid and at some point Gaussian
theory no longer applies. We have found, using realistic Al$_{1-x}$Ga$_{x}$As--AlAs
stack models (stacks I and II described in Appendix \ref{sec: stacks}),
that the 00 mode splits into two parts, so that in the side view a
{}``V'' shape is formed. Figure \ref{fig: V mode} shows a split
mode and Fig. \!\ref{fig: V mode, XY} shows the physical transverse
electric field, $\text{Re}\bm{E}_{T}\equiv\text{Re}E_{x}\hat{\bm{x}}+\text{Re}E_{y}\hat{\bm{y}}$,
for this mode near the focal region. Figure \ref{fig: folgraphs}
shows the values of $\text{Re}(kR)$ as $z_{1}$ is changed from 0.3
to about 0.63. Figure \ref{fig: zoom} shows a zoomed view of the
field at the focus of the mode at $z_{1}=0.63$, where it is qualitatively
a 00 mode. If a conducting mirror is used, the central cone simply
grows wider and wider, but does not split. The V mode is predominately
s-polarized and appears in the scalar problem as well. Thus it appears
that the V mode is a result of the non-constancy of $\arg(r_{s}(\alpha))$
for a dielectric stack. We note here that following the 00 mode as
$z_{1}$ is changed is an imperfect process: it is possible that the
following procedure may skip over narrow anticrossings that are difficult
to resolve. However, the \emph{character} of the mode is maintained
through such anticrossings.

The apparent splitting of the central cone/lobe for higher order Gaussian
modes has also been observed. In the next section we look at some
higher order Gaussian modes, focusing not on what occurs at the breaking
of the paraxial condition, but on what is allowed and observed for
modes that are very paraxial.%

\subsection{\label{sec: Laguerre} Persistent Stack-Induced Mixing of Near-Degenerate
Laguerre-Gaussian Mode Pairs}

As mentioned in the previous section, we do observe the fundamental
Gaussian mode for paraxial cavities with both dielectric and conducting
planar mirrors. The situation becomes more interesting when we look
at higher order Gaussian modes. It is true that, as modes become increasingly
paraxial, Gaussian theory must apply. However, the way in which it
applies allows the design of M1 to play a significant role. Here we
will demonstrate the ability of a dielectric stack (stack II) to mix
near-degenerate pairs predicted by Gaussian theory into new, more
complicated near-degenerate pairs of modes. We must provide considerable
background to put these modes into context. Our discussion applies
to modes with paraxial geometry, modes for which the paraxial parameter
$h\equiv\lambda/(\pi w_{0})=w_{0}/z_{R}$ is much less than 1. (Here
$w_{0}$ is the waist radius and $z_{R}$ is the Rayleigh range, as
used in standard Gaussian theory.) In this section, the solutions
we are referring to are always the $+m$ or $-m$ modes discussed
in Appendix \ref{sec: modetype}, and never mixtures of these, such
as the cosine and sine modes discussed in the same section.

\subsubsection{\label{sec: Laguerre theory} Paraxial Theory for Vector Fields}

From Gaussian theory, using the Laguerre-Gaussian (LG) basis, we expect
that the transverse electric field of any mode in the paraxial limit
is expressible in the form: \begin{align}
\bm{E}_{T}=\sum_{j=0}^{N}\Bigl[A_{j}^{+}\Bigl(\begin{smallmatrix}1\\
\imath\end{smallmatrix}\Bigr)+A_{j}^{-}\Bigl(\begin{smallmatrix}1\\
-\imath\end{smallmatrix}\Bigr)\Bigr]\text{LG}_{\min(j,N-j)}^{2j-N}(\bm{x}).\label{eqn: LG expansion}\end{align}
 Here $N\geq0$ is the order of the mode. The $A_{j}^{\pm}$ are complex
coefficients and $\bigl(\begin{smallmatrix}1\\
\imath\end{smallmatrix}\bigr)$ and $\bigl(\begin{smallmatrix}1\\
-\imath\end{smallmatrix}\bigr)$ are the Jones' vectors for right and left circular polarization,
respectively. The explicit forms of the normalized $\text{LG}_{\min(j,N-j)}^{2j-N}$
functions are given in Ref. \cite{Beijersbergen} as {}``$u_{nm}^{\text{LG}}$''
with the substitutions $n\rightarrow N-j$ and $m\rightarrow j$.
The important aspects of the $\text{LG}_{\min(j,N-j)}^{2j-N}$ functions
are that the $\phi$-dependence is $\exp(\imath(2j-N)\phi)$, and
the $\rho$-dependence includes the factor $L_{\min(j,N-j)}^{|2j-N|}(2\rho^{2}/w(z)^{2})$
where $L_{p}^{l}$ is a generalized Laguerre function and $w(z)$
is the beam radius. In the paraxial limit the vector eigenmodes of
the same order become degenerate. There are $2(N+1)$ independent
vector eigenmodes present in the expansion (\ref{eqn: LG expansion}).

\begin{table}[!htb]
\begin{center}\includegraphics[%
  width=0.55\columnwidth]{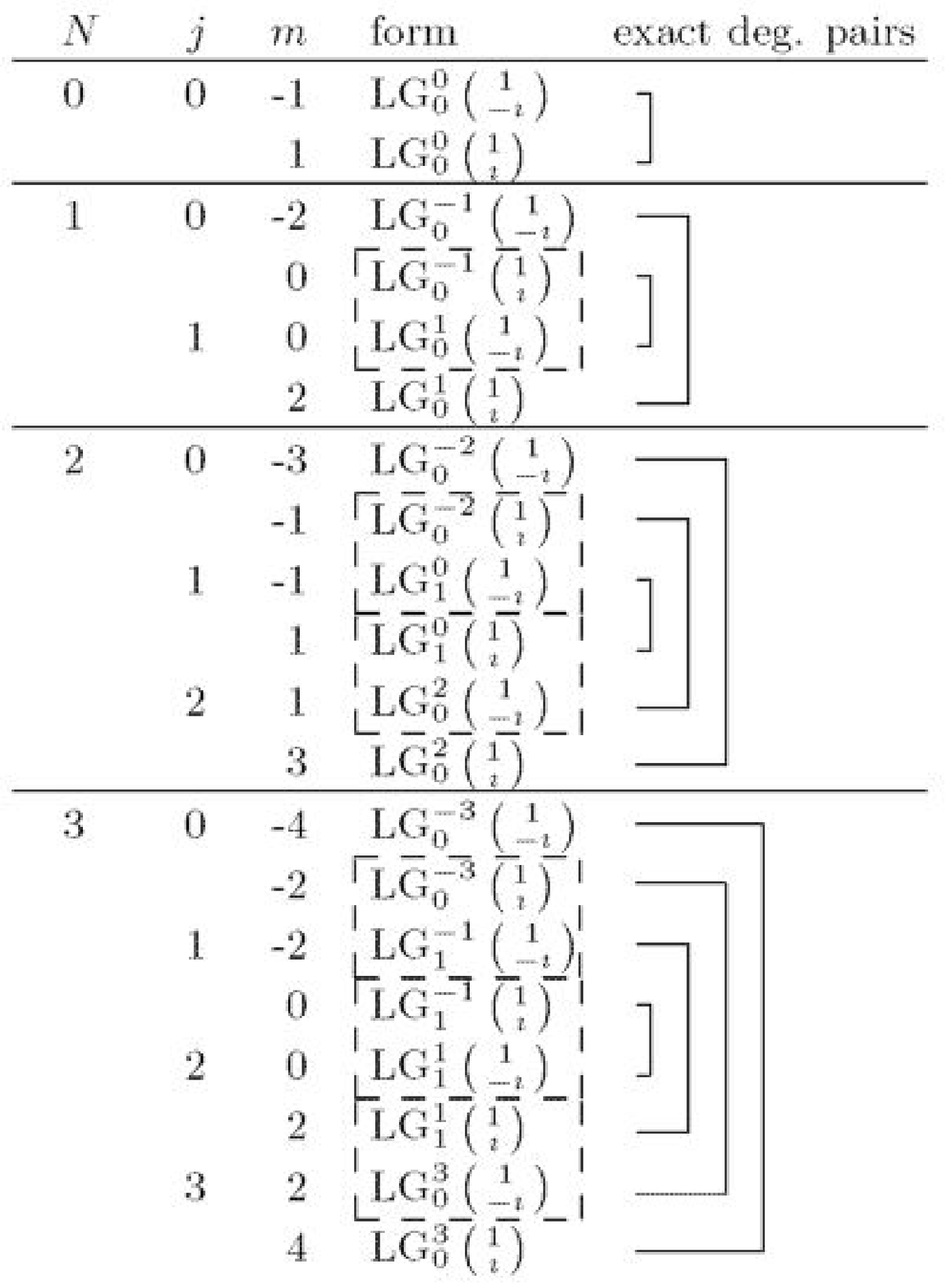}\hspace*{-0.4cm}\end{center}

\caption{\label{tab: LG modes} Vector Laguerre-Gaussian modes.}
\end{table}

Independent from the discussion above, $\bm{E}_{T}=E_{\rho}\hat{\bm{\rho}}+E_{\phi}\hat{\bm{\phi}}$
can be written as \begin{align}
\bm{E}_{T}=\frac{1}{2}\underbrace{(E_{\rho}+\imath E_{\phi})}_{\propto\,\exp(\imath m\phi)}e^{\imath\phi}\Bigl(\begin{matrix}1\\
-\imath\end{matrix}\Bigr)+\frac{1}{2}\underbrace{(E_{\rho}-\imath E_{\phi})}_{\propto\,\exp(\imath m\phi)}e^{-\imath\phi}\Bigl(\begin{matrix}1\\
\imath\end{matrix}\Bigr).\label{eqn: Et general}\end{align}
 Since the $E_{\rho}$ and $E_{\phi}$ fields have a sole $\phi$-dependence
of $\exp(\imath m\phi)$, comparing with (\ref{eqn: LG expansion})
reveals that, for a given $m$, at most two terms in (\ref{eqn: LG expansion})
are present: the $A_{j_{+}^{'}}^{-}$ and $A_{j_{-}^{'}}^{+}$ terms
where $j_{\pm}^{'}=(N+m\pm1)/2$. Explicitly, \begin{align}
(E_{\rho}\pm\imath E_{\phi})/2\approx A_{j_{\pm}^{'}}^{\mp}\text{LG}_{[N+\min(m\pm1,-m\mp1)]/2}^{m\pm1}.\label{eqn: Et explicit}\end{align}
 If the solution (\ref{eqn: Et general}) has both terms nonzero for
almost all $\bm{x}$, the constraints $0\leq j\leq N$ force $N$
to have a value given by $N=|m|+1+2\nu,\nu=0,1,2,\ldots$ . However,
if only right (left) circular polarization is present in the solution,
$N=|m|-1$ is also allowed, provided that $m\geq+1$ ($m\leq$ -1).
So, given a (paraxial) numerical solution with its $m$ value, we
can determine which orders the solutions can belong to. The reverse
procedure, picking $N$ and determining which values of $m$ are allowed
and how many independent vector solutions are associated with each
$m$, can be done by stepping $j$ in (\ref{eqn: LG expansion}) and
comparing with (\ref{eqn: Et general}) or (\ref{eqn: Et explicit}).
The results for the first four orders are summarized in Table \ref{tab: LG modes}.
The $\text{LG}_{0}^{0}\bigl(\begin{smallmatrix}1\\
\imath\end{smallmatrix}\bigr)$ and $\text{LG}_{0}^{0}\bigl(\begin{smallmatrix}1\\
-\imath\end{smallmatrix}\bigr)$ modes are the fundamental Gaussian modes.

Since the cavity modes are not perfectly paraxial, the $2(N+1)$-fold
degenerate modes from the Gaussian theory are broken into $N+1$ separate
degenerate pairs. The truly degenerate pairs for $m\neq0$ of course
consist of a $+m$ and a $-m$ mode which are related by reflection
(see Appendix \ref{sec: modetype}). The pairings are shown in the
last column of Table \ref{tab: LG modes}.

The dashed boxes in Table \ref{tab: LG modes} enclose pairs of modes
which may be mixed in a single solution for fixed $m$ (Eqn. (\ref{eqn: Et general})).
Only the mixable $m=0$ modes are exactly degenerate and may be arbitrarily
mixed. For the other mixable pairs, the degree of mixing will be fixed
by the cavity, in particular by the structure of M1. We now discuss
our results regarding the mixing of the two modes with $N=2$ and
$m=+1$.

\subsubsection{A demonstration of persistent mixing}

\begin{figure*}[!htb]
\begin{center}\includegraphics[%
  width=0.85\columnwidth]{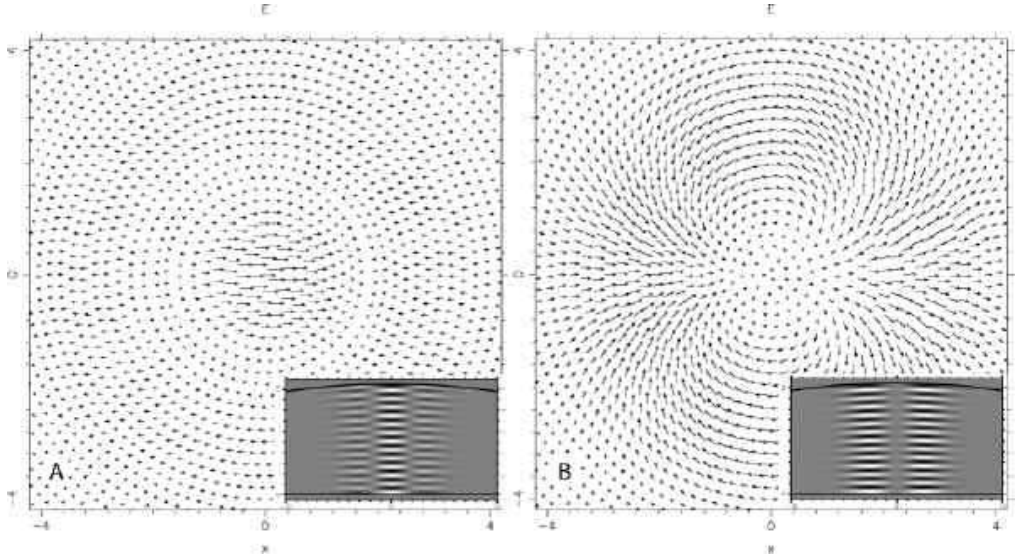}\hspace*{-1cm}\end{center}

\caption{\label{fig: LG modes A, B} 8 $\times$ 8 $\mu$m cross sections
of modes of a conducting cavity near maximum amplitude ($z=0.25$).
$k_{\text{A}}=7.89285$ and $k_{\text{B}}=7.89290$. The inset plots
show $\text{Re}E_{x}$ in the $x$-$z$ plane with horizontal and
vertical tick marks every 1 $\mu$m. }
\end{figure*}

Figure \ref{fig: LG modes A, B} shows the cross sections of two near-degenerate
$m=1$ modes found for a conducting cavity. Here $z_{1}=z_{e}=0$
and M2 is spherical with radius $R_{s}=70$ but is centered at $z=-59.5$
instead of the origin, so that $R=10.5$. The paraxial parameter $h$
is at least as small as 0.1 (see the side views in the inset plots).
Mode A corresponds very well to the pure $\text{LG}_{1}^{0}\bigl(\begin{smallmatrix}1\\
\imath\end{smallmatrix}\bigr)$ mode, while mode B corresponds very well to the pure $\text{LG}_{0}^{2}\bigl(\begin{smallmatrix}1\\
-\imath\end{smallmatrix}\bigr)$ mode. These are mixable modes, as indicated in Table \ref{tab: LG modes},
and the conducting cavity has {}``chosen'' essentially zero mixing.
Note that mode B would have zero overlap with an incident fundamental
Gaussian beam centered on the $z$ axis, while mode A would have nonzero
overlap.

\begin{figure*}[!htb]
\begin{center}\includegraphics[%
  width=0.95\columnwidth]{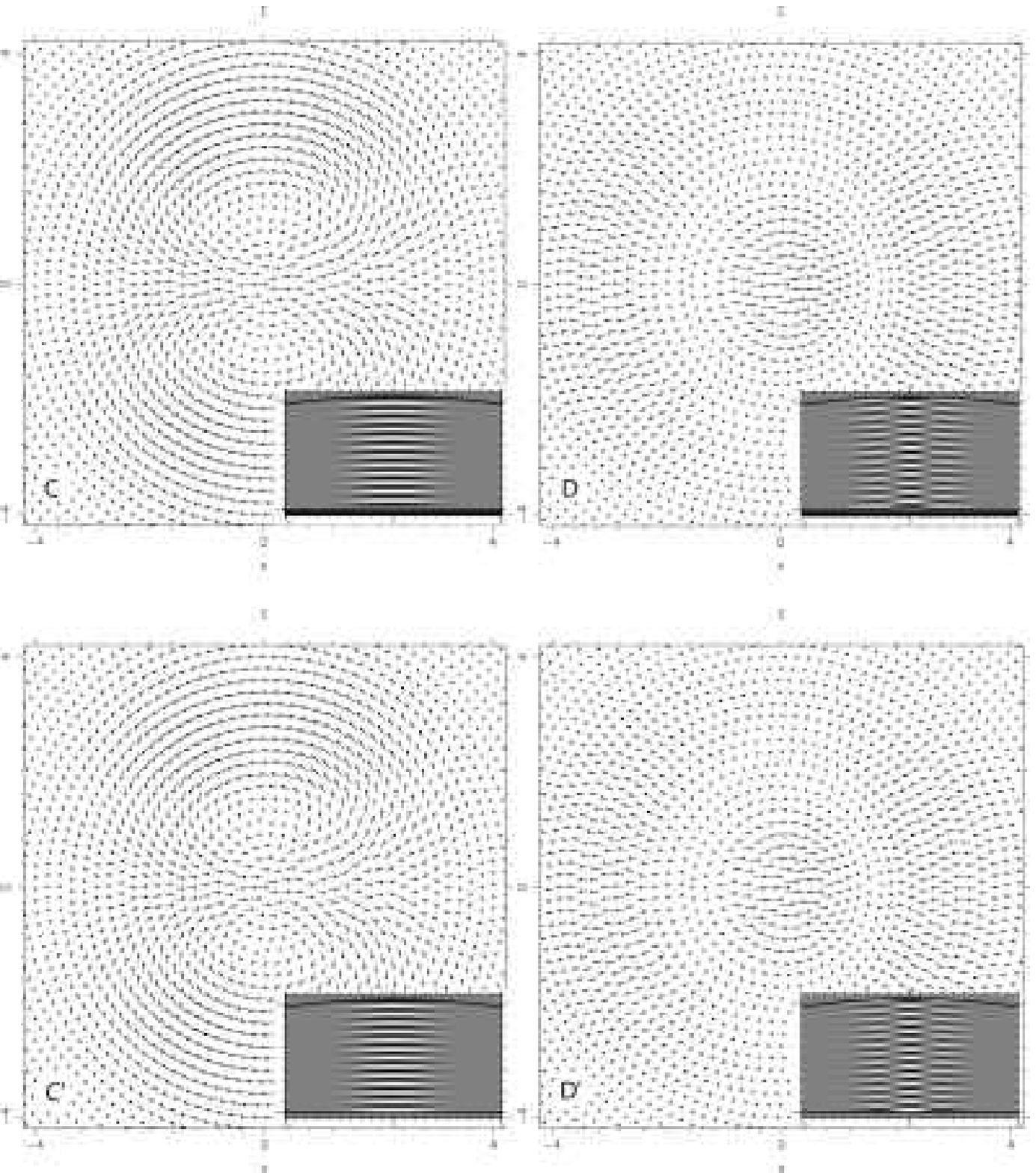}\hspace*{-1cm}\end{center}

\caption{\label{fig: LG modes C, D} Cross sections of true modes (C and D)
of a cavity with stack II near maximum amplitude ($z=0.05$), and
constructed modes (C$^{'}$ and D$^{'}$) near their maximum amplitude
($z=0.25$). The constructed cross sections are nearly identical to
the true cross section. $k_{\text{C}}=7.74685-\imath0.00005415$,
$k_{\text{D}}=7.74680-\imath0.00005516$, and the constructed modes
are plotted with $k_{\text{C}^{'}/\text{D}^{'}}=7.8929$, the average
of $k_{\text{A}}$ and $k_{\text{B}}$. The inset plots show $\text{Re}E_{x}$
in the $x$-$z$ plane. The stack parameters are $N_{s}=22$ and $k_{s}=7.746814$.}
\end{figure*}

Pictures C and D in Figure \ref{fig: LG modes C, D} are cross sections
of near-degenerate $m=1$ modes for a cavity with stack II. To show
that the transverse part of these modes are mixtures of A and B, we
have added and subtracted the solution vectors $\bm{y}_{\text{A}}$
and $\bm{y}_{\text{B}}$ to create the new (non-solution) vectors:
\begin{align}
\bm{y}_{\text{C}^{'}}=\bm{y}_{\text{A}}+\eta\bm{y}_{\text{B}}, &  & \bm{y}_{\text{D}^{'}}=\bm{y}_{\text{A}}-(0.225)\eta\bm{y}_{\text{B}},\end{align}
 and have plotted these fields (C$^{'}$ and D$^{'}$) using $k=(k_{\text{A}}+k_{\text{B}})/2$.
(The scaling factor $\eta$, here $\approx0.07$, is unphysical and
related to the effect of the seed equation on overall amplitude.)
Comparing C$^{'}$ and D$^{'}$ with C and D shows that we have reconstructed
the mode cross sections quite well (up to a constant factor). Modes
C and D are not pure Hermite-Gaussian (HG) modes, but their resemblance
to the $\text{HG}_{02}$ and $\text{HG}_{20}$ modes is not an accident.
Mode conversion formulas in Ref. \cite{Beijersbergen} show that $\text{LG}_{0}^{2}$
is made of the $\text{HG}_{02}$, $\text{HG}_{11}$, and $\text{HG}_{20}$
modes while $\text{LG}_{1}^{0}$ contains only $\text{HG}_{02}$ and
$\text{HG}_{20}$. Note that both modes C and D would couple to a
centered fundamental Gaussian beam.

\begin{figure*}[!htb]
\begin{center}\includegraphics[%
  width=0.85\columnwidth]{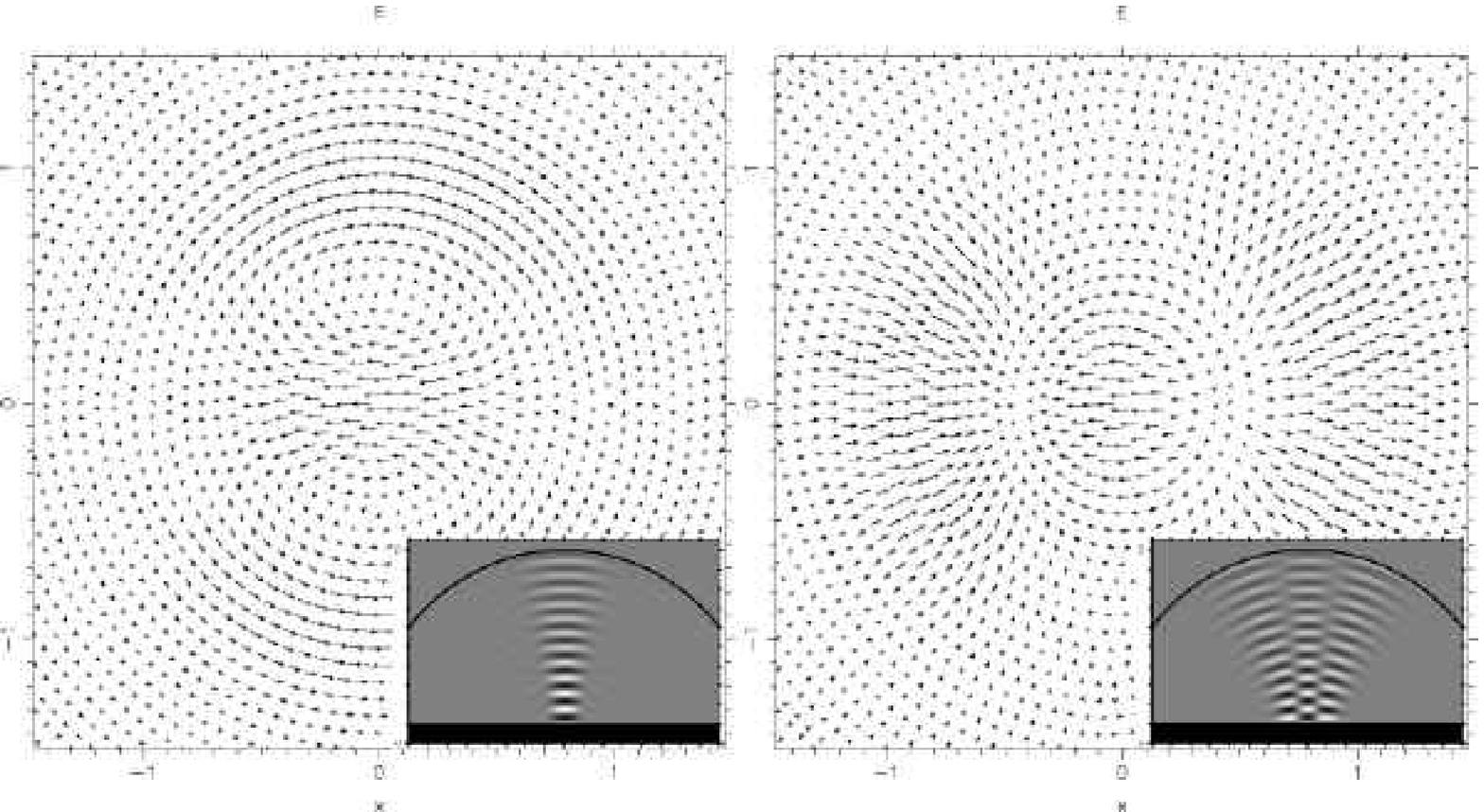}\hspace*{-1cm}\end{center}

\caption{\label{fig: LG z1} Cross sections of neighboring, poorly paraxial
modes near maximum amplitude ($z=1.33$). M2 is spherical and origin-centered
with $R=10$. M1 is stack I with $N_{s}=20$, $k_{s}=8.1600$, and
$z_{1}=z_{e}=1.0$. The mode on the left has $k=8.51160-\imath0.0002491$
and the mode on the right has $k=8.51540-\imath0.0003184$. }
\end{figure*}

An interesting property of the mixed modes C and D is that their general
forms appear to be persistent as $h$ is varied, provided that the
paraxial approximation remains sufficiently applicable. Furthermore,
we do not have to be particularly careful about setting $\lambda_{s}$
to correspond very closely to $2\pi/\text{Re}k$. The modes shown
in Figure \ref{fig: LG z1} are not nearly as paraxial as the C and
D modes. Here $k_{s}=8.1600$ which is not close to the $k$ values
of the modes. Furthermore, stack I was used which is missing the spacer
layer of stack II. Nevertheless the modes of Fig.~\ref{fig: LG z1}
bear a strong resemblance to the more carefully picked C and D modes.

\subsection{\label{sec: m neq 1 modes}Modes with $m\neq1$}

\begin{figure*}[!htb]
\begin{center}\includegraphics[%
  width=0.85\columnwidth]{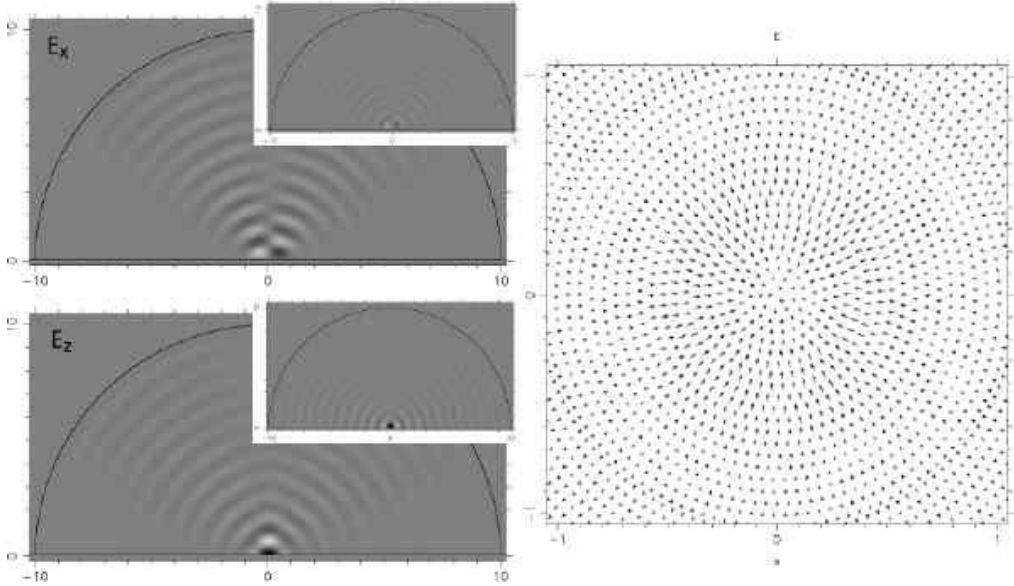}\hspace*{-1cm}\end{center}

\caption{\label{fig: m0 z 0.05} Tightly focused, radially polarized $m=0$
modes of a conducting cavity. The focusing has caused $E_{z}$ to
be greater than the $E_{\rho}$. The spot size of the dominant $E_{z}$
field for such modes can be surprisingly small, as has recently been
demonstrated experimentally by Dorn \emph{et al.}\cite{Dorn} (for
a focused beam with no cavity). Parameters of the non-inset cavity
and mode are $R=10$, $z_{1}=0.05$, and $k=7.2481$ ($\lambda=0.867$).
The cross section is at $z=0.3$. The insets show the mode after it
was followed to the perfectly hemispherical $z_{1}=0$ cavity shape;
here the mode has an even smaller central spot and $E_{z}\gg E_{\rho}$.
We expect a hemispherical mode to have only a single nonzero MB coefficient
and this was verified by the solution: $a_{1}=1.0$ and $|a_{l}|<1.1\times10^{-7}$
for $l\neq1$. The hemispherical mode has $k=7.2243$ ($\lambda=0.870$). }
\end{figure*}

The $m=0$ modes may be circularly polarized, such as the paraxial
$m=0$ modes of Table \ref{tab: LG modes}. The more interesting polarization,
perhaps, is the $m=0$ analogue of $x$ and $y$ polarization: radial
and azimuthal polarization. The mode shown in Fig.~\ref{fig: m0 z 0.05}
is radially polarized, although it is impossible to tell this from
the plots because the left and right circularly polarized modes are
radial (identical to the plot) at $t=0$ and azimuthal (like a vortex)
at $\omega t=\pi/2$. The radial or azimuthal polarizations are easier
to obtain numerically than the right or left polarizations because
radially (azimuthally) polarized modes have all of the $b_{l}$ ($a_{l}$)
MB coefficients equal to zero and thus can be selected by the seed
equation (see Appendix \ref{sec: modetype}).

\begin{figure}[!htb]
\begin{center}\includegraphics[%
  width=0.85\columnwidth]{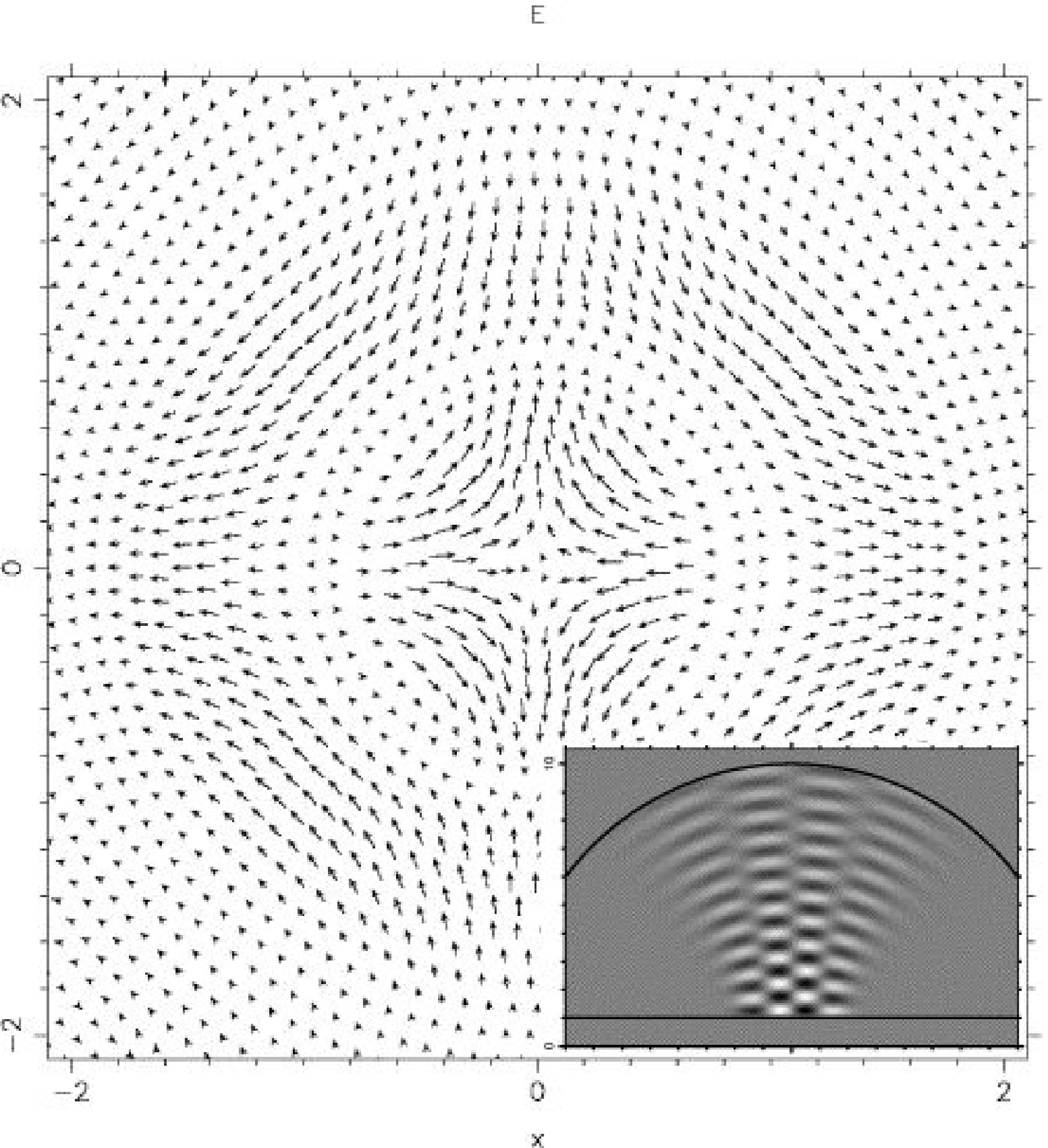}\hspace*{-1cm}\end{center}

\caption{\label{fig: m2 LG 1 1} A $m=2$ mode for a conducting cavity ($k=7.885463$).
M2 is spherical and origin-centered with $R=10$. M1 is a conductor;
$z_{1}=1.0$. The inset shows $\text{Re}E_{x}$ in the $x$-$z$ plane.}
\end{figure}

Figure \ref{fig: m2 LG 1 1} shows a $m=2$ mode that corresponds
to the $\text{LG}_{1}^{1}\bigl(\begin{smallmatrix}1\\
\imath\end{smallmatrix}\bigr)$ mode from Gaussian theory. We also have completed the $N=2$ near-degenerate
family by finding the $\text{LG}_{0}^{2}\bigl(\begin{smallmatrix}1\\
\imath\end{smallmatrix}\bigr)$ $(m=3)$ mode for both the cavity used in Fig.~\ref{fig: LG modes A, B}
($k=7.8926\approx k_{A}\approx k_{B}$) and the cavity used in Fig.~\ref{fig: LG modes C, D}
($k=7.7466-\imath0.00005466\approx k_{C}\approx k_{D}$). The cross
sections of this mode for the two cavities, which are not shown to
conserve space, appear identical, as predicted by the absence of a
$m=+3$ near-degenerate partner for this mode (Table \ref{tab: LG modes}).\vspace{0.05 in}

\subsection{\label{sec: comparing methods}$N_{\text{dirs}}$ and $l_{\text{max}}$:
Comparing the Two Primary Methods}

The authors of this paper first implemented the more complicated two-basis
method, believing that many more PWB coefficients than MB coefficients
would be required to expand dome-shaped cavity fields. When the Bessel
wave method was implemented as a check, it was discovered that the
PWB works surprisingly well, with usable values of $N_{\text{dirs}}$
being of the same order of magnitude as usable values of $l_{\text{max}}$.
The choice of seed equation, $N_{\text{dirs}}$ or $l_{\text{max}}$,
and other parameters can both affect the depth and narrowness of the
dips in the graph of $\Delta_{r}$ versus $\text{Re}k$. This makes
it difficult to perform a comprehensive and conclusive comparison
of the two methods.

One practical problem with the Bessel wave method is that its performance
is sometimes quite sensitive to the value of $N_{\text{dirs}}$. In
both methods, setting the number of basis function too high results
in a failure to solve for $\bm{y_{\text{best}}}$; the program acts
as if $A\bm{y}=\bm{b}$ were underdetermined, even though the number
of equations is several times the number of unknowns. When increasing
$l_{\text{max}}$ or $N_{\text{dirs}}$ toward its problematic value,
the $\Delta_{r}$ values for \emph{all} $k$ dramatically decrease,
sharply lowering the contrast needed to locate the eigenvalues of
$k$. The solutions that are found begin to attempt to set the field
to zero in the entire cavity region, with some regions of layer 0
outside the cavity having field intensities that are orders of magnitude
larger than the field inside the cavity. In our experience this problem
has not occurred in the two-basis method for $l_{\text{max}}$ near
the semiclassical limit of $l$, $\text{Re}(kR)$, and thus has never
been a practical annoyance. On the other hand, the problem can occur
at surprisingly low values of $N_{\text{dirs}}$. Yet taking $N_{\text{dirs}}$
to be too low can often cause solutions to simply not be found: dips
in the graph of $\Delta_{r}$ vs. $\text{Re}k$ can simply disappear.
The window of good values of $N_{\text{dirs}}$ can be at least as
narrow as $N_{\text{dirs}}/10$. The window of good $l_{\text{max}}$
values for the MB seems to be much wider (for many modes, it is sufficient
to take $l_{\text{max}}$ to be half of $\text{Re}(kR)$ or less).
In this respect, the two-basis method is easier to use than the Bessel
wave method.

On the other hand, there were situations when a scan of $\Delta_{r}$
versus $\text{Re}k$ with the Bessel wave method revealed a mode that
was skipped in the same scan with the two-basis method, due to the
narrowness of the dip feature. To the best of our recollection, we
have always been able to find a mode by both methods if we have made
an effort to search for it.

A direct comparison of the methods by looking at the solution plots
rarely reveals field value differences greater than 1\% of the maximum
value when the modes are restricted to those that do not have a large
high-angle component. The eigenvalues of $k$ located by the two methods
are usually quite close. Here we show one case in which there is a
small but visible difference between the mode plots for the Bessel
wave method and the two-basis method.

\begin{figure}[!htb]
\begin{center}\includegraphics[%
  width=0.55\columnwidth]{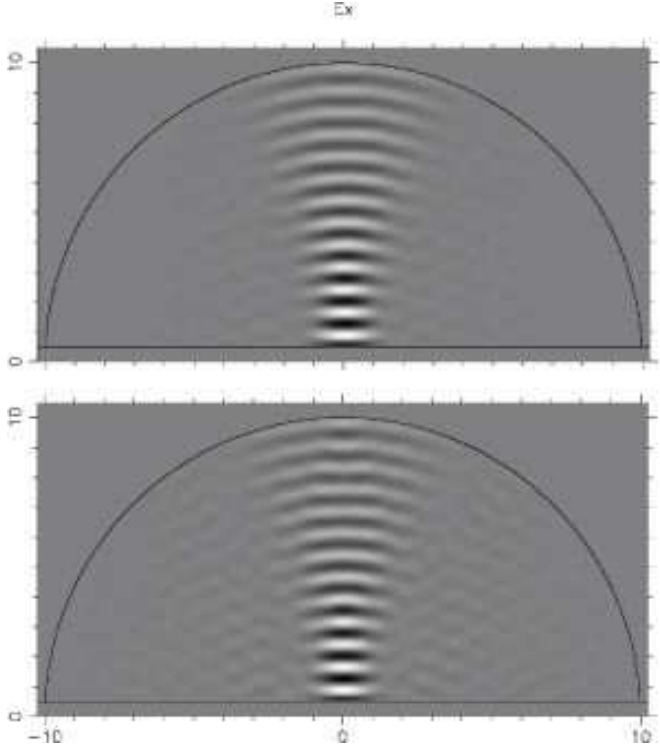}\hspace*{-1.8cm}\end{center}

\caption{\label{fig: dielectric compare} Top: Solution obtained with Bessel
wave method. $k=8.57073-\imath0.05857$; $\Delta_{n}=2.8\times10^{-5}$,
$N_{\text{dirs}}=30$; $\alpha_{k}^{(0)}$ distribution is uniform
in $\Delta(\sin\alpha_{k}^{(0)})$, not $\Delta(\alpha_{k}^{(0)})$.
Bottom: Solution obtained with two-basis method. $k=8.57086-\imath0.05900$;
$\Delta_{n}=2.9\times10^{-5}$; $l_{\text{max}}=86$. The field in
layer X is not plotted.}
\end{figure}

Figure \ref{fig: dielectric compare} shows $E_{x}$ in the $x$-$z$
plane for a mode in a cavity with $n_{0}=1.0$ and $n_{X}=0.5$ and
no stack layers at all. This situation is meant to model a dielectric-filled
dome cavity surrounded by air. (Our program assumes that layer 0 is
free space ($n_{0}=1$), so we have set $n_{X}<n_{0}$ to achieve
this effect.) Here M2 is spherical and centered at the origin with
$R=10$ and $z_{e}=z_{1}=0.5$. The reason for the disagreement between
the two methods for this case is not known. This was the only low
finesse cavity we have tried, as well as being the only cavity with
a {}``high'' index of refraction in layer 0. Another unusual property
of this mode solution, which looks similar to a fundamental Gaussian,
is that its electric field in the $x$-$y$ plane actually spirals:
its instantaneous linear direction rotates with $z$ as well as with
$t$. The other modes we show throughout the demonstrations section
give extremely good agreement between the two primary methods.

In our implementation, computation time to set up and solve $A\bm{y}=\bm{b}$
is of the same order for the Bessel wave method and the two-basis
method (with variant 1). For the large cavity shown in Fig. \!\ref{fig: V mode},
the Bessel wave method took about 10 s with $N_{\text{dirs}}=100$
and about 600 locations chosen on M2. Variant 1 with $l_{\text{max}}=200$,
600 M2 locations, and 600 uniformly spaced $\alpha_{k}^{(0)}$, took
about 15 s. For comparison, Variant 2 took about 2 hours (achieving
a $10^{-6}$ relative accuracy for each $\alpha_{k}^{(0)}$ integral).

\subsection{\label{sec: real coefficients}Almost-Real MB Coefficients}

When M1 is a conductor we find that the imaginary parts of $a_{l}$
and $b_{l}$ are essentially zero. Furthermore, we often find that
for dielectric M1 mirrors, the imaginary parts of $a_{l}$ and $b_{l}$
are one or more orders of magnitude smaller than their real parts.
We show one way in which this tendency simplifies the interpretation
of mode polarization in Appendix \ref{sec: modetype}.

There is a an argument that suggests that $a_{l}$ and $b_{l}$ should
be real for a conducting cavity. A conducting cavity has a real eigenvalues,
$k$, and real reflection functions, $r_{s/p}=-1$. It is not difficult
to see that for real $k$ the three types of M2 equations, $E_{\phi}=0$,
$E_{\parallel}=0$, and $H_{\perp}=0$, can be satisfied with real
$a_{l}$ and $b_{l}$. The seed equation of course can also be chosen
so that it can be satisfied by real variables. The M1 equations, because
of the complex nature of $\bar{r}_{s/p}=r_{s/p}\exp(-\imath2z_{1}kn_{0}\cos\alpha_{k}^{(0)})$
when $z_{1}\neq0$, are not obviously satisfied by real $a_{l}$ and
$b_{l}$. However, for a conducting cavity, the M1 boundary condition
can be given as a real space boundary condition (like M2) instead
of a Fourier space boundary condition. In this case, locations on
M1 are chosen just as they are on M2. The same argument that the M2
equations can be satisfied by real $a_{l}$ and $b_{l}$ now applies.
Thus the M1 boundary condition for a conducting cavity should not,
in any basis, force $a_{l}$ and $b_{l}$ to be complex.

\section{\label{sec: conclusions}Conclusions}

The two methods presented in this communication have proved to be
useful tools for modeling small optical dome-shaped cavities. As yet
we have no general verdict on which method is best. Overall, it was
quite helpful to have both methods available for finding all of the
modes. Once a mode was found, agreement of both eigenvalue and eigenmode
between the two methods was quite good.

The primary advantage of both of these methods is the treatment of
the dielectric stack through its transfer matrices. First of all,
including the dielectric stack in the model is essential to obtain
a number of qualitative features of the modes. A model with a mirror
of constant phase shift is inadequate, except in special cases such
as the fundamental Gaussian mode in its paraxial regime. In addition,
including the stack in the model allows one to calculate the field
inside the stack, which may be the location of interest for the engineer
of an optical device. Secondly, the use of the transfer matrices is
the most natural way to include a dielectric stack that is of large
lateral extent compared to the wavelength. The only alternative that
we know of would involve a separate basis for each stack layer, resulting
in an increase of \emph{both} dimensions of the matrix $A$ by about
$N$ times, where $N$ is the number of stack layers.

Another advantage of the methods we have presented is their speed.
Each point in Fig.~\ref{fig: folgraphs}, which was generated in
a few hours or less, required $A\bm{y}=\bm{b}$ to be constructed
and solved 20--60 times for a cavity with $R\approx50\lambda$. The
combination of using a basis expansion method, taking advantage of
cylindrical symmetry, using one basis set for all the layers, avoiding
numerical integrals, and using C++ \texttt{inline} functions for complex
arithmetic has resulted in a fast implementation.

Several {}``stack-induced'' effects that we have demonstrated here
will be more thoroughly treated in a separate publication. The splitting
of the fundamental Gaussian (and other Gaussian modes) into a non-axisymmetric
{}``V'' shape may be of practical interest to workers who wish to
achieve a tight focus. The $m=0$ modes, which have the greatest potential
for tight focusing, are also of interest. The ability to analyze the
persistent stack-induced mixing is a result that stands in its own
right. This stack effect persists arbitrarily far into the paraxial
limit; it exists as long as the resonance widths of both the cavity
and the laser source are smaller than the splitting of the near-degenerate
pair. This effect should exist for a wide range of cavity lengths,
and is of interest to anyone engineering higher-order Gaussian modes.

Perhaps the greatest limitation to the practical application of our
methods is its rigid treatment of the curved mirror as a conductor.
As mentioned in the Introduction, it is reasonable to expect highly
focused modes, such as those shown in Figs.~\ref{fig: V mode}, \ref{fig: zoom}
and \ref{fig: m0 z 0.05}, to undergo little change when the curved
conducting mirror is exchanged for a dielectric one (with an appropriate
change of mirror radius by $-\lambda/4<\delta<\lambda/4$ to account
for a phase shift). This is because the local wave fronts are primarily
perpendicular to the curved mirror for such modes (imagine a Wigner
function evaluated at the surface of M2). For paraxial geometries
such as those in our discussion of stack-induced mixing, this argument
says nothing about how replacing M2 with a curved dielectric stack
will affect the modes. A pursuit of this question may require a more
brute force approach.

Despite all limitations, we find our methods to be extremely versatile
and powerful (fast and allowing for relatively large cavities). The
full vector electromagnetic field is used. Exactly degenerate modes
can be separated. The shape of the curved mirror is arbitrary and
we have implemented parabolic shapes as well as spherical shapes with
no difficulties. A first-try scan of $\Delta_{r}$ versus $\text{Re}k$
using either method with a non-contrived seed equation will locate
the vast majority of modes that exist. Reasonable results are obtained
when the interior of the cavity is a dielectric. Hopefully, the explicit
development of the two methods given here can benefit a number of
workers in optics-related fields.

\section*{Note Added in Proof}

We note that we have not found significant stack-induced effects (the
V mode of Section \ref{sec: V} and the mixing of Section \ref{sec: Laguerre})
for a {}``standard'' quarter-wave dielectric stack which has design
ABAB$\ldots$AB with front surface A and $n_{A}>n_{B}$. If $n_{A}<n_{B}$,
we do find both types of stack-induced effects. The standard ($n_{A}>n_{B}$)
quarter-wave layer structure exhibits less variation in $\arg(r_{s/p}(\alpha_{k}))$
than our mirrors (cf.~Fig.~\ref{fig: phase II}), and therefore
behaves much like a perfect conductor except at high angles of incidence.
The methods presented here, which model the curved dome mirror (M2)
as a conductor, should therefore carry over to standard dielectric
M2 mirrors. The planar mirror will in general have a {}``functional''
layer design, leading to the stack-induced effects reported above.

\section*{Acknowledgments}

We would like to thank Prof. \!Michael Raymer for getting us interested
in the realistic cavity problem. This work is supported by the National
Science Foundation through CAREER Award No. \!0239332. \appendix

\section{\label{sec: model limitations}Further Explanations and Limitations
of the Model}

\subsection{\label{sec: exclusion high-angle} Exclusion of High-Angle Plane
Waves}

As mentioned in the Introduction, in the usual application of a basis
expansion method, the field is expanded in each dielectric region
separately and henceforth each region gets its own complete basis
and its own set of coefficients. However, our methods, when {}``cast''
into the PWB, use a single set of plane waves, complete in layer 0,
which are propagated down into the other layers via the transfer matrices.
There are several advantages to this approach. The Bessel wave method
becomes simpler, as there are far fewer unknowns. This approach is
also ideal for incorporating the use of the MB for layer 0, as in
the two-basis method, although a straightforward use of the MB for
layer 0 and a separate PWB for each layer $q$, $q>0$, could be implemented.

One drawback to our approach is that certain high-angle or evanescent
plane waves are not included. No plane wave basis vectors are allowed
for which $n_{q}\sin\theta_{k}^{(q)}>n_{0}$. If the true quasimode
expansion in any layer has significant weight for these high-angle
or evanescent waves, the calculated solution of the field could be
erroneous.

Probably the best way to determine whether or not such intrinsic error
is present at a significant level is to look at the plane wave distribution
in layer 0 (using (\ref{eqn: scalar MB to PWB conversion}) or (\ref{eqn: vector MB to PWB conversion})
to get this if one is using the two-basis method) and to see whether
the distribution dies off as $\alpha_{k}^{(0)}$ approaches $\pi/2$.
If it does, the solution should be reasonably error free.

We note here that Berry \cite{Berry} has shown how evanescent waves,
in a finite region near the origin, can be expressed in the PWB (using
plane waves with real-valued directions, as usual). Thus the fact
that no evanescent waves are included in layer 0 may not be a problem
in itself for depicting the field in layer 0.

\subsection{\label{sec: hat brim} The Hat Brim}

\subsubsection{The infinitesimal hat brim}

\begin{figure}[!htb]
\begin{center}\includegraphics[%
  width=0.75\columnwidth]{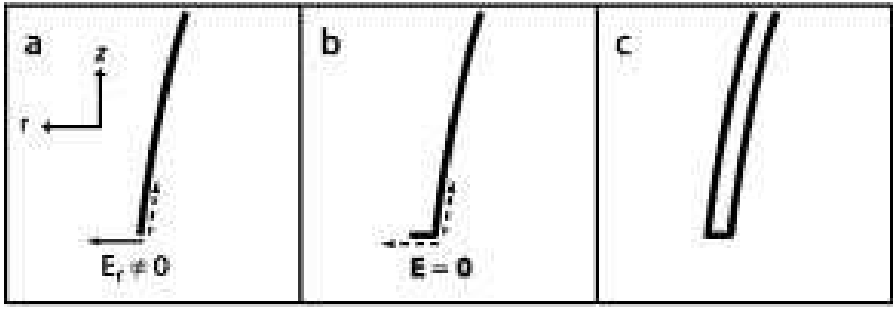}\hspace*{-1cm}\end{center}

\caption{\label{fig: tiny hat brim} Modeling the edge of the dome.}
\end{figure}

For the vector field solution, a tiny hat brim must be added to M2
to correctly model a conducting edge. Figure \ref{fig: tiny hat brim}
shows a cross section of the edge of M2 for increasingly better approximations
of a real conducting mirror. In (a), $E_{\rho}$ can be nonzero at
the edge, which is unphysical. In (b), a small hat brim has been added
and $\bm{E}=\bm{0}$ at the inside edge, as it should be. In (c),
locations are chosen on a (closed) conductor with finite thickness.
As the thickness of M2 becomes insignificant compared to $\lambda$,
model (b) should approximate model (c) arbitrarily closely. Usually
2--10 locations are chosen on the hat brim (the density of course
is \emph{much} greater on the hat brim than on the dome). We have
taken $\omega_{b}\approx0.0001\mu\text{m}$ for our demonstrations.

\subsubsection{The infinite hat brim and the 1-D half-plane cavity}

The way in which our model includes the interior of the cavity and
the entire $z>z_{1}$ half-plane in a single region, layer 0, is somewhat
unorthodox. We believe our method produces the correct field in a
finite region surrounding the cavity, but not at $z\rightarrow\infty$.

A somewhat less strange model is obtained if one imagines an infinite
hat brim. The upper half-plane is then no longer in the problem. Layer
0 still extends infinitely in $\rho$ as do the other layers, but
the vertical confinement makes it easier to see that this is an eigenproblem
and not a scattering problem. Unfortunately our implementation does
not allow an infinite hat brim (unless M1 is a conductor and $z_{e}=z_{1}$,
for which the hat brim condition is already set in the M1 equations).
At least for some solutions, taking $w_{b}\approx R$ and giving the
hat brim roughly the same number of locations as the dome, produces
no visible changes in the mode structure (from the solution obtained
with a tiny hat brim). The mode shown in Fig. \!\ref{fig: V mode}
is one of these.

\begin{figure}[!htb]
\begin{center}\includegraphics[%
  width=0.75\columnwidth]{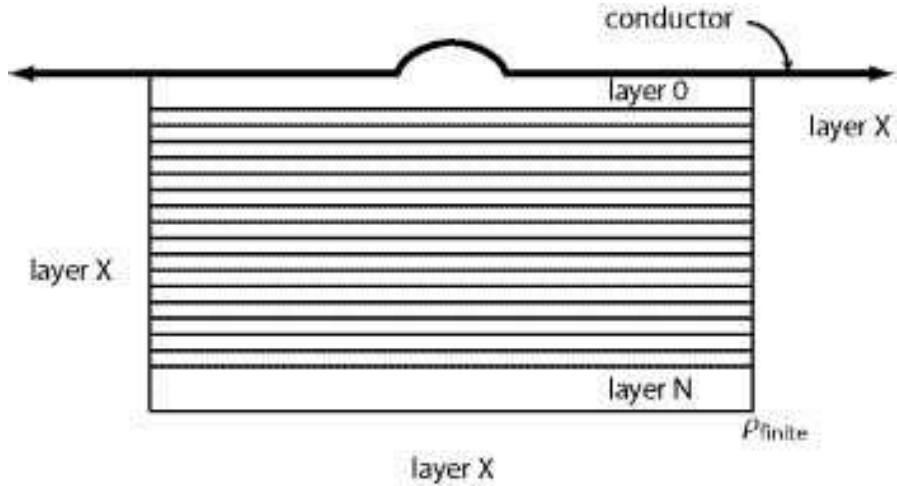}\hspace*{-1cm}\end{center}

\caption{\label{fig: QNM complete} Including the sides of the stack at radius
$\rho_{\text{finite}}$.}
\end{figure}

If we allow that some solutions are not affected by the value of $w_{b}$,
we can take the infinite hat brim more seriously. The problem becomes
similar in some respects to the 1-D half-plane problem, in which $x<0$
is a conductor, $x>L$ has a constant refractive index, and the $[0,L]$
region is segmented into several 1-D dielectric layers. The solutions
of the 1-D problem form a set of quasimodes, (or quasinormal modes),
which are \emph{complete} in $[0,L]$ and obey an orthogonality condition
on the same interval. The conditions for completeness, and the characterization
of the incompleteness of the quasinormal mode for problems which do
not meet the completeness conditions, are discussed in Ching \emph{et
al.} \cite{Ching} and Leung \emph{et al.} \cite{Leung}. The model
with the infinite hat brim does not rigorously meet the the completeness
condition (for 3-D cavity resonators). We note that it appears from
the references mentioned above that the situation depicted in Fig.
\!\ref{fig: QNM complete} would meet the completeness condition,
with the quasinormal modes being complete in layers $0$--$N$. We
do not attempt to resolve this further, as the use of our mode solutions
as a basis for time-dependent problems is beyond the scope of what
we have been trying to accomplish.

\section{\label{sec: modetype}Negative $m$ Modes and Sine and Cosine Modes}

The cylindrical symmetry group consists of $\phi$-rotations and reflections
about the $x$-$z$ or $y$-$z$ planes. For solutions with a fixed,
nonzero $m$, rotations are equivalent to multiplying the field by
a complex phase, which is of course equivalent to a translation in
time. (This means that, when looking at any cross section plot in
Section \ref{sec: demonstrations} for a fixed $m\neq0$, the time
evolution can be realized by simply rotating the figure: counter-clockwise
for $m>0$ and clockwise for $m<0$.) Thus we say that for $m\neq0$,
reflections generate a new solution but rotations do not. (If we were
considering the sine and cosine modes discussed later in the section,
instead of the $\pm m$ modes, we would say that a $\pi/(2m)$ rotation
generates a new solution but a reflection does not.) Hence the symmetry
of the cavity causes modes to come in truly degenerate pairs. It can
be shown that a reflection is equivalent to taking $m$ to $-m$ (up
to a $\pi$ rotation). For $m\neq0$, general complex linear superpositions
of the $\pm m$ pair of modes are equivalent to arbitrary complex
superpositions of any number of $\phi$-rotations, reflections, and
combinations of these acting on a single $+m$ mode. We note that
since our methods solve $\bm{y}$ for a fixed $m$, we are able to
separate the $\pm m$ pairs.

The $m=0$ modes also come in degenerate pairs, but the interrelation
can be more complicated than a reflection. The $m=0$ pairs can be
separated easily in the two-basis method by choosing a seed equation
that contains only $a_{l}$ or only $b_{l}$ coefficients, since the
M1 and M2 equations do not couple $a_{l}$ and $b_{l}$ coefficients
if $m=0$. Thus all of the true (non-accidental), exact (not obtained
only in a limit, such as the paraxial limit) degeneracies that exist
can be separated.

In constructing the M1 and M2 equations, it is easiest to write code
that assumes that $m\geq0$. For presentations and movies, it is nice
to be able to plot both $+m$ and $-m$ solutions, as well as the
linear combinations adding and subtracting these modes. All four of
these modes can be plotted from a solution $\bm{y}_{+m}$ that has
been been found using the positive azimuthal quantum number $+m$.
Here we briefly discuss how to do this (for $m\neq0$).

\subsection{Plotting with \emph{-}\textrm{\emph{m}}}

To plot the $-m$ modes, we use simple rules, allowed by the M1 and
M2 equations, to create a new solution vector $\bm{y}_{-m}$ from
the solution $\bm{y}_{+m}$. To obtain the field, the new coefficients
from $\bm{y}_{-m}$ are simply inserted into the various expansion
(or basis conversion) equations from Sections \ref{sec: PWB} and
\ref{sec: MB}, using $(-m)$ as the {}``$m$'' argument in these
equations. (All of these equations work for negative values of their
$m$ argument.) The relations $Y_{l,-m}=(-1)^{m}Y_{lm}^{*}$, $J_{-m}=(-1)^{m}J_{m}$,
$f_{l,-m}=(-1)^{m}f_{lm}$, and $g_{l,-m}=-(-1)^{m}g_{lm}$ are useful.
The rules for the new solution vectors are given in Table \ref{tab: m to -m rules}.
(There is a certain arbitrariness to these rules, since $(\alpha\bm{y})$,
with $\alpha$ being an arbitrary constant, satisfies the M1 and M2
equations.)

\begin{table}[!htb]

\caption{\label{tab: m to -m rules} Rules for taking $\bm{y}_{+m}\rightarrow\bm{y}_{-m}$.}

\begin{tabular}{l|c|c}
&
 PWB &
 MB \tabularnewline
\hline
scalar: \,&
 \,$\psi_{u,-m}=(-1)^{m}\psi_{u,+m}$\,&
 \,$c_{l,-m}=c_{l,+m}$\tabularnewline
vector: \,&
 \,$S_{u/d,-m}=-(-1)^{m}S_{u/d,+m}$\,&
 \,$a_{l,-m}=a_{l,+m}$\tabularnewline
&
 $P_{u/d,-m}=(-1)^{m}P_{u/d,+m}$&
 \,$b_{l,-m}=-b_{l,+m}$\\
\tabularnewline
\end{tabular}
\end{table}

\subsection{Plotting cosine and sine modes}

We can define the {}``cosine'' and {}``sine'' modes as $X_{(c)}\equiv(X_{(+m)}+(-1)^{m}X_{(-m)})/2$,
and $X_{(s)}\equiv(X_{(+m)}-(-1)^{m}X_{(-m)})/2$, where $X$ stands
for $\psi$, $\bm{E}$, or $\bm{H}$. By adding explicit expansion
expressions for the $+m$ and $-m$ modes, one can obtain expressions
for $X_{(c)}$ and $X_{(s)}$. For example, using (\ref{eqn: VSH expansion of E, H}),
(\ref{eqn: M and N explicit}), and the above rules one finds that
\begin{multline}
\bm{E}_{(c)}^{(0)}(\bm{x},t)=e^{-\imath\omega t}\sum_{l}\Biggl\{\hat{\bm{r}}\biggl[a_{l}\frac{-l(l+1)}{kn_{0}r}j_{l}\text{Re}Y_{lm}\biggr]\\
+\hat{\bm{\theta}}\biggl[a_{l}\frac{-1}{kn_{0}r}\frac{\partial}{\partial r}(rj_{l})\frac{\partial}{\partial\theta}(\text{Re}Y_{lm})+b_{l}\frac{-m}{\sin\theta}j_{l}\text{Re}Y_{lm}\biggr]\\
+\hat{\bm{\phi}}\biggl[a_{l}\frac{m}{kn_{0}r\sin\theta}\frac{\partial}{\partial r}(rj_{l})\text{Im}Y_{lm}+b_{l}j_{l}\frac{\partial}{\partial\theta}(\text{Im}Y_{lm})\biggr]\Biggr\},\notag\end{multline}
 where $a_{l}$ and $b_{l}$ come from the original $\bm{y}_{+m}$.
When $a_{l}$ and $b_{l}$ are predominately real (see Section \ref{sec: real coefficients}),
we can see from the above equation that the real-valued physical fields
$E_{\rho}$ and $E_{z}$ are proportional to $\cos m\phi\cos\omega t$
while $E_{\phi}$ is proportional to $\sin m\phi\cos\omega t$. (The
cosine time-dependence of $\bm{E}$ is why we call this the cosine
mode.) Thus for the cosine mode with $|m|=1$, the transverse portion
of $\bm{E}$ has an average linear polarization along the $x$ axis
(note $\langle E_{y}\rangle_{\phi}=0\;\forall t$), as opposed to
the average circular polarization the $m=\pm1$ modes would have.
(For the $m=+1$ mode, $\langle\bm{E\cdot}\hat{\bm{q}}\rangle_{\phi}=0\;\forall t$
where $\hat{\bm{q}}\equiv-\hat{\bm{x}}\sin\omega t+\hat{\bm{y}}\cos\omega t$.)
Due to their separating of time and $\phi$-dependence for the physical
fields, the sine and cosine modes are very useful final forms of the
field (when $a_{l}$ and $b_{l}$ are predominately real).

\section{\label{sec: stacks} Stacks used in Section \ref{sec: demonstrations}}

\begin{figure}[!htb]
\begin{center}\includegraphics[%
  width=0.70\columnwidth]{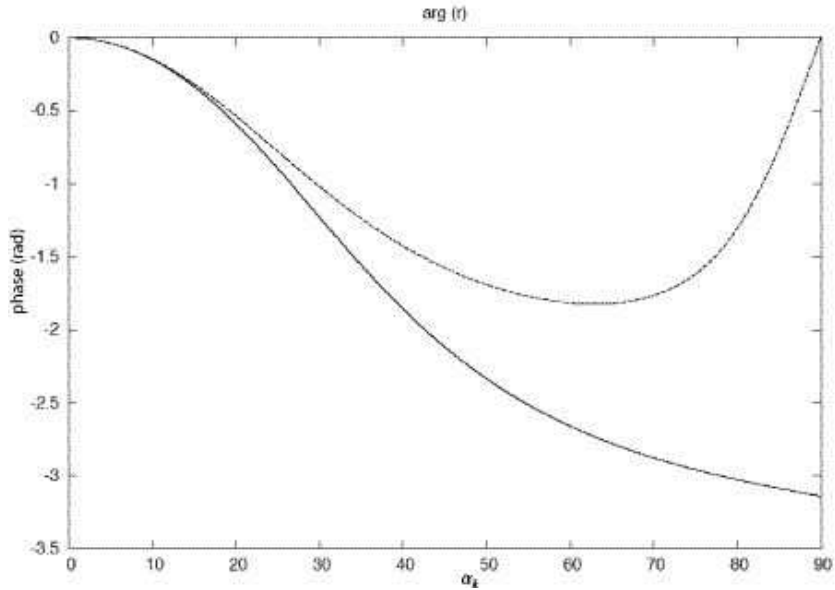}\hspace*{-1cm}\end{center}

\caption{\label{fig: phase II} Plane wave reflection phases (rad.) of stack
II ($N_{s}=20$) as a function of $\alpha_{k}$ (deg.). The solid
(dashed) line is for s (p) polarization. The wavelength of the plane
wave, $\lambda_{\text{test}}$, is set at $\lambda_{s}$. The graph
for stack I is similar.}
\end{figure}

Stacks I and II are similar to Al$_{1-x}$Ga$_{x}$As--AlAs stacks
that Raymer has used experimentally \cite{Raymer}. Figure \ref{fig: phase II}
shows the reflection phases for stack II. The stack parameters that
are varied in our demonstrations are $N_{s}$ and $k_{s}\equiv2\pi/\lambda_{s}$.
The meaning of these parameters can be inferred from the stack definitions
below. The normal reflection coefficient for stack II with $N_{s}=20$
is $|r_{s/p}(\alpha_{k}=0)|=0.9964$. For $N_{s}=22$, $|r_{s/p}|=0.9981$.

Stack I: $n_{0}=n_{X}=1$; $n_{1}=n_{3}=\ldots=n_{(2N_{s}-1)}=3.003$;
$n_{2}=n_{4}=\ldots=n_{(2N_{s})}=3.51695$. Layers 1--$(2N_{s})$
are quarter-wave layers (optical thickness $=\lambda_{s}/4$).

Stack II: $n_{0}=n_{X}=1$; $n_{1}=n_{3}=\ldots=n_{(2N_{s}+1)}=3.51695$;
$n_{2}=n_{4}=\ldots=n_{(2N_{s})}=3.003$. Layers 2--$(2N_{s}+1)$
are quarter-wave layers. Layer 1 is a spacer layer that has optical
thickness $1\lambda_{s}$.


\begin{thebibliography}{10}
\bibitem{Hall}P. L. Greene and D. G. Hall, {}``Focal shift in vector beams'',
Opt. Expr. $\bm{4}$ (1999) 411. 
\bibitem{Hall2}P. L. Greene and D. G. Hall, {}``Properties and diffraction of vector
Bessel-Gauss beams'', J. Opt. Soc. Am. A $\bm{15}$ (1998) 3020. 
\bibitem{Sheppard}C. J. R. Sheppard and P. T\"{o}r\"{o}k, {}``Efficient calculation
of electromagnetic diffraction in optical systems using a multipole
expansion'', J. Mod. Opt. $\bm{44}$ (1997) 803. 
\bibitem{Dorn}R. Dorn, S. Quabis, and G. Leuchs, {}``Smaller, sharper focus for
a radially polarized light beam'', Phys. Rev. Lett. $\bm{91}$ (2003)
233901. 
\bibitem{Jens}J. U. N\"{o}ckel, G. Bourdon, E. Le Ru, R. Adams, I. Robert, J-M.
Moison, I. Abram, {}``Mode structure and ray dynamics of a parabolic
dome microcavity'', \mbox{Phys. Rev. E} \textbf{62} (2000) 8677. 
\bibitem{MEPchapter}J.U. N\"{o}ckel and R.K. Chang, ''2-d Microcavities: Theory and
Experiments'', in \emph{Cavity-Enhanced Spectroscopies}, Roger D.
van Zee and John P. Looney, eds. (Experimental Methods in the Physical
Sciences \textbf{40}), Academic Press, San Diego (2002), 185--226. 
\bibitem{Siegman}A. E. Siegman, \emph{Lasers}, University Science Books, Sausalito,
CA, 1986, pp. 626--97, 744--76. 
\bibitem{Raymer}M. G. Raymer (colloquium, University of Oregon, 4/24/03, and private
communication). 
\bibitem{Meissner}M. Aziz, J. Pfeiffer, and P. Meissner, {}``Modal Behaviour of Passive,
Stable Microcavities'', Phys. Stat. Sol. (a) \textbf{188} (2001)
979. 
\bibitem{Jackson}J. D. Jackson, \emph{Classical Electrodynamics, Third Edition}, John
Wiley \& Sons, Inc., New York, 1999, pp. 107--10, 113--4, 426--32,
471--73. 
\bibitem{Yeh}P. Yeh, \emph{Optical waves in layered media}, Wiley, New York, 1988,
pp. 60--64, 102--15. 
\bibitem{recipes}W. H. Press, S. A. Teukolsky, W. T. Vetterling, B. P. Flannery, \emph{Numerical
Recipes in C: The Art of Scientific Computing, Second Edition}, Cambridge
University Press, Cambridge, 1992, pp. 59--70, 359--62. 
\bibitem{Arfken}G. B. Arfken and H. J. Weber, \emph{Mathematical Methods for Physicists,
Fifth Edition}, Harcourt Academic Press, San Diego, 2001, p. 767. 
\bibitem{Videen}G. Videen, {}``Light Scattering from a particle on or near a perfectly
conducting surface'', Optics Communications \textbf{115} (1995) 1. 
\bibitem{Beijersbergen}M. W. Beijersbergen, L. Allen, H. E. L. O. van der Veen, and J. P.
Woerdman, \char`\"{}Astigmatic laser mode converters and transfer
of orbital angular momentum\char`\"{}. Optics Communications \textbf{96}
(1993) 123. 
\bibitem{Berry}M. V. Berry, {}``Evanescent and real waves in quantum billiards and
Gaussian beams'', \mbox{J. Phys. A:} \mbox{Math. Gen.} \textbf{27}
(1994) L391. 
\bibitem{Ching}E. S. Ching, P. T. Leung, A. Maassen van den Brink, W. M. Suen, S.
S. Tong, K. Young, {}``Quasinormal-mode expansion for waves in open
systems'', Rev. Mod. Phys. \textbf{70} (1998) 1545. 
\bibitem{Leung}P. T. Leung, S. Y. Liu, and K. Young, {}``Completeness and orthogonality
of quasinormal modes in leaky optical cavities'', \mbox{Phys. Rev.
A} \textbf{49} (1994) 3057.\end{thebibliography}
\end{document}